\documentclass[journal]{IEEEtran}
\usepackage{amsmath,etoolbox}
\usepackage{mathtools, cuted}
\usepackage{setspace}
\usepackage{flushend, cuted}
\usepackage{lipsum}
\usepackage{lingmacros}
\usepackage{tree-dvips}
\usepackage{varioref}
\usepackage{colortbl}
\usepackage{url}
\usepackage{hyperref}
\usepackage{balance}

\usepackage{graphicx}
\usepackage{epstopdf}
\usepackage{color,soul}
\usepackage{amsfonts}
\usepackage{subfigure}
\usepackage{multicol}
\usepackage{multirow}
\usepackage{hhline}
\usepackage{amssymb}
\usepackage{newlfont}
\usepackage{mathrsfs}
\usepackage{mathtools}
\usepackage{mathrsfs}
\usepackage{cite}
\usepackage{multirow}
\usepackage{nomencl}   
\usepackage{fancyhdr}
\usepackage{caption}
\usepackage{slashbox}
\usepackage{float}
\usepackage{placeins}
\usepackage{tabularx}
\usepackage[table]{xcolor}
\usepackage[linesnumbered,ruled]{algorithm2e}
\SetKwRepeat{Do}{do}{while}%

\DeclareRobustCommand{\frac}[3][0pt]{%
	{\begingroup\hspace{#1}#2\hspace{#1}\endgroup\over\hspace{#1}#3\hspace{#1}}}

\ifCLASSINFOpdf
\else
\fi
\begin{document}

\title{Incorporating Gradient Similarity for\\ Robust Time Delay Estimation in\\ Ultrasound Elastography}
%
%
\author{Md~Ashikuzzaman, \textit{Graduate Student Member}, \textit{IEEE},
	    Timothy~J.~Hall, \textit{Member}, \textit{IEEE},
        and~Hassan~Rivaz, \textit{Senior Member}, \textit{IEEE}
\thanks{Md Ashikuzzaman and Hassan Rivaz are with the Department
of Electrical and Computer Engineering, Concordia University, Montreal,
QC, H3G 1M8, Canada.
 Email: m\_ashiku@encs.concordia.ca~and~hrivaz@ece.concordia.ca}
\thanks{Timothy~J.~Hall is with the Department of Medical Physics, University of Wisconsin–Madison, Madison, WI 53706, USA.
	Email: tjhall@wisc.edu}
}

%

\maketitle

\begin{abstract}
Energy-based ultrasound elastography techniques minimize a regularized cost function consisting of data and continuity terms to obtain local displacement estimates based on the local time-delay estimation (TDE) between radio-frequency (RF) frames. The data term associated with the existing techniques takes only the amplitude similarity into account and hence is not sufficiently robust to the outlier samples present in the RF frames under consideration. This drawback creates noticeable artifacts in the strain image. To resolve this issue, we propose to formulate the data function as a linear combination of the amplitude and gradient similarity constraints. We estimate the adaptive weight concerning each similarity term following an iterative scheme. Finally, we optimize the non-linear cost function in an efficient manner to convert the problem to a sparse system of linear equations which are solved for millions of variables. We call our technique rGLUE: \textbf{r}obust data term in \textbf{GL}obal \textbf{U}ltrasound \textbf{E}lastography. rGLUE has been validated using simulation, phantom, \textit{in vivo} liver, and breast datasets. In all of our experiments, rGLUE substantially outperforms the recent elastography methods both visually and quantitatively. For simulated, phantom, and \textit{in vivo} datasets, respectively, rGLUE achieves $107\%$, $18\%$, and $23\%$ improvements of signal-to-noise ratio (SNR) and $61\%$, $19\%$, and $25\%$ improvements of contrast-to-noise ratio (CNR) over GLUE, a recently-published elastography algorithm.              
\end{abstract}


\IEEEpeerreviewmaketitle
                    
\section{Introduction}
Ultrasound is one of the most frequently used medical imaging modalities since it is non-invasive, low-cost, and portable. Among numerous applications of ultrasound imaging, elastography~\cite{ophir1}, which refers to mapping the mechanical properties of the tissue, is prominent. Ultrasound elastography has successfully been applied to breast health monitoring~\cite{hall2003vivo,hybrid,jiang_2015}, characterization of breast cancer-related lymphedema~\cite{hoda_lymph}, liver tissue classification~\cite{DPAM,guest,tang2015ultrasound}, ablation monitoring~\cite{varghese2002elastographic,varghese2003elastographic,rivaz2008ablation,mariani2014real}, and cerebral imaging~\cite{Selbekk_2009}, and can broadly be classified into two classes: ``dynamic" and ``quasi-static". Dynamic elastography techniques such as shear-wave elastography (SWE)~\cite{Gallot_2011,shear_wave_2018,horeh_2019} and acoustic radiation force imaging (ARFI)~\cite{Nightingale_2002} often provide quantitative values for the tissue properties. Quasi-static elastography~\cite{ophir_91} entails generating a comparatively larger deformation with slower velocity. An advantage of quasi-static methods is that tracking larger deformation generally has higher values of signal to noise ratio (SNR) compared to small deformations in dynamic methods. However, this large deformation field increases signal-decorrelation. This work concerns free-hand palpation quasi-static elastography~\cite{hall2003vivo} which entails acquisition of ultrasound radio-frequency (RF) data while the operator creates slow deformations simply with a hand-held probe. The displacement field between the pre- and post-deformed frames is spatially differentiated to obtain the strain map which reveals the pathological tissue by showing a contrast with the healthy tissue.               

Time-delay estimation (TDE) between pre- and post-deformed frames is a non-trivial task and usually accomplished by one of the three mainstream techniques: window-based, machine learning-based, and regularized optimization-based. The window-based, otherwise known as block matching techniques~\cite{hall2003vivo,kk,hybrid,intro11,intro30,intro12}, divide the entire RF frame into several data windows and assume that all of the samples in a particular window undergo the same amount of displacement. The displacement of a window is estimated by the location of the peak normalized cross-correlation~\cite{ophir1,intro15,kuzmin2015} or zero phase-crossing~\cite{intro12,intro18}. Window-based tracking can be performed either in only axial direction~\cite{ophir1,intro24} or both axial and lateral~\cite{intro26,intro27} directions. $2D$ tracking improves the displacement estimate in the axial direction, but the lateral displacement field is less accurate due to several reasons. First, the lateral ultrasound echo-signal lacks a carrier~\cite{intro26,mirzaei20203d}. Second, the lateral direction often exhibits a wider point spread function (PSF)~\cite{qiong_2017,mirzaei20203d}. Third, the sampling rate in the lateral direction is substantially lower than the axial direction~\cite{jianwen_2009,mirzaei20203d}. A compromise made in window-based techniques is between the displacement accuracy and the spatial resolution. Higher accuracy can be achieved by dividing the RF data into large windows. A typical selection of window size is at least ten times the ultrasound wavelength. Although a large correlation window facilitates the tracking technique to provide better accuracy by reducing the estimation variance commonly known as the ``jitter error", it induces signal decorrelation due to the non-stationary nature of the RF data. This issue can be alleviated by choosing smaller $2D$ windows, with a potential loss of estimation accuracy. Recently, TDE using $3D$ correlation windows (with time as the third dimension) has been proposed by our group~\cite{mirzaei20203d} for $2D$ displacement estimation. In addition, some techniques have coupled adaptive Bayesian optimization framework with block-matching approach to facilitate better displacement tracking~\cite{al2020locally,McCormick_2011,byram2012bayesian,dumont2016improving}. Furthermore, beam-steered ultrasound datasets have been taken into account to obtain high-quality lateral strain image~\cite{beam_steer1}. Since tumors and cysts can be of different complicated shapes, $2D$ elastograms do not provide the complete information regarding the shape or size of the pathological tissue. To resolve this issue, $3D$ elastography techniques that utilize volumetric ultrasound data have been developed~\cite{fisher2010volumetric,rivaz2008ablation,WANG20181638,papadacci20163d,wang_3d}. 

The second TDE class utilizes machine learning-based approaches. In~\cite{kibria2018gluenet}, our group has addressed the well-known issue of signal decorrelation by adapting FlowNet 2.0~\cite{flownet2}, a convolutional neural network (CNN) for ultrasound displacement estimation. An implicit strain reconstruction technique based on convolutional neural network has been proposed in~\cite{gao2019learning,wu2018direct}. In addition, a neural network-based technique for the automatic selection of the suitable frames for ultrasonic strain estimation has been introduced in~\cite{zayed2019automatic}. Another technique~\cite{peng2020neural} has retrained three existing networks named FlowNet 2.0~\cite{flownet2}, PWC-Net~\cite{pwcnet}, and LiteFlow-Net~\cite{liteflownet} with simulation datasets and tested the performance on real datasets. Along this line, to enable the multi-resolution pyramidal framework work for ultrasound RF tracking, two deep networks called MPWC-Net and RFMPWC-Net stemmed from the original PWC-Net~\cite{pwcnet} have been proposed in~\cite{ali_2020}. Although the recently introduced CNN-based techniques are promising for high-quality strain imaging, they are highly data-dependent which is a major drawback due to the limited availability of medical imaging data.

The third TDE class entails optimization of a cost function~\cite{rpca_glue,intro14,DPAM,mglue,soul,soulmate} that includes amplitude similarity and regularization terms. The displacement estimation is performed by optimizing a cost function consisting of data amplitude similarity term and a regularization term. The most attractive attribute of this approach is its ability to obtain spatially smooth displacement map. However, this tracking scheme is known to be computationally burdensome. This issue of computational complexity has been resolved by the Dynamic Programming (DP)~\cite{DP,intro14} technique. Since the DP integer displacement estimate is not sufficient to provide an accurate and spatially smooth displacement map, Dynamic Programming Analytic Minimization~(DPAM)~\cite{DPAM} has been proposed to refine the DP estimate by an efficient optimization technique where TDE of all samples of an RF line is calculated simultaneously. Due to the discontinuity between the displacement estimates of two laterally adjacent samples, DPAM exhibits vertical striking artifacts in the displacement image. To address this issue, GLobal Ultrasound Elastography~(GLUE)~\cite{GLUE} has been proposed where the DP initial estimate is refined by optimizing a non-linear cost function for all samples of the entire RF frame. Although GLUE is capable of producing spatially smooth displacement map, it does not take the temporal dimension into account. To investigate the prior information of temporal continuity, Spatio-Temporal Global Ultrasound Elastography~(GUEST)~\cite{guest} has been proposed where we consider three RF frames instead of two to formulate the cost function. In addition, to avoid the over-smoothing induced by the quadratic regularization function used in GLUE and GUEST, total variation regularization has been proposed in~\cite{overwind}. Furthermore, principal components of the displacement field have been investigated in~\cite{pohlman2018dictionary,abdel_embc} to reduce the execution time of GLUE. In~\cite{tauhidul_2018}, multi-scale pyramidal approach has been adopted to refine the DP integer estimate.

The aforementioned regularized optimization-based elastography techniques set the sample amplitude similarity as the data term, and the gradient similarity remains unexploited. Ultrasound RF frames often contain outlier samples stemming from numerous sources. Thermal noise can create outliers especially at low electronic SNR regimes. Another source of outlier samples is speckle decorrelation stemming from various sources such as out-of-plane motion of the sonographer’s hand and target deformation during strain elastography. In addition, the high local temperatures during ablation therapy can create microbubbles, which can create nonlinear outlier samples. Moreover, outliers can be observed due to attenuation of ultrasound beam~\cite{bora_2017} while travelling through an attenuating media such as rib bone or skull. Similar phenomenon is observed in synthetic aperture imaging where acoustic energy is low. Since the outlier samples can create large artifacts, they need to be handled during the optimization procedure to obtain an accurate displacement map. However, the sample amplitude alone is not capable of devising a robust data term. Computer vision literature~\cite{brox_2004,bruhn_2005,xu_2012} suggests that incorporation of image gradient similarity along with the sample amplitude similarity makes the data function robust to outliers. In this paper, we propose a novel technique for regularized optimization-based ultrasound elastography where the total data cost has been defined as the adaptively weighted linear combination of the contributions from sample amplitude and image gradient dissimilarities to facilitate a robust TDE. Inspired by~\cite{xu_2012}, we estimate the weights associated with the amplitude and gradient related costs in an iterative fashion. We name the proposed technique rGLUE- \textbf{GLUE} with \textbf{r}obust data function. We have validated rGLUE using simulation, experimental phantom, \textit{in vivo} liver and breast datasets, and compared with two recently published elastography techniques. Preliminary results of this work have been presented in IEEE International Ultrasonics Symposium (IEEE IUS 2020)~\cite{rglue_ius}. This manuscript represents more detailed explanation of the methods and additional results obtained from simulated, phantom, and \textit{in vivo} datasets. Similar to our previous work~\cite{guest,RAPID_TMI,soul,soulmate}, we have published the rGLUE code at \url{ http://code.sonography.ai}.    

 \begin{figure}
	\centering
	{\includegraphics[width=0.45\textwidth]{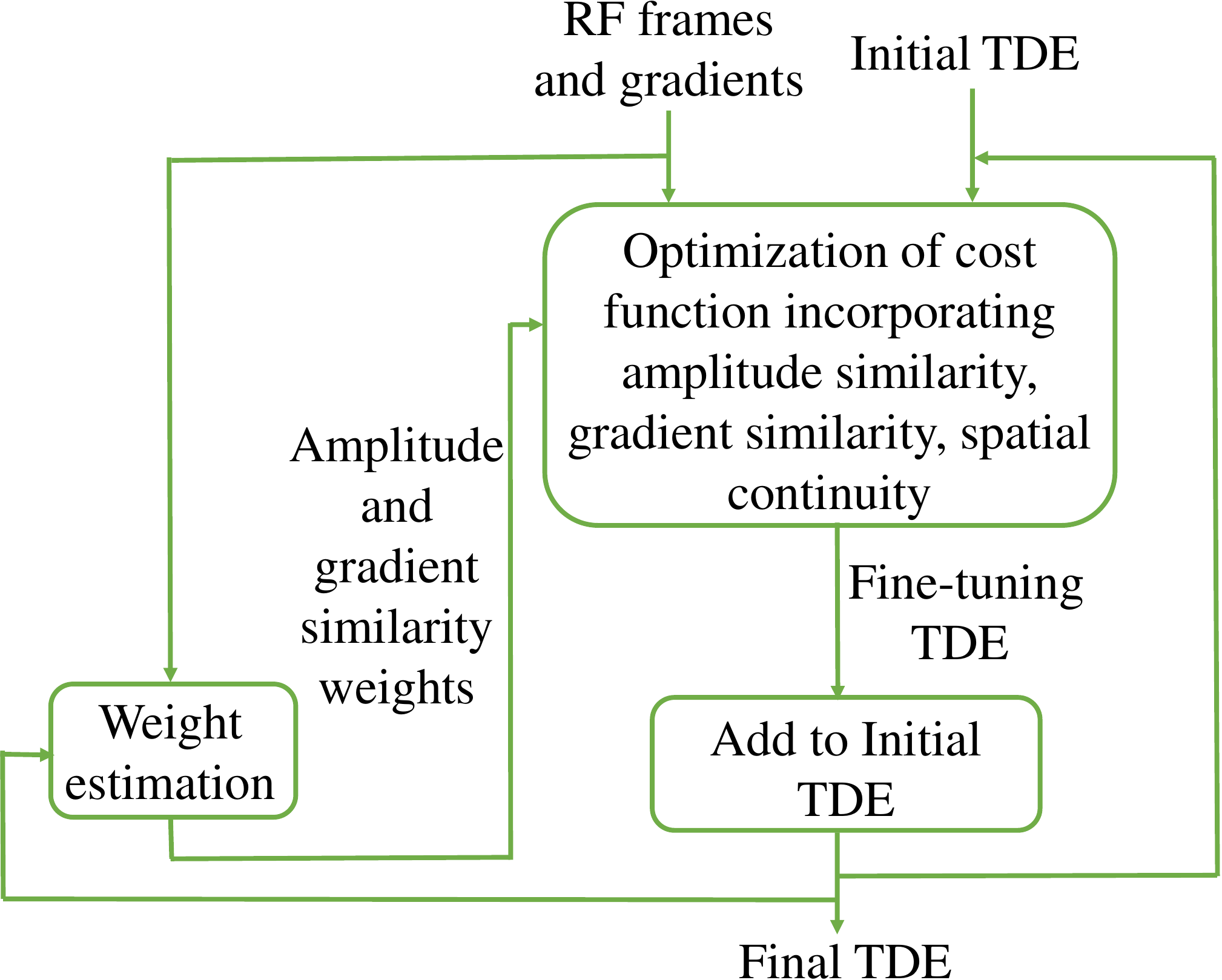} }%
	\qquad
	\caption{An illustration of the proposed rGLUE algorithm.}
	\label{depict_method}
\end{figure} 

\begin{figure}
	\centering
	\subfigure[Rectangular traget]{{\includegraphics[width=0.2\textwidth,height=0.224\textwidth]{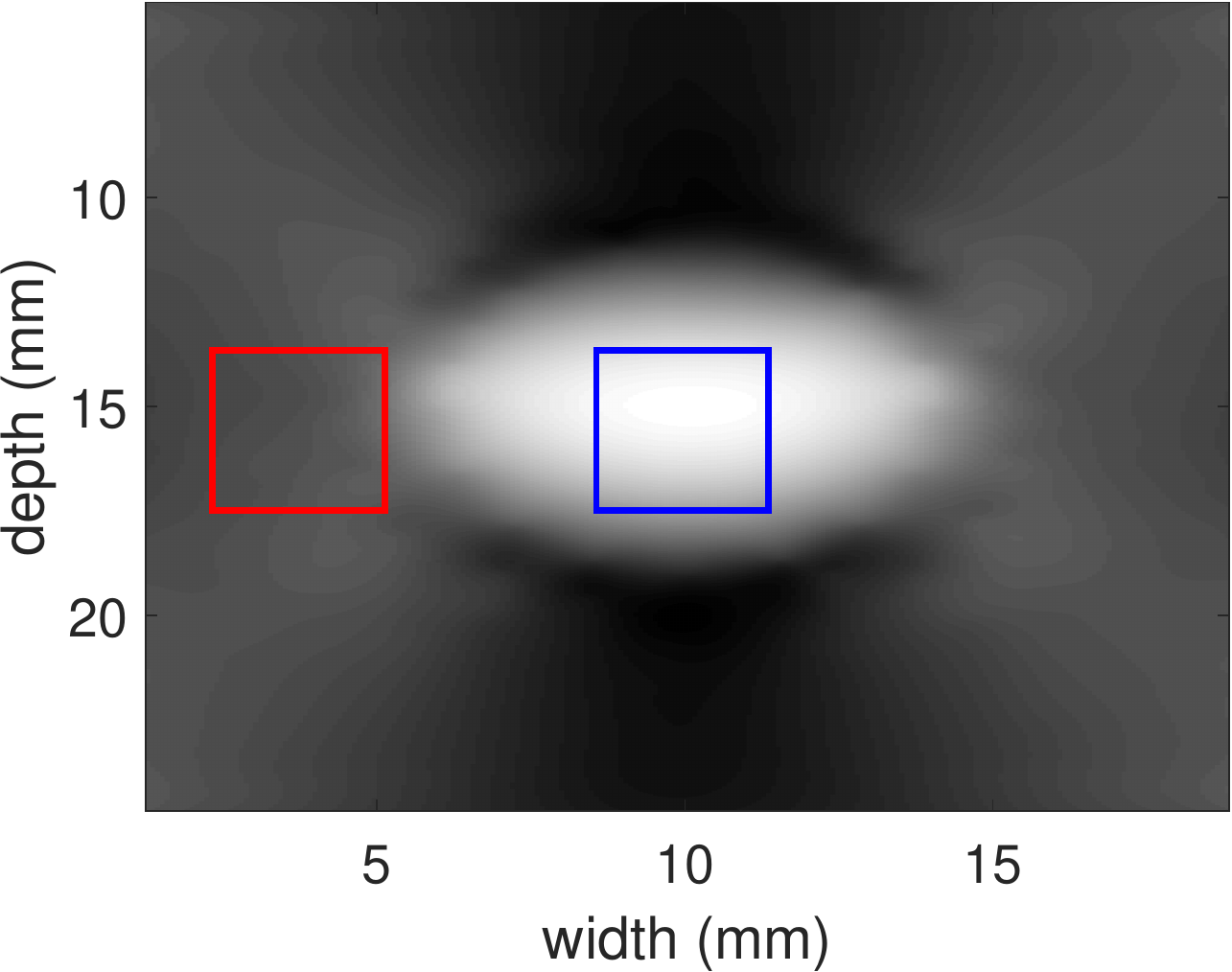}}}%
	\subfigure[Circular traget]{{\includegraphics[width=0.2\textwidth,height=0.224\textwidth]{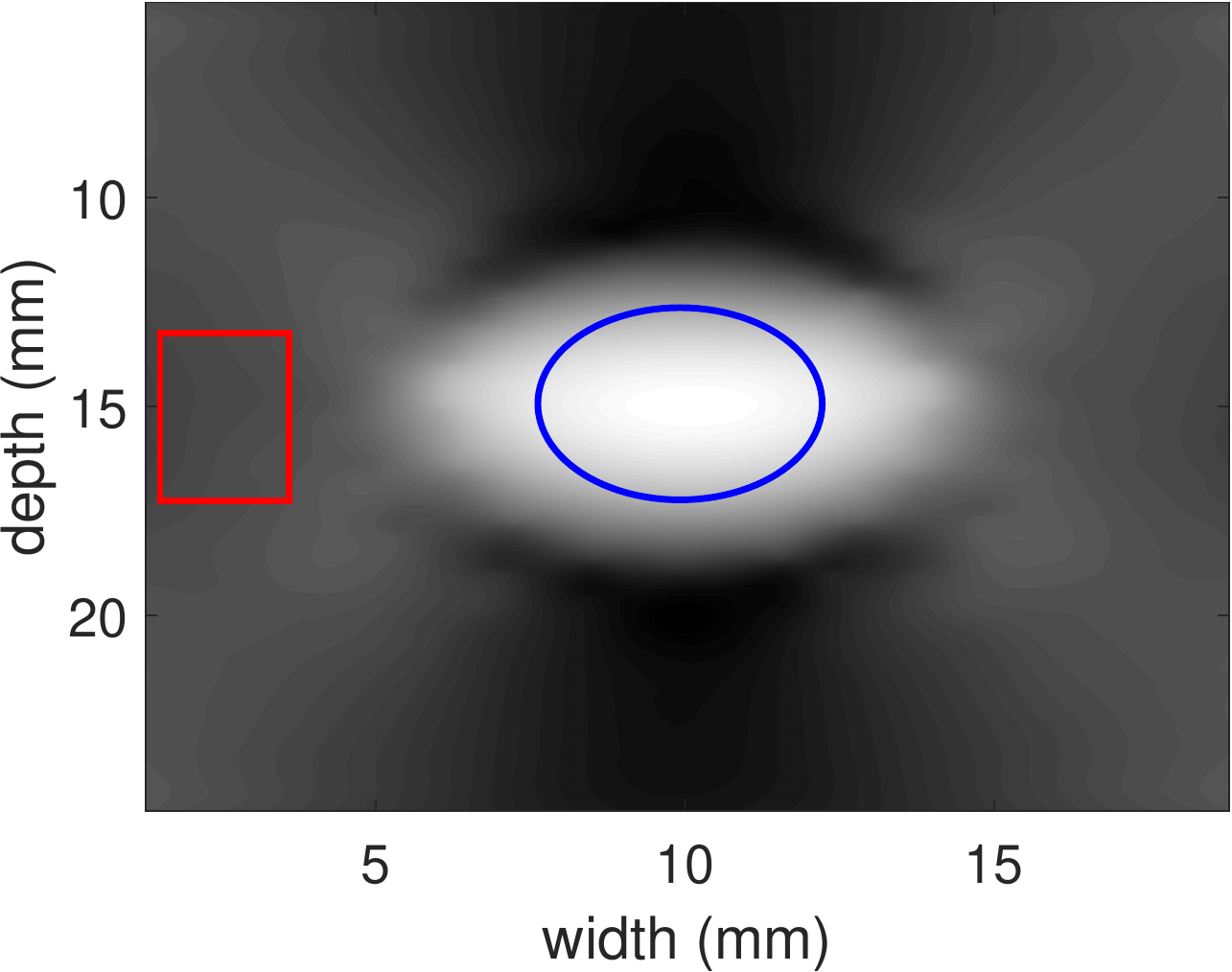}}}
	\caption{Ground truth axial strain for the soft-inclusion simulation data. The blue and red colored windows show foreground and background windows, respectively.}
	\label{ground}
\end{figure}

\begin{figure*}
	\centering
	\subfigure[Hybrid, $3.5\%$ compression]{{\includegraphics[width=0.33\textwidth,height=0.37\textwidth]{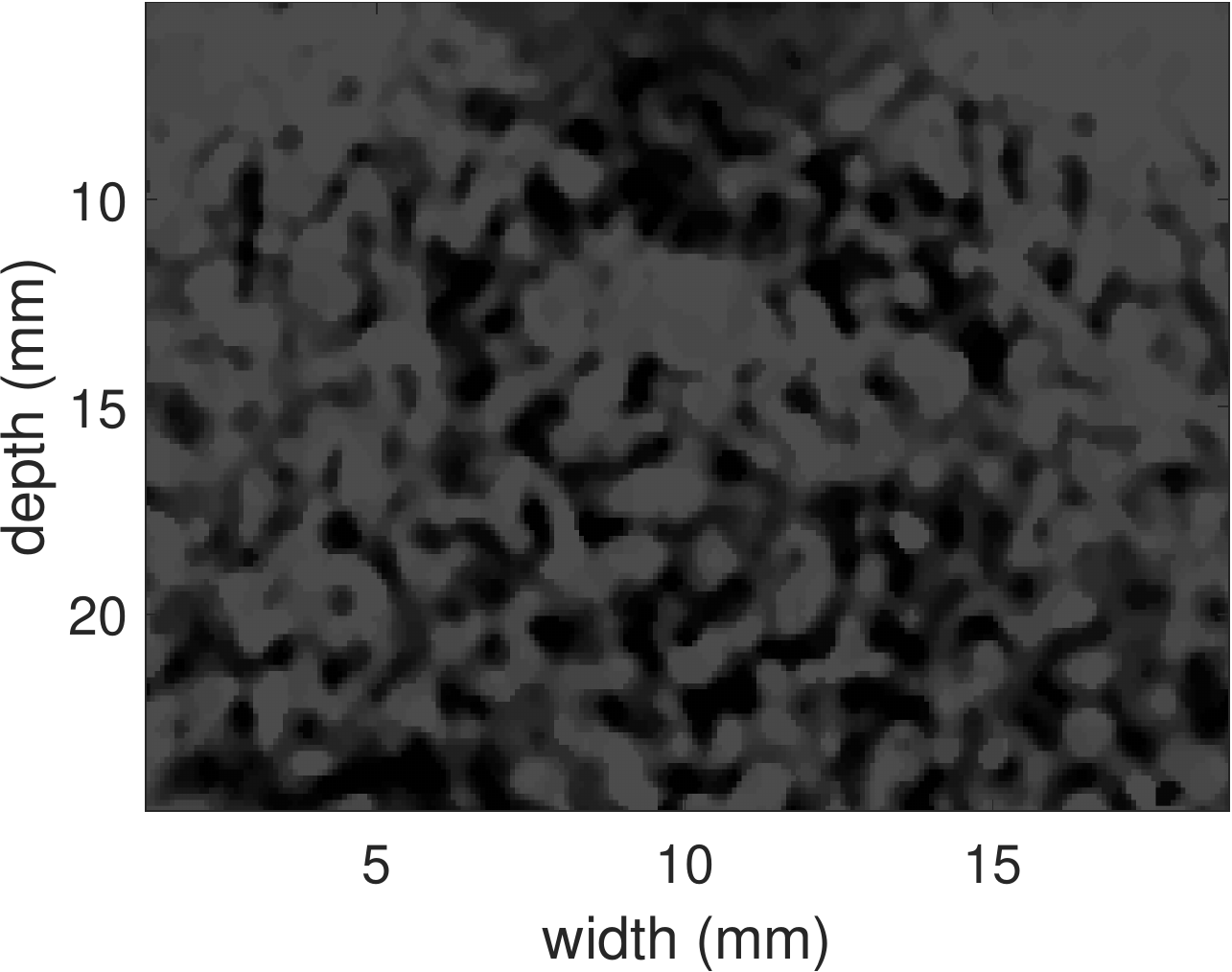} }}%
	\subfigure[GLUE, $3.5\%$ compression]{{\includegraphics[width=0.33\textwidth,height=0.37\textwidth]{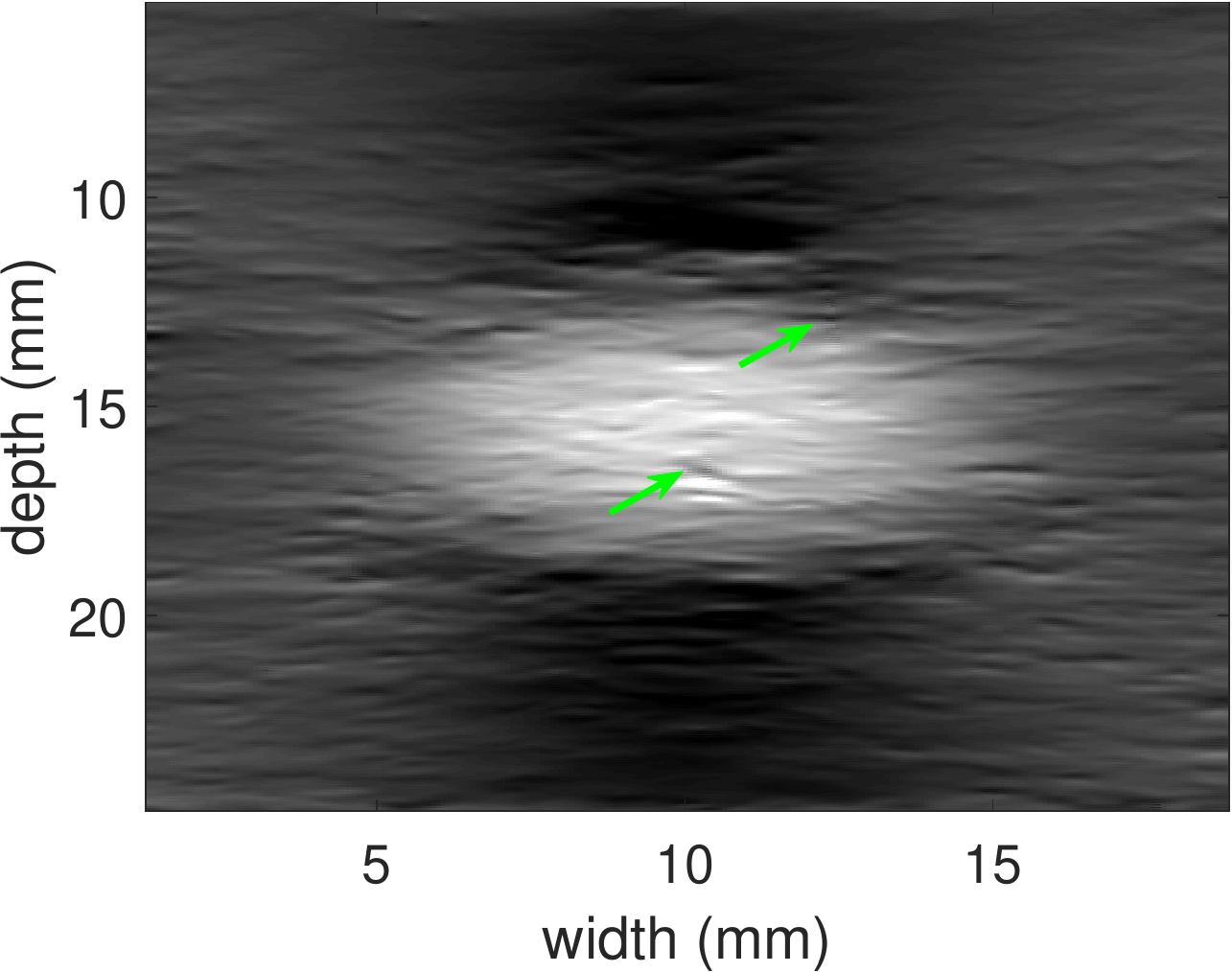} }}%
	\subfigure[rGLUE, $3.5\%$ compression]{{\includegraphics[width=0.33\textwidth,height=0.37\textwidth]{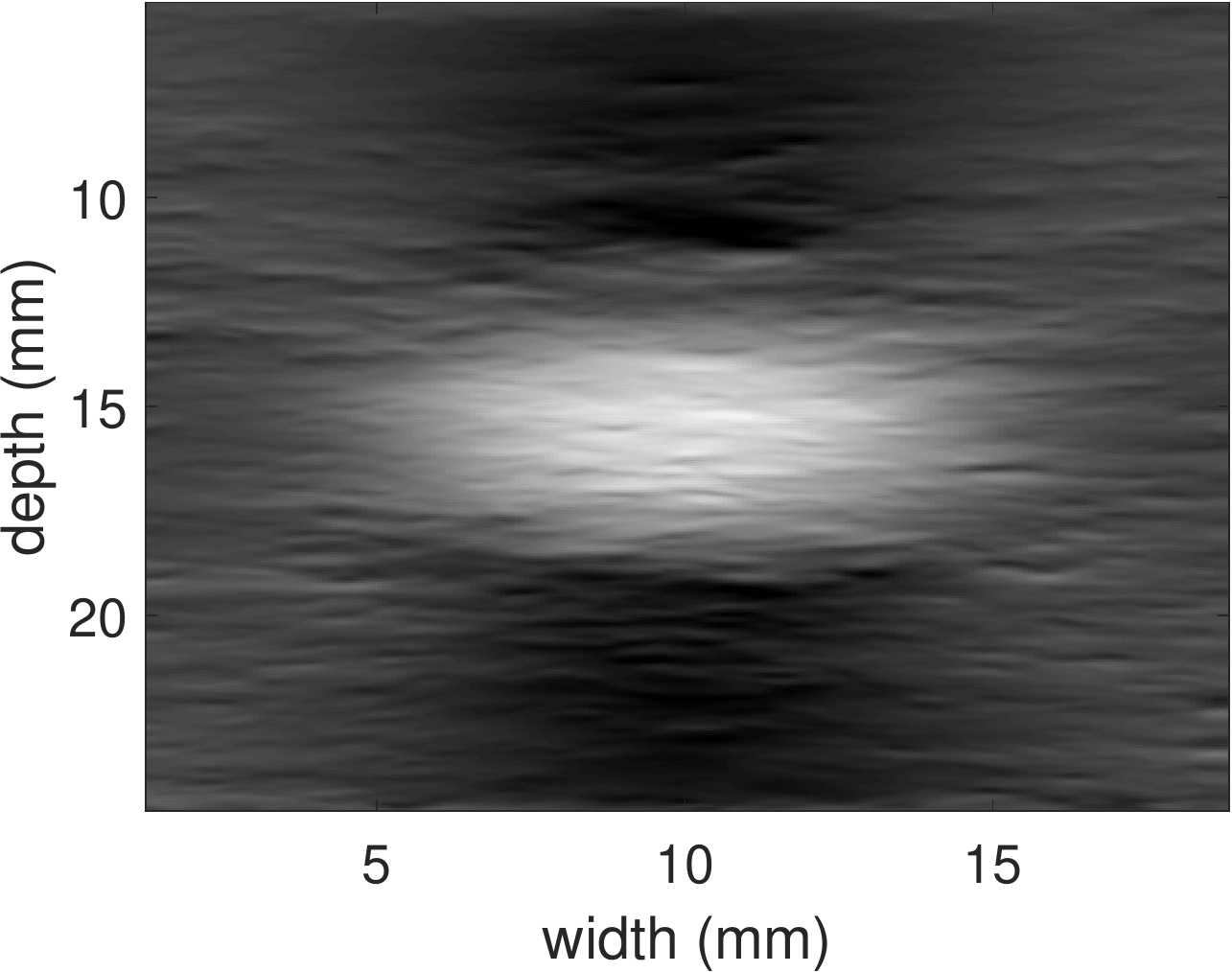} }}
	\subfigure[Hybrid, $4\%$ compression]{{\includegraphics[width=0.33\textwidth,height=0.37\textwidth]{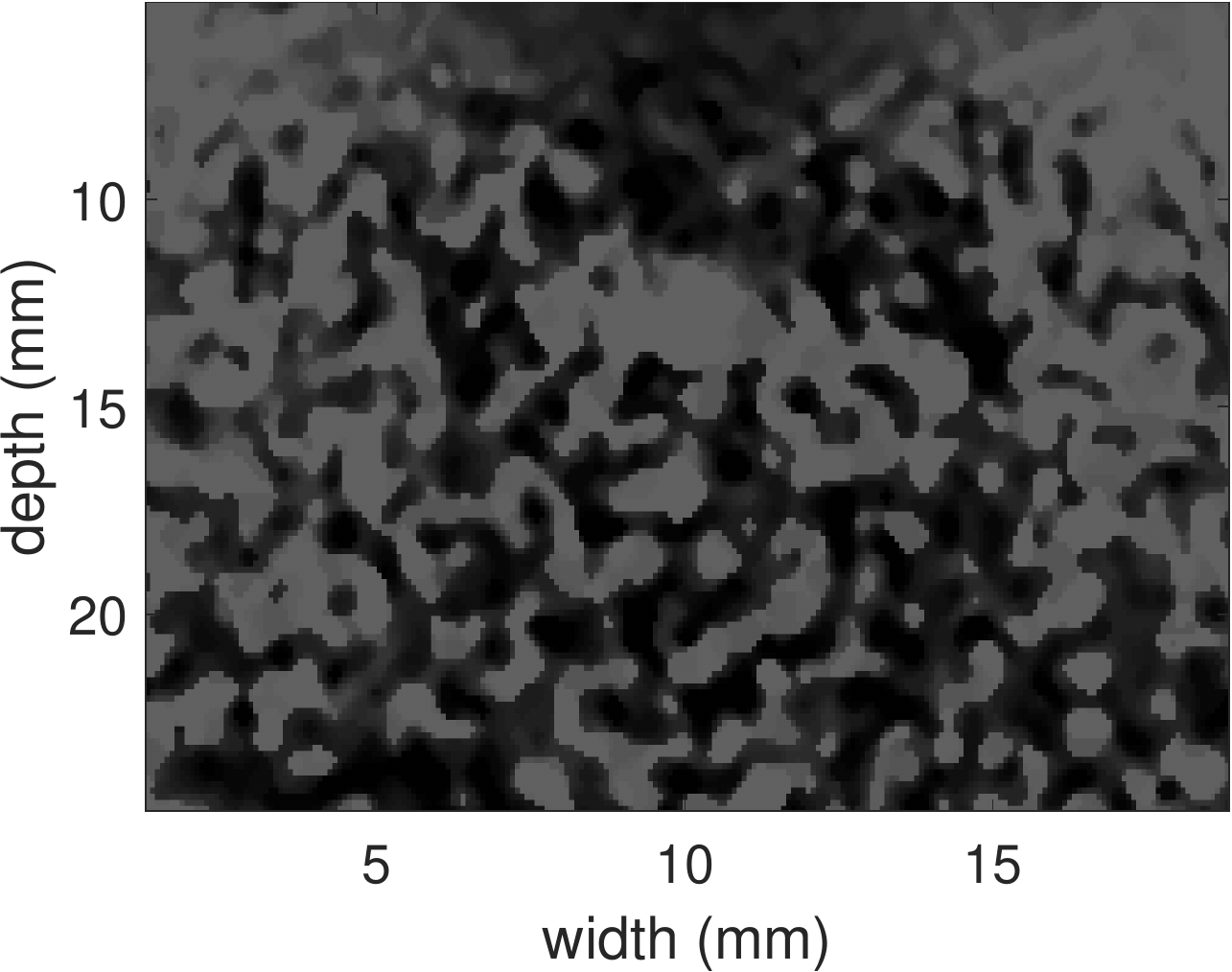} }}%
	\subfigure[GLUE, $4\%$ compression]{{\includegraphics[width=0.33\textwidth,height=0.37\textwidth]{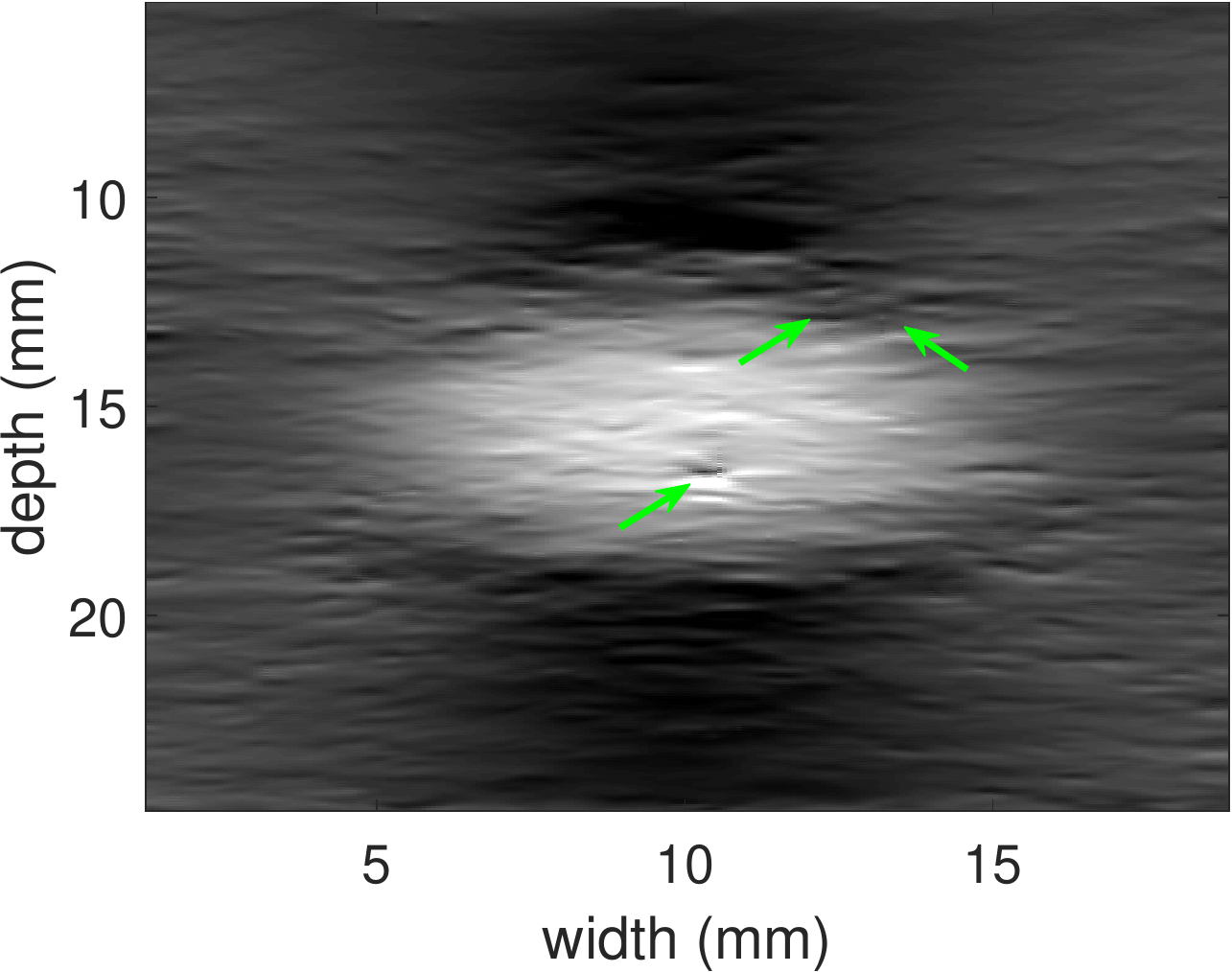} }}%
	\subfigure[rGLUE, $4\%$ compression]{{\includegraphics[width=0.33\textwidth,height=0.37\textwidth]{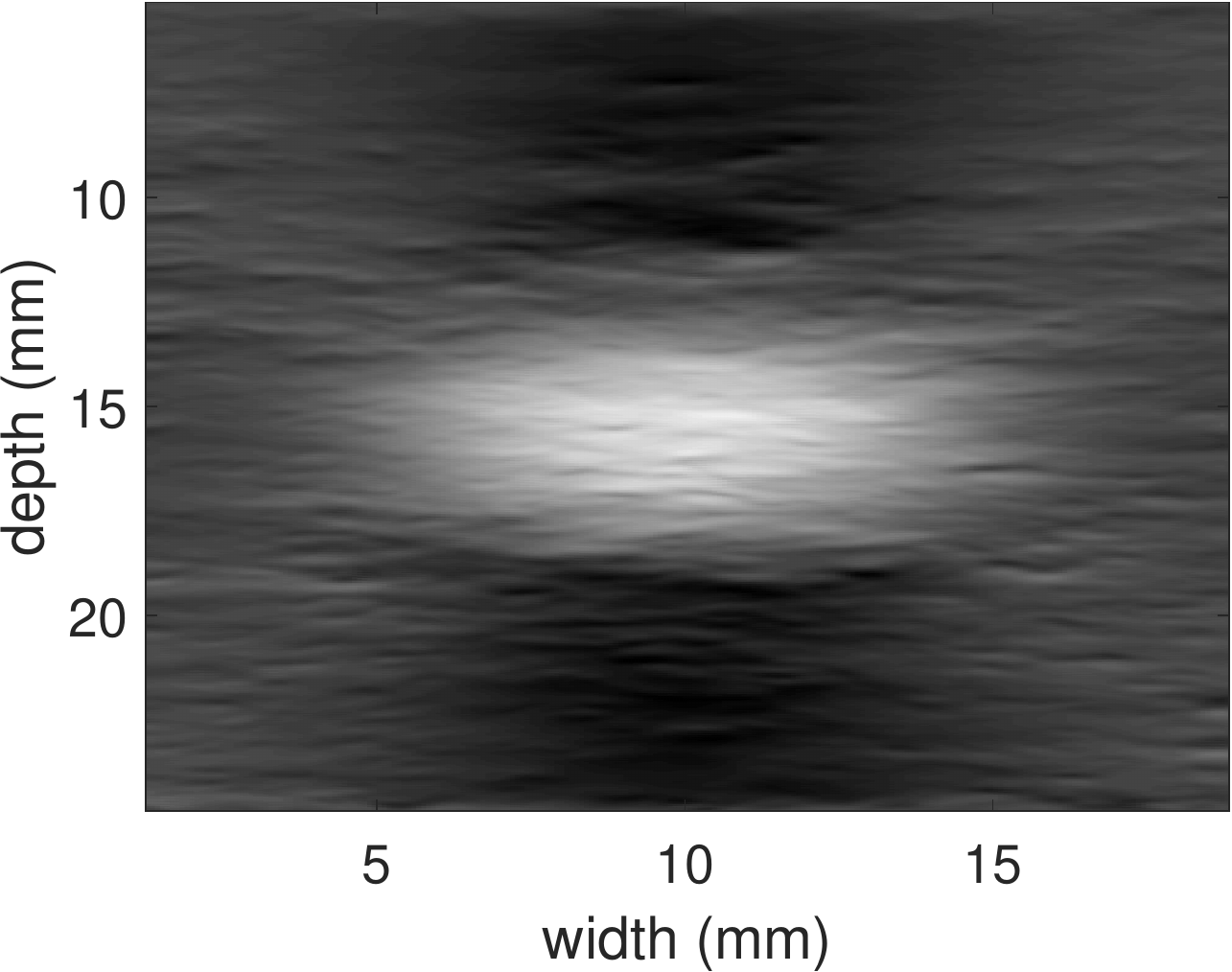} }}
	\subfigure[Strain, $3.5\%$ compression]{{\includegraphics[width=0.3\textwidth]{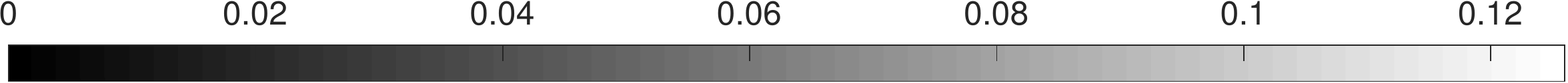} }}%
	\subfigure[Strain, $4\%$ compression]{{\includegraphics[width=0.3\textwidth]{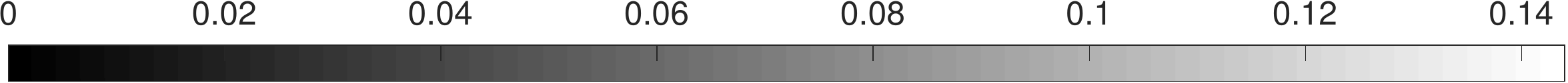} }}
	\caption{Axial strain images for the soft-inclusion simulation phantom. Rows 1 and 2 correspond to $3.5\%$ and $4\%$ compression levels, respectively. Columns 1, 2, and 3 correspond to Hybrid, GLUE, and rGLUE, respectively. Row 3 shows the color bars. The green arrow marks indicate the strain artifacts due to outlier samples.}
	\label{simulation1}
\end{figure*}

\begin{figure}
	\centering
	\subfigure[SNR]{{\includegraphics[width=0.23\textwidth]{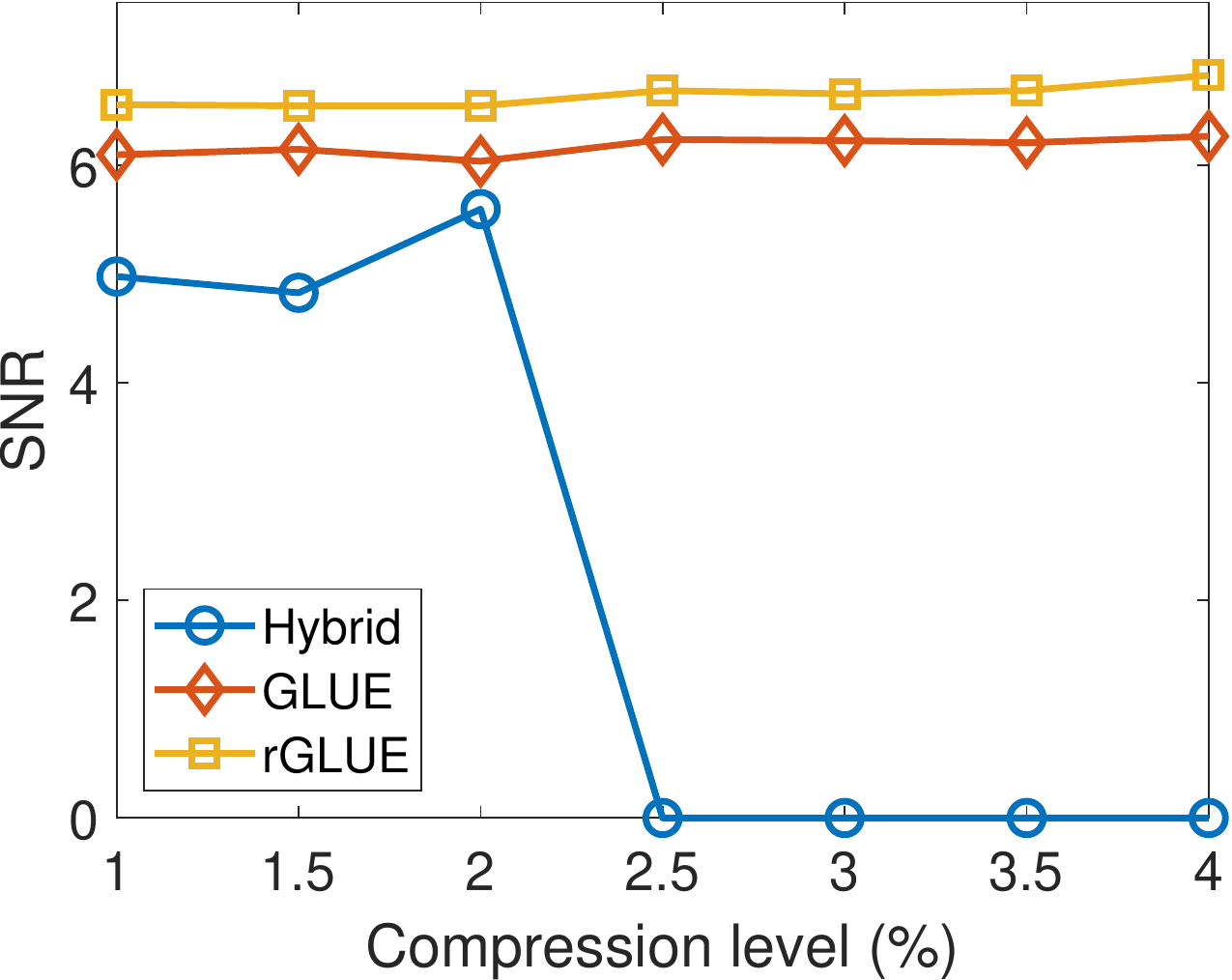} }}%
	\subfigure[CNR]{{\includegraphics[width=0.23\textwidth,height=0.185\textwidth]{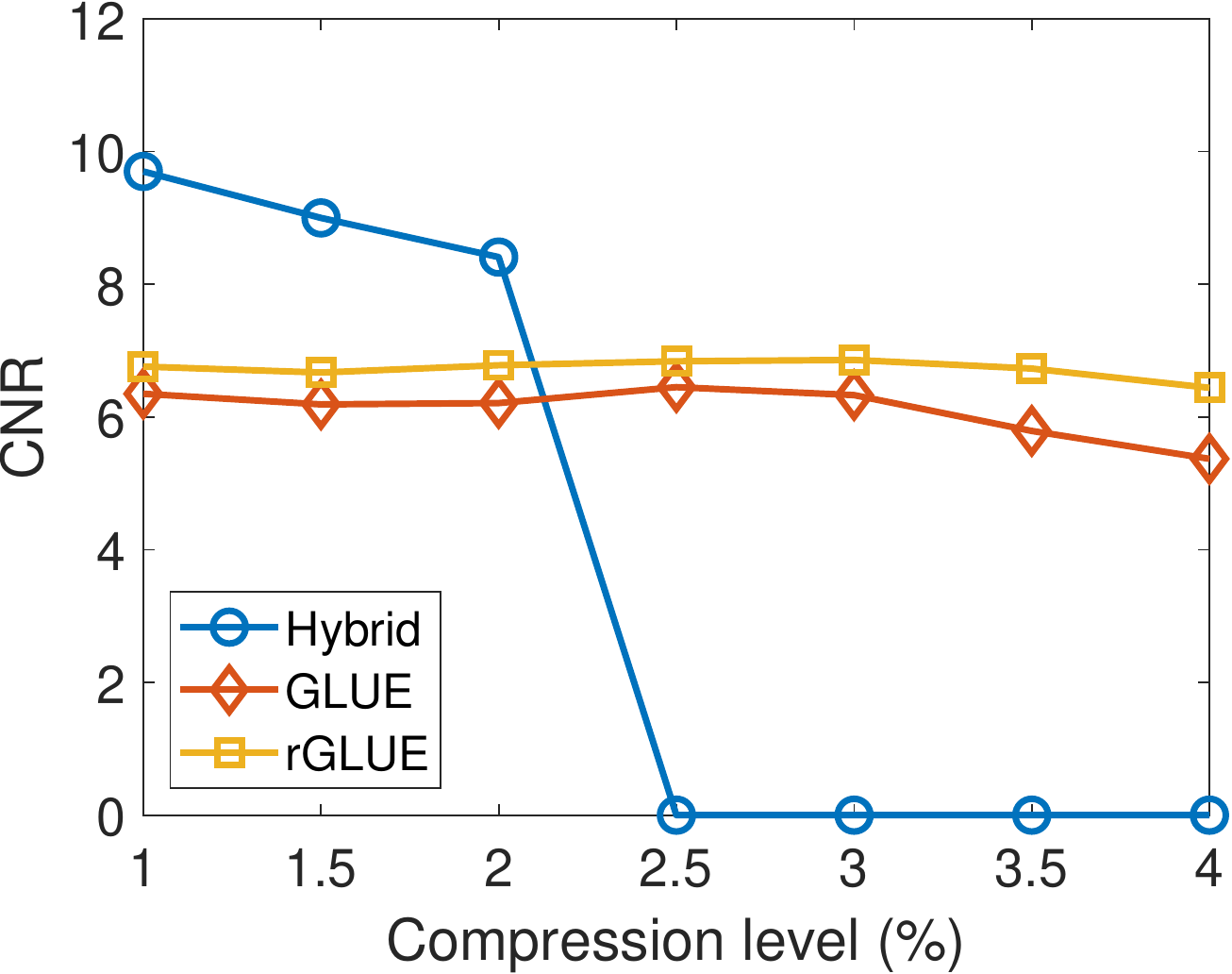} }}%
	\caption{SNR and CNR for the soft-inclusion simulation data. The left and right figures show the SNR and CNR values corresponding to different compression levels. The SNR values are calculated on the red colored background window shown in Fig.~\ref{ground}(a). The locations of the target (blue) and the background windows (red) for calculating the CNR values are shown in Fig.~\ref{ground}(a).}
	\label{snr_cnr}
\end{figure}

\section{Methods}
Let $I_{1}(i,j)$ and $I_{2}(i,j)$, $1 \leq i \leq m$, $1 \leq j \leq n$ denote two ultrasound RF frames acquired before and after tissue deformation, respectively. We aim to calculate the $2D$ displacement map between $I_{1}$ and $I_{2}$ and spatially differentiate the obtained displacement field to find the strain image. We first describe GLUE~\cite{GLUE}, a closely-related previous technique. Then we provide a detailed discussion on rGLUE, the proposed method.

\subsection{Global ultrasound elastography (GLUE)}
GLUE obtains the initial axial and lateral displacement fields $a_{i,j}$ and $l_{i,j}$ using DP~\cite{DP}. It is necessary to refine this discrete estimate to obtain a continuous and smooth displacement field. The refinement fields $\Delta a_{i,j}$ and $\Delta l_{i,j}$ are estimated by optimizing the following non-linear cost function $C_{g}$ consisting of data amplitude similarity term and a spatial regularization term:

\begin{equation}
\begin{aligned}
&C_{g} (\Delta a_{1,1},...,\Delta a_{m,n},\Delta l_{1,1},...,\Delta l_{m,n}) =\\
&\sum\limits_{j=1}^n \sum\limits_{i=1}^m \{D_{I}(i,j,a_{i,j},l_{i,j},\Delta a_{i,j},\Delta l_{i,j})+R\}
\end{aligned}
\label{eq:glue_c}
\end{equation} 

\noindent
where $D_{I}$ and $R$ denote the data and regularization terms, respectively, and are defined as follows:

\begin{equation}
\begin{aligned}
&D_{I}(i,j,a_{i,j},l_{i,j},\Delta a_{i,j},\Delta l_{i,j})=\\
&[I_{1}(i,j)-I_{2}(i+a_{i,j}+\Delta a_{i,j},j+l_{i,j}+\Delta l_{i,j})]^{2}
\end{aligned}
\label{eq:glue_d}
\end{equation}

\begin{equation}
\begin{aligned} 
&R=\alpha_{1} (a_{i,j}+\Delta a_{i,j}-a_{i-1,j}- \Delta a_{i-1,j})^{2} \\ 
&+ \alpha_{2} (a_{i,j}+\Delta a_{i,j}-a_{i,j-1}- \Delta a_{i,j-1})^{2} \\
& + \beta_{1} (l_{i,j}+\Delta l_{i,j}-l_{i-1,j}- \Delta l_{i-1,j})^{2} \\
&+ \beta_{2} (l_{i,j}+\Delta l_{i,j}-l_{i,j-1}- \Delta l_{i,j-1})^{2} \}
\end{aligned}
\label{eq:glue_r}
\end{equation}          

\noindent
where $\alpha_{1}$, $\alpha_{2}$ and $\beta_{1}$, $\beta_{2}$ indicate the axial and lateral regularization parameters, respectively.

\subsection{Robust time delay estimation (rGLUE)}
As described above, the data term in GLUE imposes only the sample amplitude similarity constraint. However, taking only the sample amplitude similarity into account makes the optimization framework too sensitive to the change in amplitude values~\cite{brox_2004}. This high sensitivity to slight amplitude changes adversely affects the optimization procedure's ability to handle the outlier samples in the RF data. Therefore, an additional constraint such as gradient similarity, which is not susceptible to amplitude changes~\cite{brox_2004}, is required to efficiently detect and suppress the outliers. To be precise, adding the gradient constancy assumption to the penalty function introduces relaxation to the optimization scheme making it less sensitive to amplitude alteration. However, studies~\cite{xu_2012} show that considering both amplitude and gradient similarity constraints at the same time is less accurate than taking one of them into account. Therefore, it is more reasonable to put binary weights on the two constancy assumptions and estimate the weights iteratively. However, estimating the binary weights and continuous displacement fields simultaneously is a computationally demanding task. Consequently, in rGLUE, we adopt an approach similar to the one devised in~\cite{xu_2012} where the binary process is transformed to a continuous estimation scheme by employing mean field approximation~\cite{mf_1989}.

Similar to GLUE, we employ DP~\cite{DP} to estimate the initial displacement field. In order to refine the DP initial estimate, we introduce a novel cost function $C$ where the data function consists of both sample amplitude and gradient similarities as follows:

\begin{equation}
\begin{aligned}
&C (\Delta a_{1,1},...,\Delta a_{m,n},\Delta l_{1,1},...,\Delta l_{m,n}) =\\
&\sum\limits_{j=1}^n \sum\limits_{i=1}^m \{\theta(i,j) D_{I}(i,j,a_{i,j},l_{i,j},\Delta a_{i,j},\Delta l_{i,j})+\\
&(1-\theta(i,j)) \sum \nolimits_{p \in \{y,x\}}D_{\nabla I,p}(i,j,a_{i,j},l_{i,j},\Delta a_{i,j},\Delta l_{i,j})+R\}
\end{aligned}
\label{eq:glue_r}
\end{equation}

\noindent
where subscripts $y$ and $x$, respectively, refer to axial and lateral directions. $D_{\nabla I,p}$ denotes the gradient similarity term(s) and is given by:

\begin{equation}
\begin{aligned}
&D_{\nabla I,p}(i,j,a_{i,j},l_{i,j},\Delta a_{i,j},\Delta l_{i,j})=\\
&\gamma[\nabla I_{1,p}(i,j)-\nabla I_{2,p}(i+a_{i,j}+\Delta a_{i,j},j+l_{i,j}+\Delta l_{i,j})]^{2}
\end{aligned}
\label{eq:rglue_dy}
\end{equation}


\noindent
where $\gamma$ denotes a matching parameter and $\nabla$ stands for the discrete gradient operator. $\theta(i,j)$ refers to a data driven weight map which adaptively controls the contributions of amplitude and gradient dissimilarities to the total cost. Inspired by~\cite{xu_2012}, we estimate $\theta(i,j)$ using $\theta(i,j)=\frac{1}{1+\exp(\lambda \delta(i,j))}$ where

\begin{equation}
\begin{aligned}
&\delta(i,j)=D_{I}(i,j,a_{i,j},l_{i,j},\Delta a_{i,j},\Delta l_{i,j})-\\
&\sum \nolimits_{p \in \{y,x\}} D_{\nabla I,p}(i,j,a_{i,j},l_{i,j},\Delta a_{i,j},\Delta l_{i,j})
\end{aligned}
\label{eq:theta}
\end{equation}

Here, $\lambda$ denotes a parameter which is tuned to obtain a balanced weight map. We perform $2D$ Taylor series expansions of $I_{2}$ and $\nabla I_{2,p}$ around $(i+a_{i,j},j+l_{i,j})$ to remove the non-linearities present in $D_{I}$ and $D_{\nabla I,p}$:

\begin{equation}
\begin{aligned}
&I_{2}(i+a_{i,j}+\Delta a_{i,j},j+l_{i,j}+\Delta l_{i,j}) \approx \\
&I_{2}(i+a_{i,j},j+l_{i,j})+\Delta a_{i,j}I_{2,a}^{'}+\Delta l_{i,j}I_{2,l}^{'} 
\end{aligned}
\label{eq:i2_taylor}
\end{equation}

\begin{equation}
\begin{aligned}
&\nabla I_{2,p}(i+a_{i,j}+\Delta a_{i,j},j+l_{i,j}+\Delta l_{i,j}) \approx \\
&\nabla I_{2,p}(i+a_{i,j},j+l_{i,j})+\Delta a_{i,j}\nabla I_{2,p,a}^{'}+\Delta l_{i,j}\nabla I_{2,p,l}^{'} 
\end{aligned}
\label{eq:i2y_taylor}
\end{equation}


\noindent
where $I_{2,a}^{'}$ and $I_{2,l}^{'}$ denote axial and lateral derivatives of $I_{2}$, respectively. $\nabla I_{2,p,a}^{'}$ and $\nabla I_{2,p,l}^{'}$ refer to axial and lateral derivatives of $p$ component of the gradient of $I_{2}$. After the aforementioned expansions, Eq.~\ref{eq:glue_r} becomes quadratic in unknowns and we set $\frac{\partial C_{i,j}}{\partial \Delta a_{i,j}}=0$, $\frac{\partial C_{i,j}}{\partial \Delta l_{i,j}}=0$. After some algebraic manipulation, we get:

\begin{equation}
(H_{I}+H_{\nabla I_{y}}+H_{\nabla I_{x}}+D)\Delta d=P_{I}\mu_{1}+P_{\nabla I_{y}}\mu_{2}+P_{\nabla I_{x}}\mu_{3}-Dd 
\label{eq:rglue_axb}
\end{equation}

\noindent
where $d \in \mathbb{R}^{2mn \times 1}$ contains the DP initial estimates: $d=[a_{1,1},l_{1,1},a_{1,2},l_{1,2},\dots,a_{m,n},l_{m,n}]^T$ whereas $\Delta d$ denotes a vector of size $2mn$ where the unknown fine-tuning displacement estimates are stacked: $\Delta d=[\Delta a_{1,1},\Delta l_{1,1},\Delta a_{1,2},\Delta l_{1,2},\dots,\Delta a_{m,n},\Delta l_{m,n}]^T$. $D \in \mathbb{R}^{2mn \times 2mn}$ contains the spatial regularization parameters. It should be noted that the regularization term of the proposed technique is same as GLUE. Therefore, $D$ is also the same as described in~\cite{GLUE}. To keep this paper succinct, we define matrices $H_{I}$, $H_{\nabla I_{y}}$, $H_{\nabla I_{x}}$, $P_{I}$, $P_{\nabla I_{y}}$, and $P_{\nabla I_{x}}$ and vectors $\mu_{1}$, $\mu_{2}$, and $\mu_{3}$ in the Supplementary Material.
 
Once the fine-tuning fields $\Delta a$ and $\Delta l$ are estimated, we add them to the DP initial estimates $a$ and $l$ to find the final TDE. Least-squares fitting approach is incorporated to differentiate the axial displacement field for obtaining a high-quality strain map. An illustrative diagram explaining the TDE work-flow of rGLUE has been provided in Fig.~\ref{depict_method}.

\begin{figure*}
	\centering
	\subfigure[Hybrid]{{\includegraphics[width=0.3\textwidth,height=0.336\textwidth]{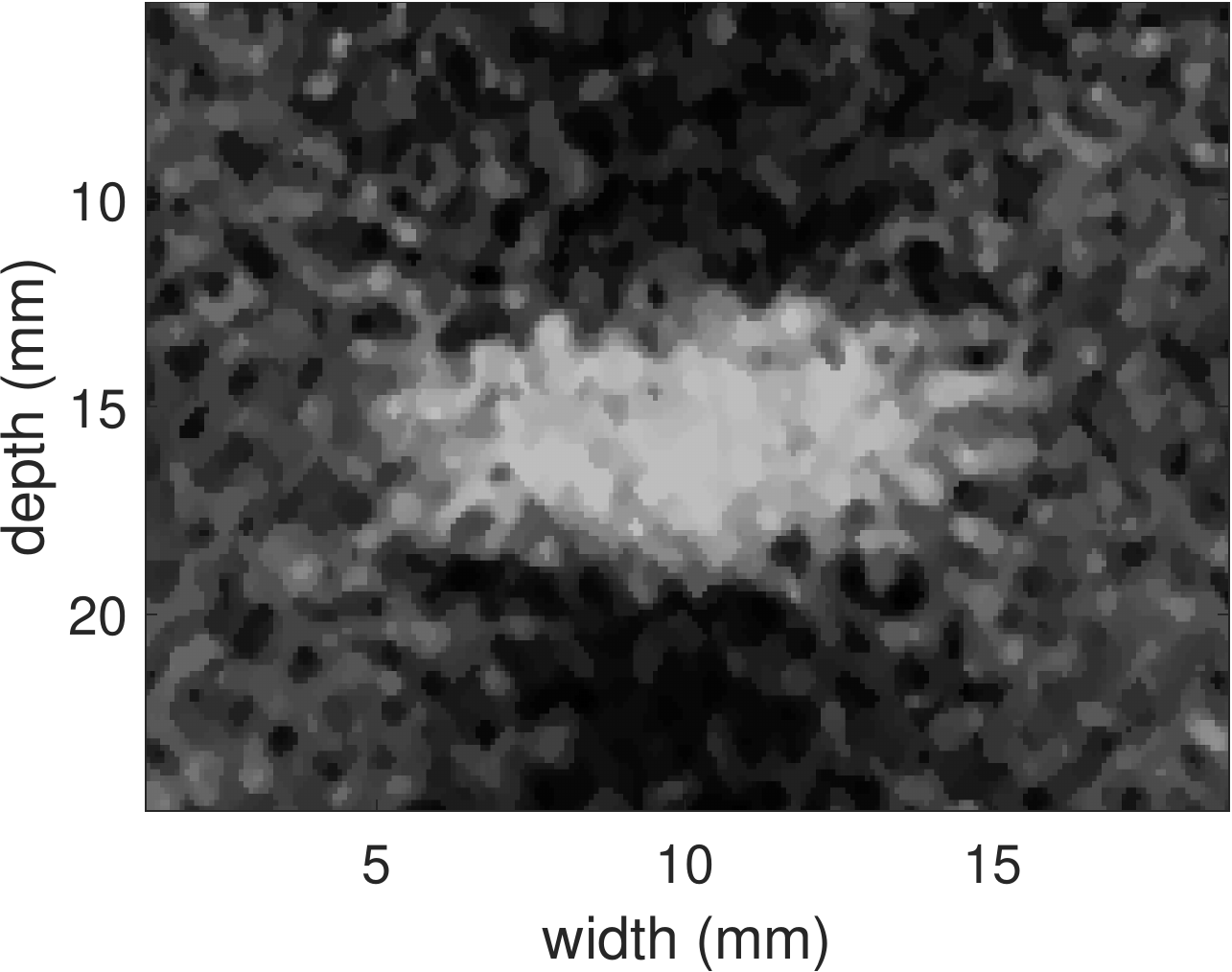} }}%
	\subfigure[GLUE]{{\includegraphics[width=0.3\textwidth,height=0.336\textwidth]{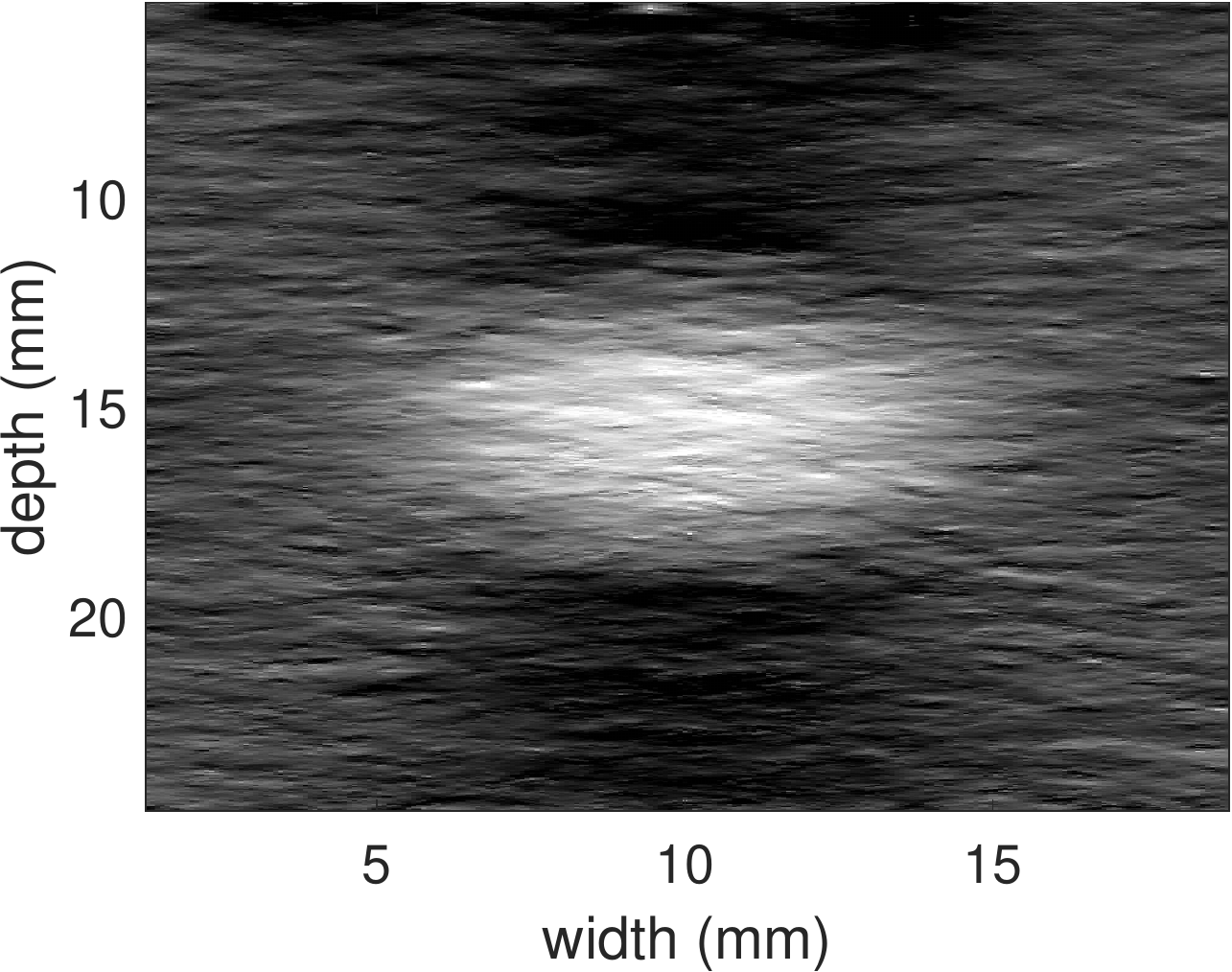}}}%
	\subfigure[rGLUE]{{\includegraphics[width=0.3\textwidth,height=0.336\textwidth]{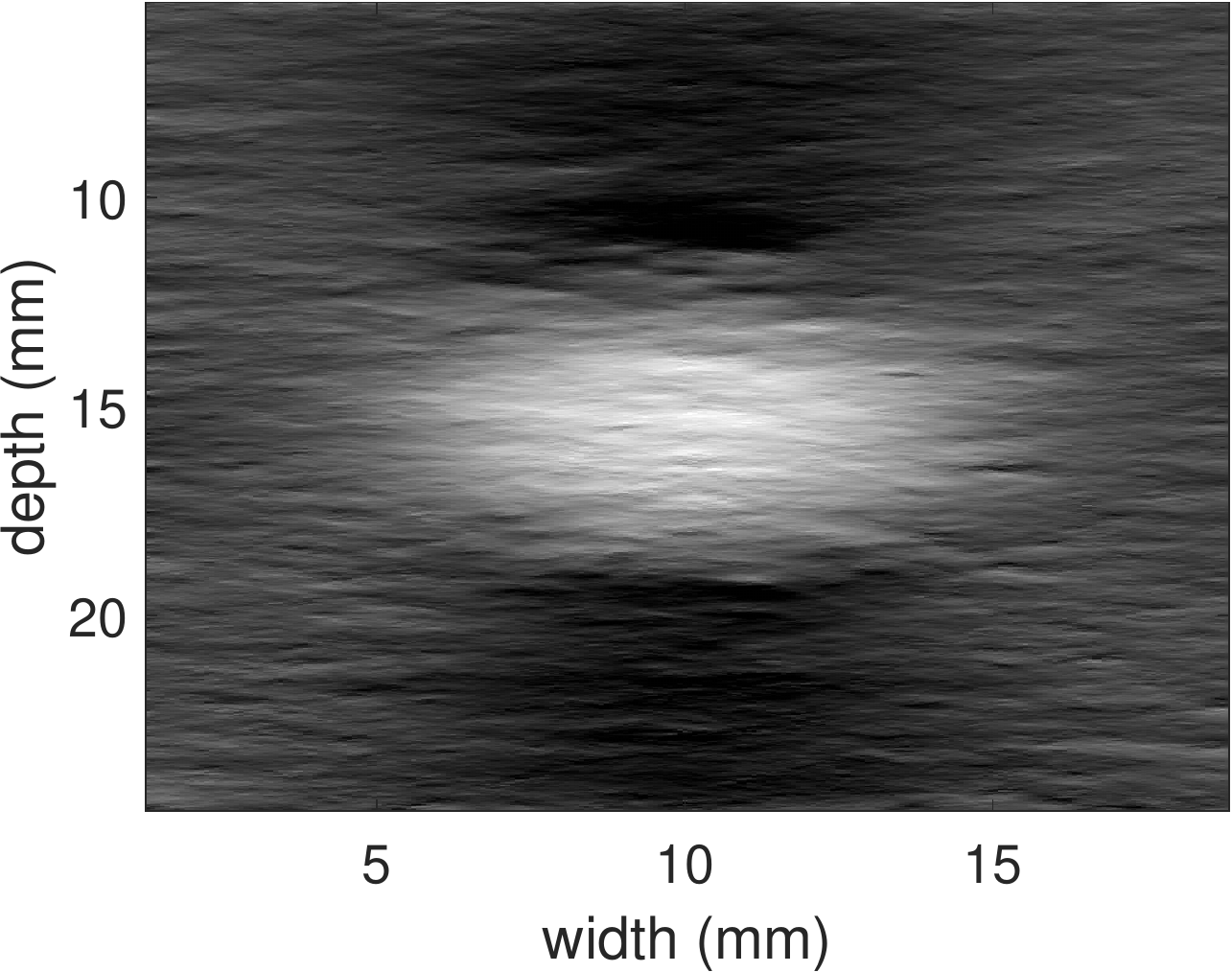} }}
	\subfigure[Hybrid]{{\includegraphics[width=0.3\textwidth,height=0.336\textwidth]{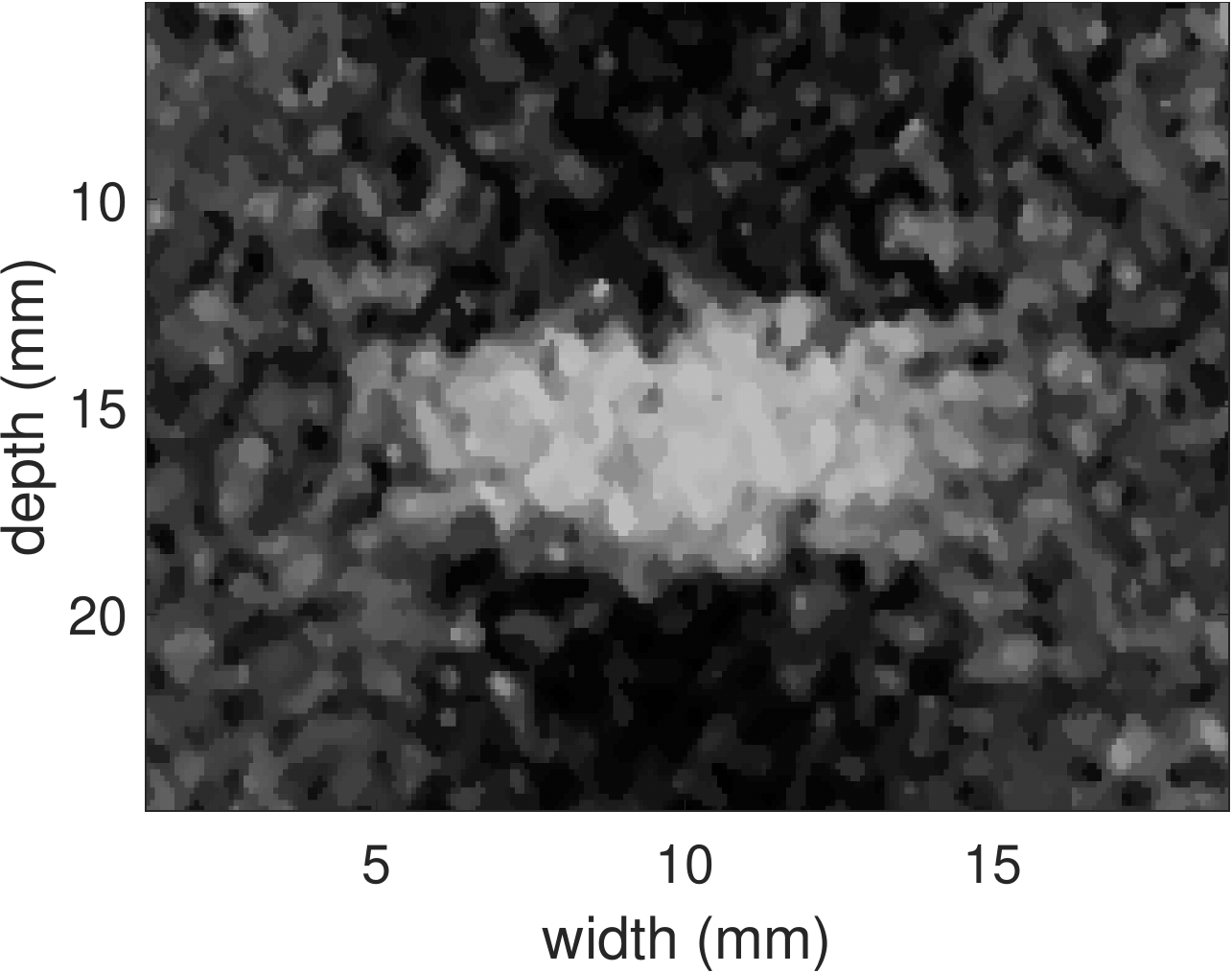} }}%
	\subfigure[GLUE]{{\includegraphics[width=0.3\textwidth,height=0.336\textwidth]{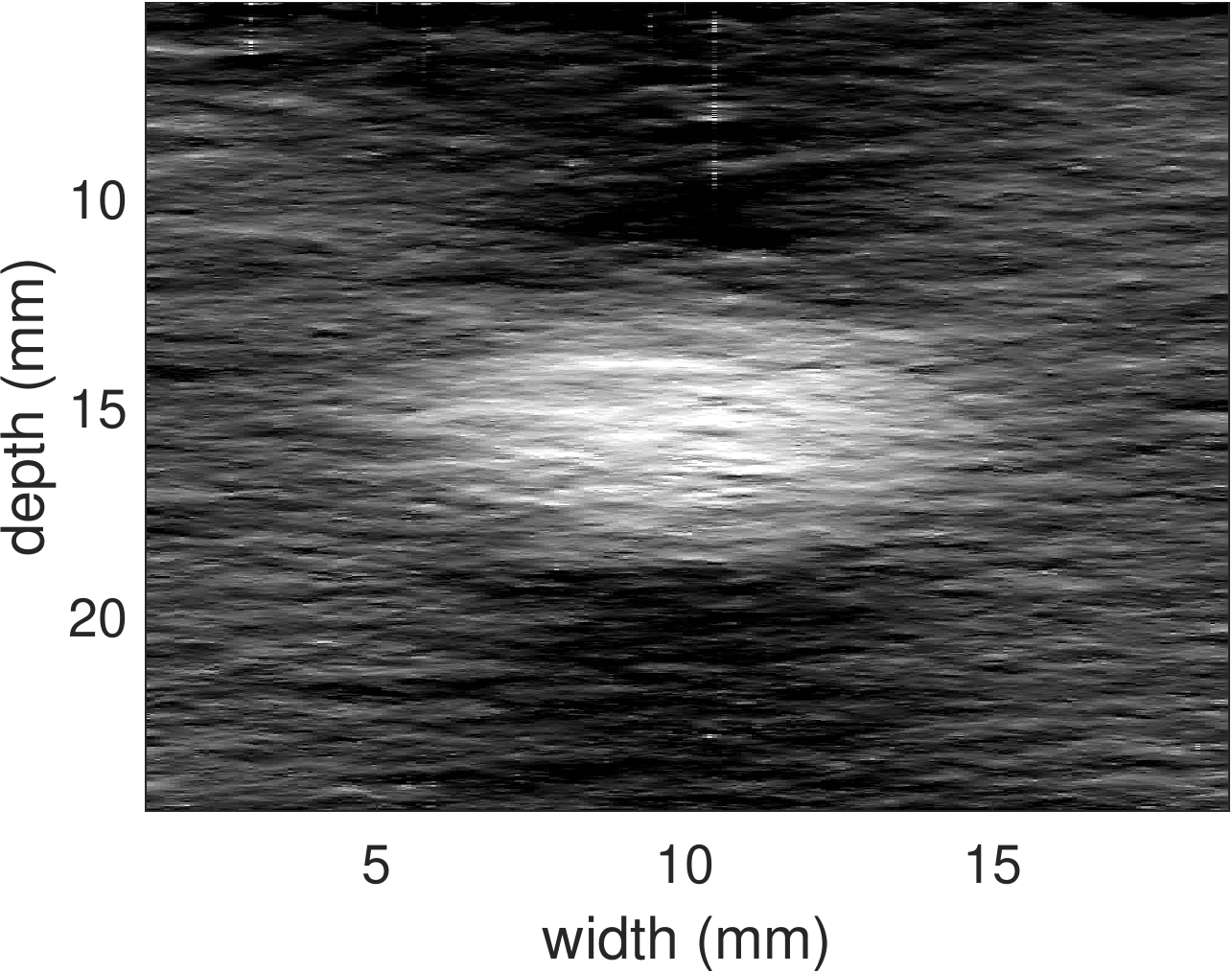}}}%
	\subfigure[rGLUE]{{\includegraphics[width=0.3\textwidth,height=0.336\textwidth]{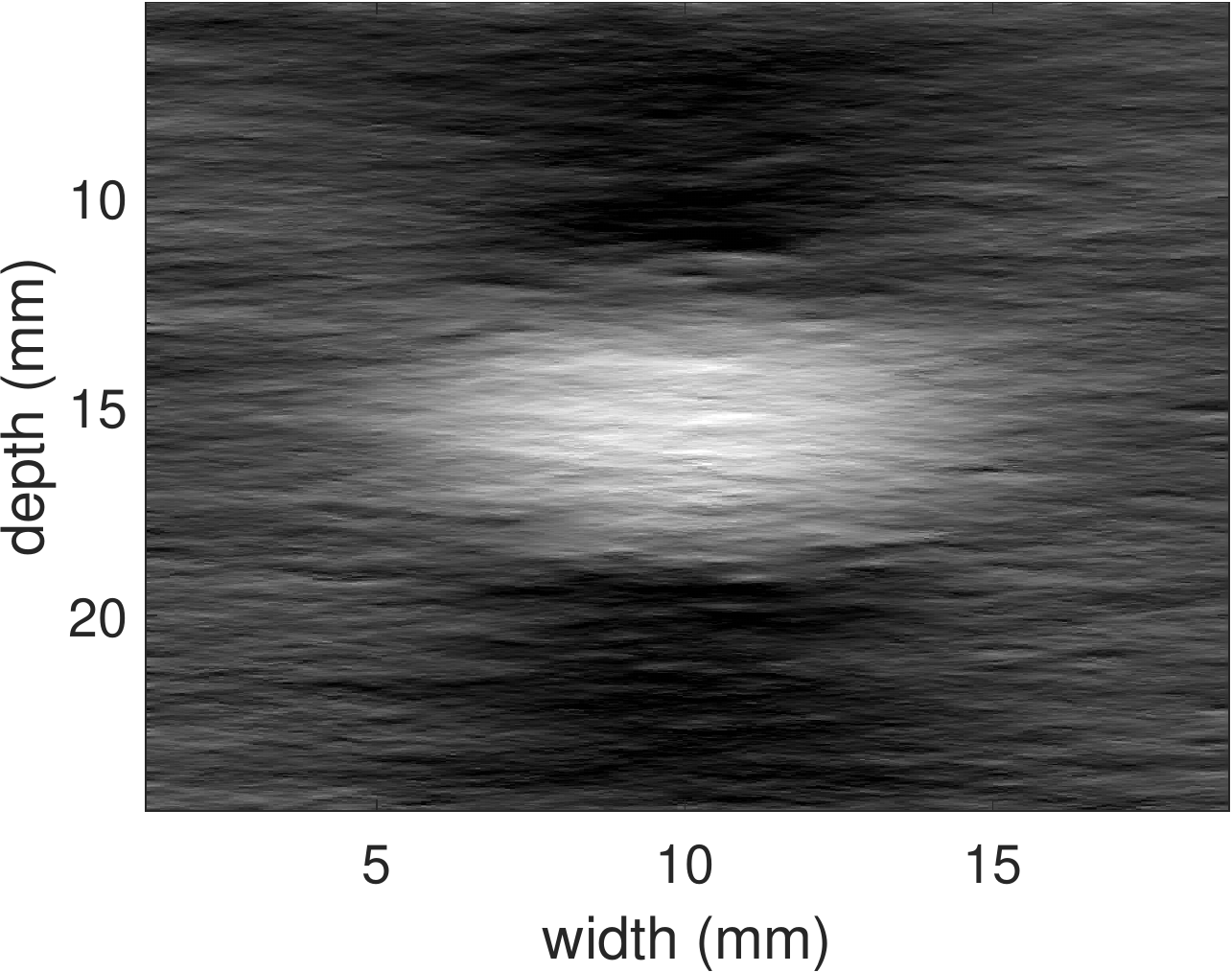} }}
	\subfigure[Hybrid]{{\includegraphics[width=0.3\textwidth,height=0.336\textwidth]{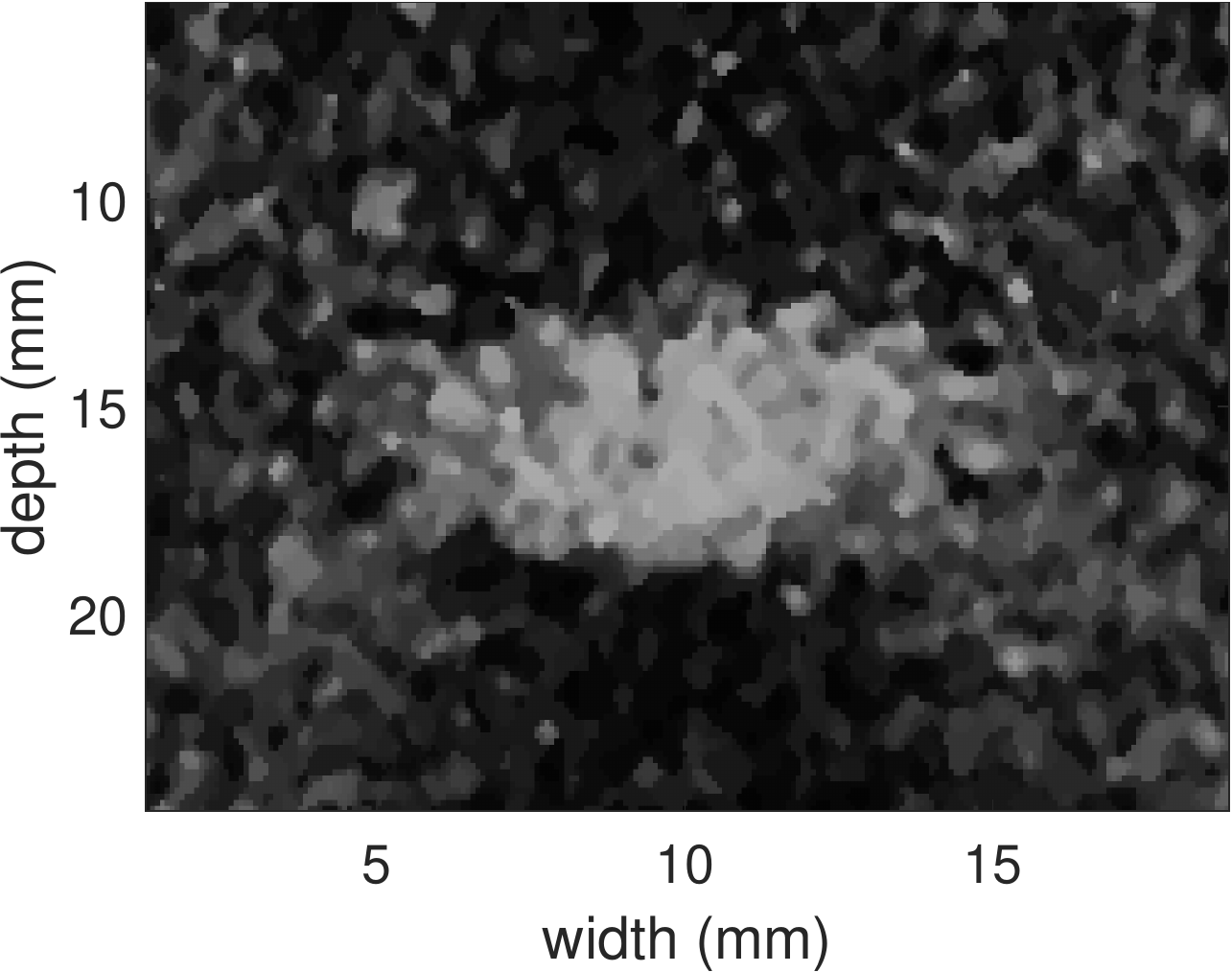} }}%
	\subfigure[GLUE]{{\includegraphics[width=0.3\textwidth,height=0.336\textwidth]{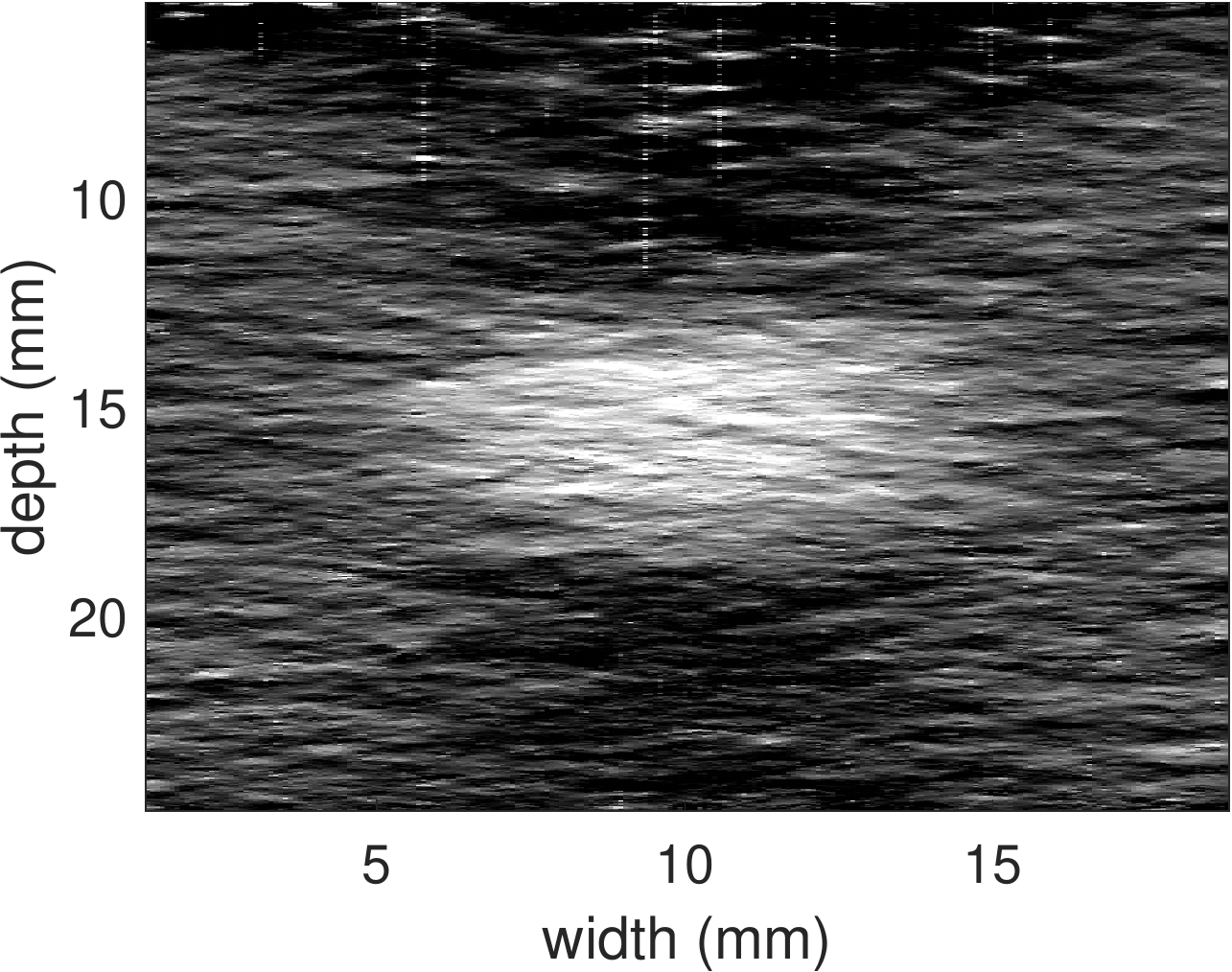}}}%
	\subfigure[rGLUE]{{\includegraphics[width=0.3\textwidth,height=0.336\textwidth]{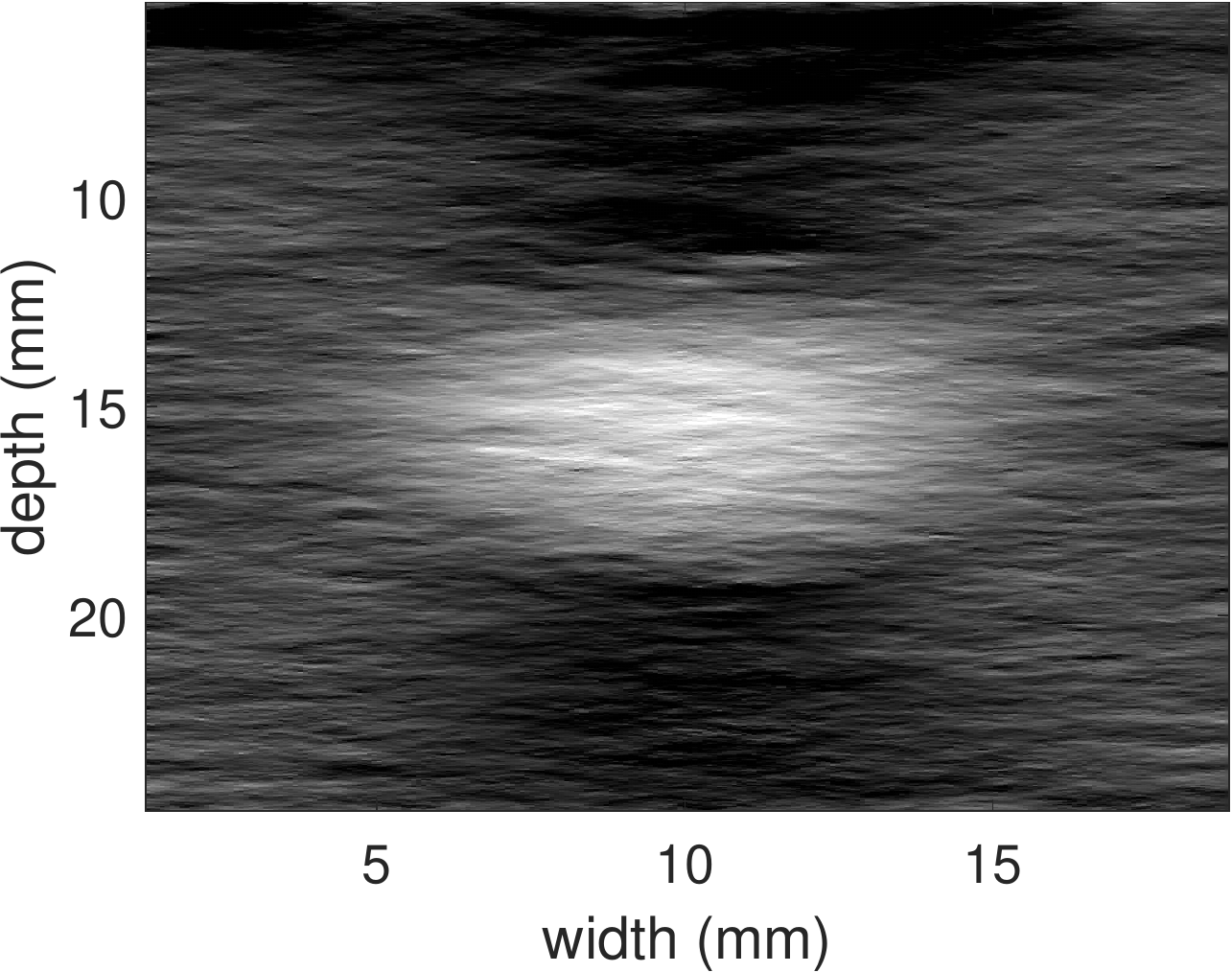} }}
	\subfigure[Strain]{{\includegraphics[width=0.33\textwidth]{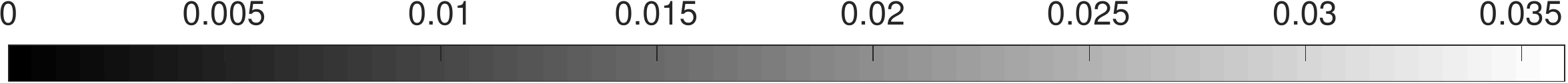}}}%
	\caption{Results for soft-inclusion simulated phantom with added Gaussian noise. Rows 1-3 correspond to 20, 18, and 16 dB PSNR, respectively, whereas columns 1-3 represent the strain images obtained from Hybrid, GLUE, and rGLUE, respectively.}
	\label{circular_noisy}
\end{figure*}


\begin{table*}[tb]  
	\centering
	\caption{SNR and CNR of the strain images for the simulated FEM phantom with added Gaussian noise. CNR is calculated utilizing the blue colored target and red colored background windows depicted in Fig.~\ref{ground}(b). SNR values are calculated from the background window only. rGLUE's improvements over GLUE are also shown.}
	\label{table_circular}
		\begin{tabular}{c c c c c c c c c c c c} 
			\hline
			\multicolumn{1}{c}{} &
			\multicolumn{2}{c}{PSNR = 20 dB} &
			\multicolumn{1}{c}{} &
			\multicolumn{2}{c}{PSNR = 18 dB} &
			\multicolumn{1}{c}{} &
			\multicolumn{2}{c}{PSNR = 16 dB}\\
			\cline{2-3} 
			\cline{5-6}
			\cline{8-9}
			$ $  $ $&    SNR & CNR $ $  $ $&$ $  $ $ &$ $  $ $ SNR & CNR$ $  $ $&$ $  $ $ &$ $  $ $ SNR & CNR\\
			\hline
			Hybrid & 3.23 &  4.66 && 2.53 & 3.76 && 1.89 & 3.52\\
			GLUE &  3.25 &  4.18 && 2.29 & 3.36 && 1.67 & 2.65\\
			rGLUE &  \textbf{5.39}  & \textbf{5.19}  && \textbf{4.55} & \textbf{4.78} && \textbf{3.46} & \textbf{4.27}\\
			Improvement &  65.85\% &  24.16\% && 98.69\% & 42.26\% && 107.19\% & 61.13\%\\
			\hline
		\end{tabular}
\end{table*}

\begin{figure*}
	\centering
	\subfigure[B-mode]{{\includegraphics[width=0.25\textwidth,height=0.25\textwidth]{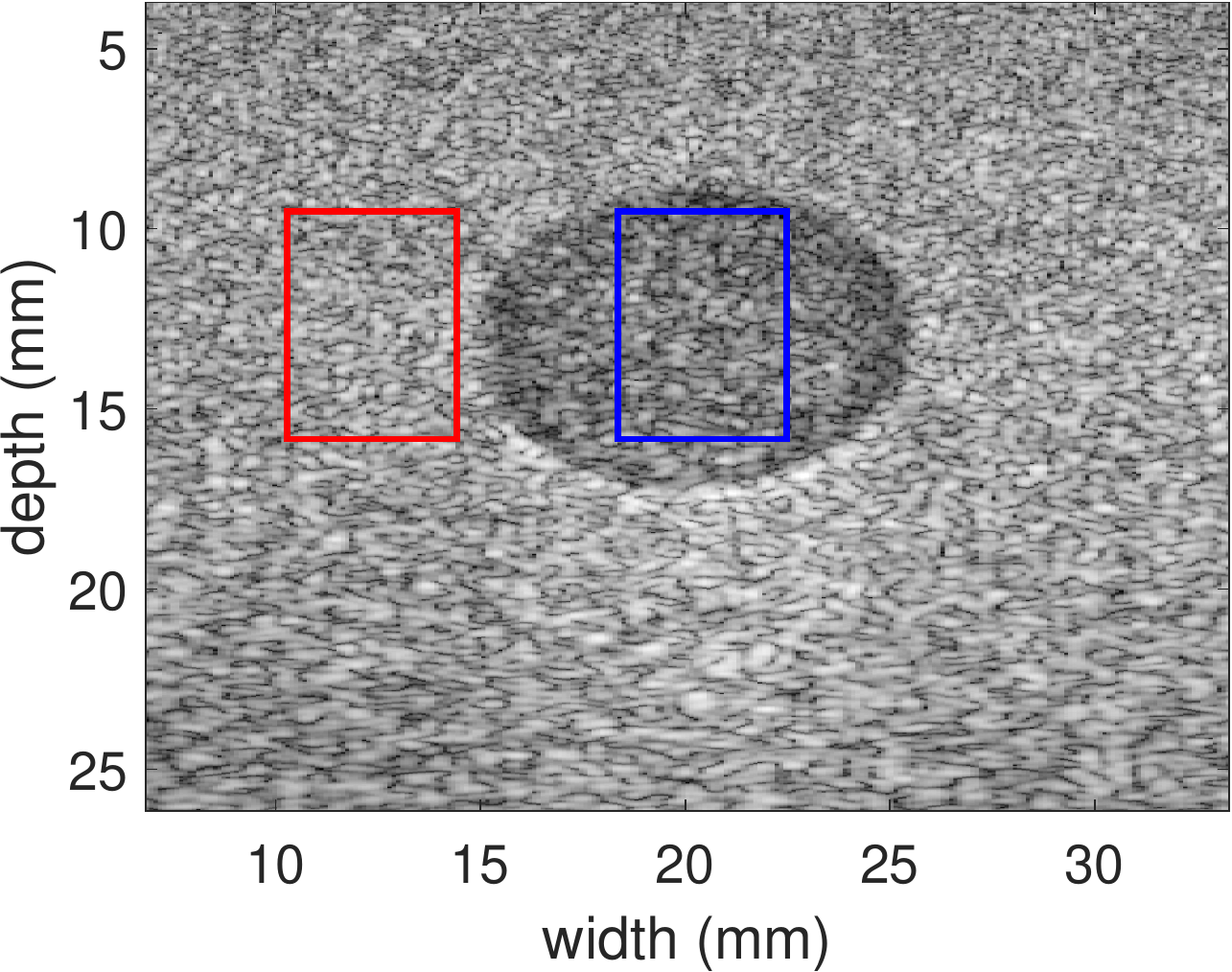} }}%
	\subfigure[Hybrid]{{\includegraphics[width=0.25\textwidth,height=0.25\textwidth]{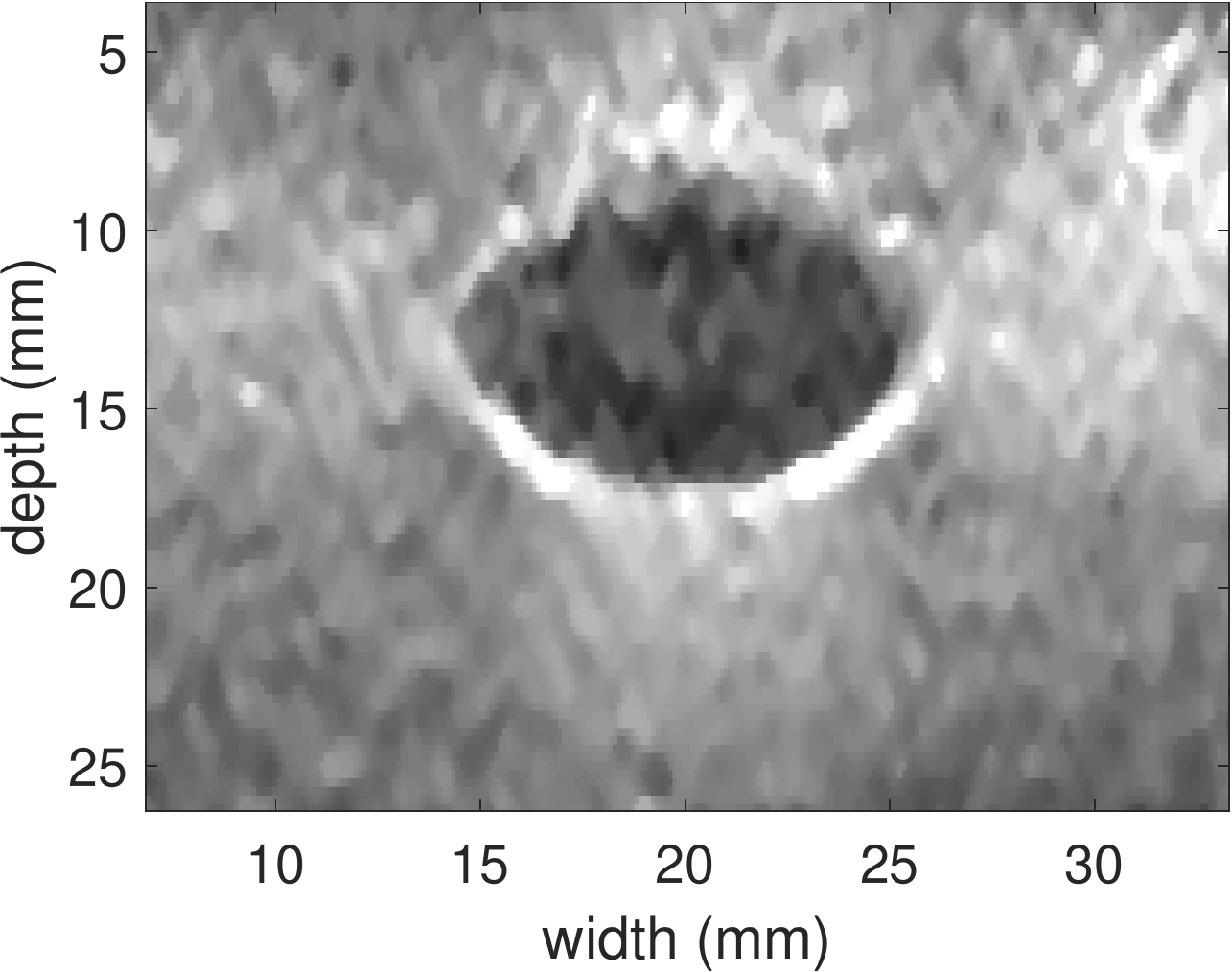} }}%
	\subfigure[GLUE]{{\includegraphics[width=0.25\textwidth,height=0.25\textwidth]{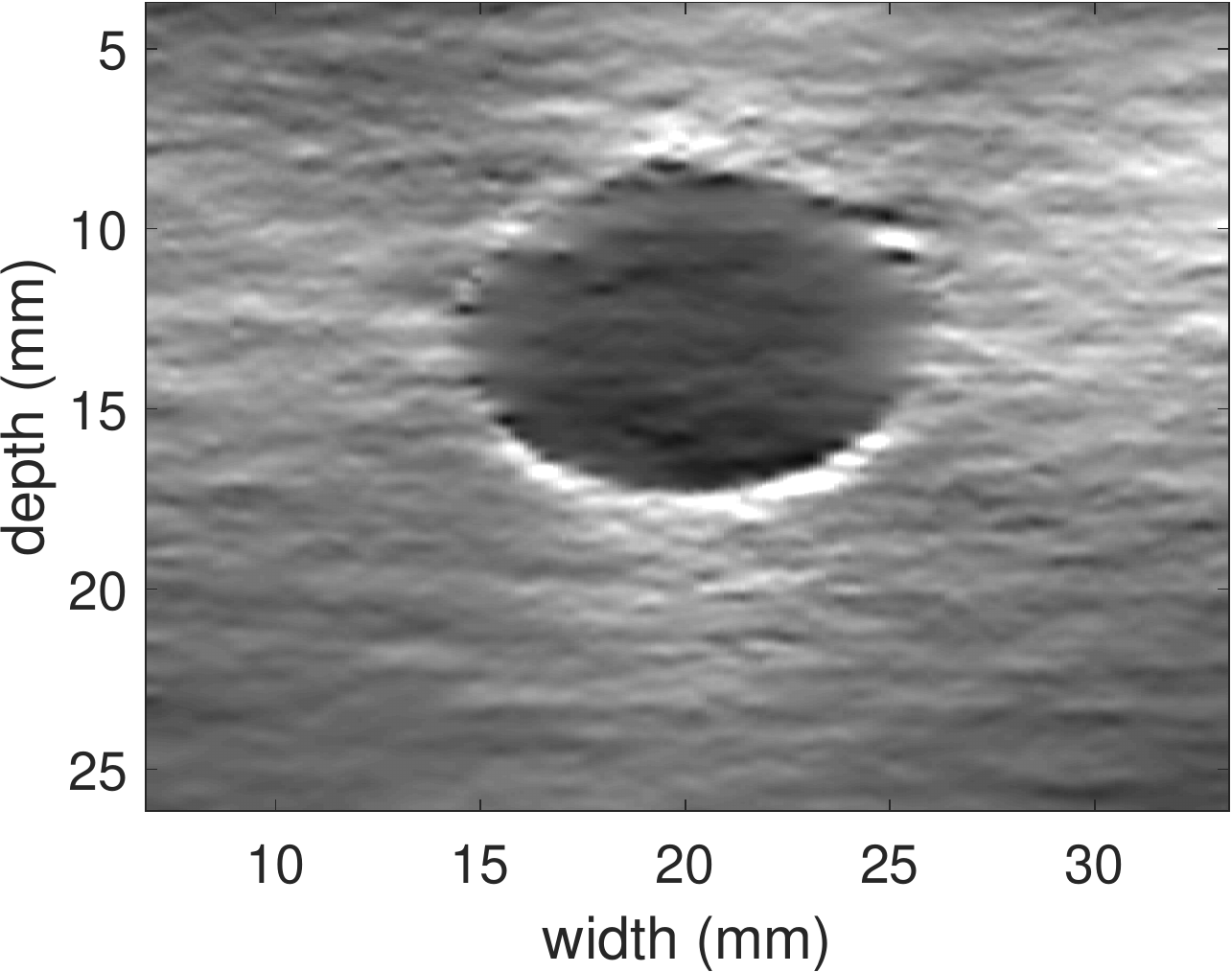}}}%
	\subfigure[rGLUE]{{\includegraphics[width=0.25\textwidth,height=0.25\textwidth]{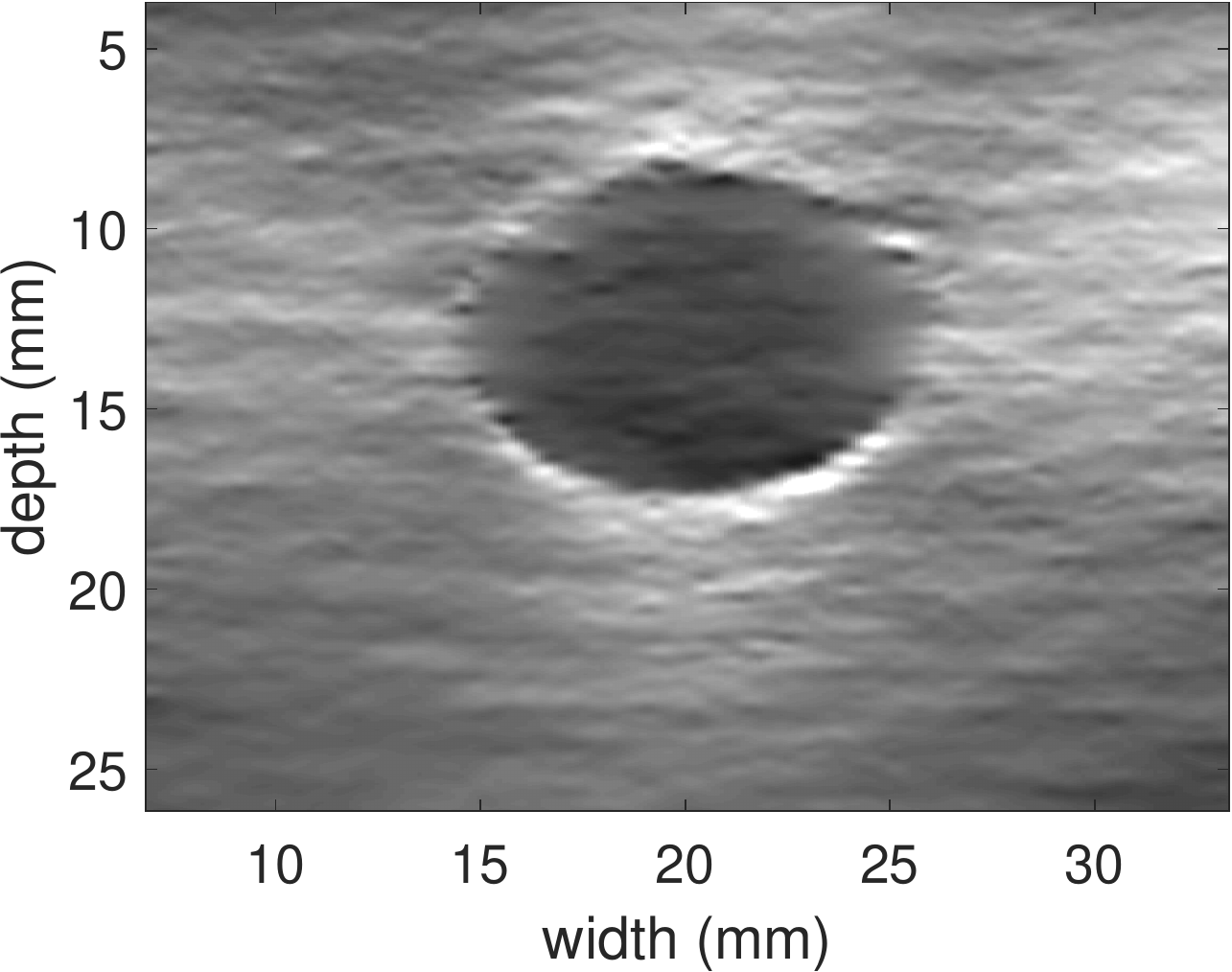} }}
	\subfigure[Strain]{{\includegraphics[width=0.33\textwidth]{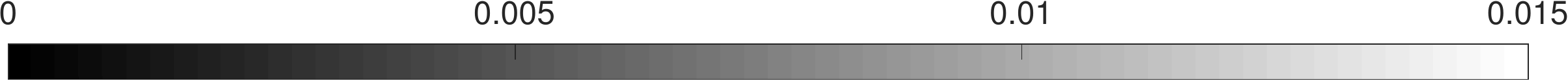}}}%
	\caption{Results for the CIRS breast elastography phantom. Columns 1-4 represent the B-mode image and the strain images obtained from Hybrid, GLUE, and rGLUE, respectively. The blue and red colored windows in (a) show foreground and background windows, respectively.}
	\label{phan_1}
\end{figure*}

\begin{table}[tb]  
	\centering
	\caption{SNR and CNR of the strain images for experimental phantom. CNR is calculated from blue colored target windows and red colored background windows depicted in Fig.~\ref{phan_1}(a). SNR is calculated on the red colored background window. The improvement of rGLUE over GLUE is also shown.}
	\label{table_phan}
	\begin{tabular}{c c c c c c c} 
		\hline
		$ $  $ $&    SNR & CNR\\
		\hline
		Hybrid & 8.78 & 4.81\\
		GLUE & 7.64 &  4.63\\
		rGLUE & \textbf{9.04}  & \textbf{5.50}\\
		Improvement & 18.32\% &  18.79\%\\
		\hline
	\end{tabular}
\end{table}


\subsection{Simulation and data acquisition}
\subsubsection{Simulated data with soft inclusion}
A homogeneous phantom containing an easily deformable $8$ mm diameter cylindrical vessel was designed. The Young's moduli of the cylinder and the background were set to $0$ kPa and $4$ kPa, respectively. The phantom was compressed by different levels from $1\%$ to $4\%$ using ABAQUS~(Providence, RI, USA), a Finite Element software. The pre- and post-compressed RF frames were generated using Field~II~\cite{field2,field22}. The center frequency and the sampling rate were set to $7.27$ MHz and $100$ MHz, respectively, whereas the width and height of the elements were considered to be $0.2$ mm and $5$ mm, respectively. The fractional bandwidth and the number of active elements for beamforming were set to $60\%$ and $64$, respectively.

To analyze the TDE techniques’ robustness to noise, we added three levels of random Gaussian noise with 20 dB, 18 dB, and 16 dB peak SNR (PSNR) to RF data. We swept the noise levels in the aforementioned range to emulate real data collection environment.

\subsubsection{Phantom experiment}
The phantom experiment was conducted at Concordia University's PERFORM Centre with an E-Cube R12 research ultrasound system, Alpinion, USA. RF data were acquired using an L3-12H linear array probe with transmit and sampling frequencies of $10$ MHz and $40$ MHz, respectively, from a tissue-mimicking breast phantom~(Model 059, CIRS; Tissue Simulation \& Phantom Technology, Norfolk, VA, USA) while undergoing compression. The phantom is made of Zerdine$^{\textregistered}$ which represents the reflective properties of human breast with an elasticity modulus of $20 \pm 5$ kPa corresponding to the background. The Young's modulus of the hard inclusion is at least twice that of the background.

\subsubsection{\textit{In vivo} liver datasets}
The \textit{in vivo} liver datasets were collected from two liver cancer patients at The Johns Hopkins Hospital, Baltimore, MD, USA using a Siemens Antares research ultrasound machine with an VF 10-5 linear array probe. The center frequency of the probe and the temporal sampling rate were set to $6.67$ MHz and $40$ MHz, respectively. The tissue deformation was performed by pushing the probe against the liver. The institutional review board approved this \textit{in vivo} study and informed consent was obtained from both patients. More detailed information about this study can be found in~\cite{DPAM}.

\subsubsection{\textit{In vivo} breast data}
The \textit{in vivo} breast experiment was conducted at The University of Kansas Medical Center, KS, USA. RF datasets were acquired from one patient with breast lesion using a commercial ultrasound imaging platform~(SONOLINE Elegra, Siemens Medical Solutions, Ultrasound Group) with a 7.5L40 linear array probe, setting the center and sampling frequencies to $7.2$ MHz and $36$ MHz, respectively. The diagnosis of Invasive Ductal Carcinoma (IDC) was confirmed using biopsy. The data acquisition was carried out following the conventional breast examination procedure with the patients in the supine position. A compression of around $1-1.5\%$ was achieved by pushing the probe towards the chest wall with real-time strain image feedback. All procedures associated with the scan were performed with informed consent of the patient and according to the ethics approval obtained from the institutional review board. Interested readers are encouraged to refer to~\cite{hall2003vivo} for more details regarding this data collection.

\subsection{Quantitative metrics}
A comprehensive comparative study was performed to investigate the superiority of rGLUE over two recently published strain imaging techniques: Hybrid~\cite{hybrid}, a window-based method and GLUE~\cite{GLUE}, a regularized optimization-based technique. Hybrid is chosen as the window-based comparison technique since it masks the outlier samples taking the neighborhood information into account. In addition, Hybrid facilitates the detection of both solid- and fluid-filled lesions combining the results from NCC and speckle-tracking techniques.

Quantitative performance was measured based on root-mean-square error (RMSE), signal-to-noise ratio (SNR), and contrast-to-noise ratio (CNR), three conventional quality metrics. RMSE is given by:

\begin{equation}
\textrm{RMSE}=\sqrt{\frac{\sum\limits_{j=1}^n \sum\limits_{i=1}^m (\hat{s}_{i,j}-s_{i,j})^{2}}{mn}}
\end{equation}

\noindent
where $\hat{s}_{i,j}$ and $s_{i,j}$ stand for the estimated and ground truth strains at $(i,j)$. SNR and CNR are defined as:
\begin{equation}
\textrm{SNR}=\frac{\bar{s_{b}}}{\sigma_{b}} \qquad
\textrm{CNR}=\frac{C}{N}=\sqrt{\frac{2(\bar{{s_{b}}}-\bar{{s_{t}}})^2}{{\sigma_b}^2+{\sigma_t}^2}} 
\end{equation}    
\noindent
where $\bar{s_{b}}$ and $\bar{s_{t}}$ denote the averages on background and target strain windows, respectively. $\sigma_b$ and $\sigma_t$ indicate the standard deviations of the strain values associated with the background and target windows, respectively.

\section{Results}
We evaluated the performance of Hybrid, GLUE, and rGLUE on simulation phantoms, CIRS breast elastography phantom, \textit{in vivo} liver, and breast datasets. The parameters corresponding to each method were carefully tuned to obtain the best attainable result. The regularization parameters of GLUE namely \{$\alpha_{1}$, $\alpha_{2}$, $\beta_{1}$, $\beta_{2}$\} were set to \{4, 0.4, 4, 0.4\}, \{5, 1, 5, 1\}, \{3, 0.6, 3, 0.6\}, \{27.5, 1.375, 27.5, 1.375\}, and \{5, 0.125, 5, 0.125\} for soft-inclusion simulated phantom, simulated layer phantom, experimental phantom, liver, and breast data, respectively. Optimal results from the Hybrid method were obtained by setting the nearest neighbor factors and the weighting factors to 3 and 0.4, respectively for all experiments. Optimal performance of the proposed rGLUE technique was obtained by setting \{$\alpha_{1}$, $\alpha_{2}$, $\beta_{1}$, $\beta_{2}$, $\lambda$\} to \{3, 0.3, 3, 0.3, 20\}, \{4.5, 0.9, 4.5, 0.9, 20\}, \{3, 0.6, 3, 0.6, 20\}, \{25, 1.25, 25, 1.25, 30\}, and \{2.5, 0.0625, 2.5, 0.0625, 100\} for soft-inclusion simulation, layer simulation, breast phantom, liver, and breast data, respectively. For all of our validation experiments, $\gamma$ was set to 0.5.               

\subsection{Simulation Results}
\subsubsection{Soft-inclusion phantom}
The axial strain images for $1-3\%$ compression levels are shown in Fig. 1 of the Supplementary Material which shows that all three algorithms are capable of distinguishing the soft inclusion from the homogeneous background. Although the Hybrid method generates almost uniform strain in the inclusion region, the background suffers from large estimation variance resulting in a noisy strain image. The SNR and CNR plots in Fig.~\ref{snr_cnr} also support our visual perception. For $1-2\%$ compression levels, GLUE and rGLUE exhibit similar strain images and outperform Hybrid. For the strain levels higher than $2\%$, Hybrid fails to generate acceptable strain images. For $2.5\%$ and $3\%$ compression levels, rGLUE marginally outperforms GLUE. Since simulated data does not contain many outlier samples for low strain levels, rGLUE does not yield large improvement over GLUE. Therefore, the results for $1-3\%$ compression levels have been included in the Supplementary Material. However, for $3.5\%$ and $4\%$ strain levels, GLUE suffers from large artifacts which are indicated by green arrow marks in Fig.~\ref{simulation1} of the current document. The proposed rGLUE technique shows its robustness to the outliers by generating a high-quality strain map without any noticeable artifact. Consistent high values of SNR and CNR in Fig.~\ref{snr_cnr} also demonstrate the proposed technique's robustness to outliers and compression levels. To further assess the robustness of quantitative performance to window placement, we select 6 target and 20 background windows to obtain a total of 120 target-background combinations. The histogram of 120 CNR values (for $1\%$ compression) calculated utilizing the aforementioned small window pairs has been reported in Fig.~\ref{cnr_histograms}(a). rGLUE exhibits the highest frequency of high CNR values. The average CNR values obtained from Hybrid, GLUE, and rGLUE are 6.91, 8.35, and 8.67, respectively. This experiment reveals that Hybrid's CNR performance declines when the window locations are sweeped throughout the image. It is worth mentioning that all three techniques exhibit slight edge-blurring. This might happen due to negligible elasticity in the soft inclusion region.

To examine the techniques' robustness to noise, we report the strain images corresponding to this soft-inclusion phantom with three different levels of added Gaussian noise in Fig.~\ref{circular_noisy}. A gradual degradation of Hybrid and GLUE strain quality with the increase of noise power is noticed. rGLUE efficiently handles the noise components to preserve the strain image quality which is proven by the SNR and CNR values reported in Table~\ref{table_circular}. It is worth mentioning that the quantitative values in this case have been calculated on circular target and rectangular background windows shown in Fig.~\ref{ground}(b) which demonstrates the assessment criteria's robustness to window shape.

Figs. 2 and 3 of the Supplementary Material depict that Hybrid and GLUE suffer from noticeable artifacts while working with a dataset containing multiplicative or additive outliers. In contrast, rGLUE efficiently handles the outlier samples to minimize the strain artifacts.


\begin{figure*}
	\centering
	\subfigure[B-mode]{{\includegraphics[width=0.25\textwidth,height=0.28\textwidth]{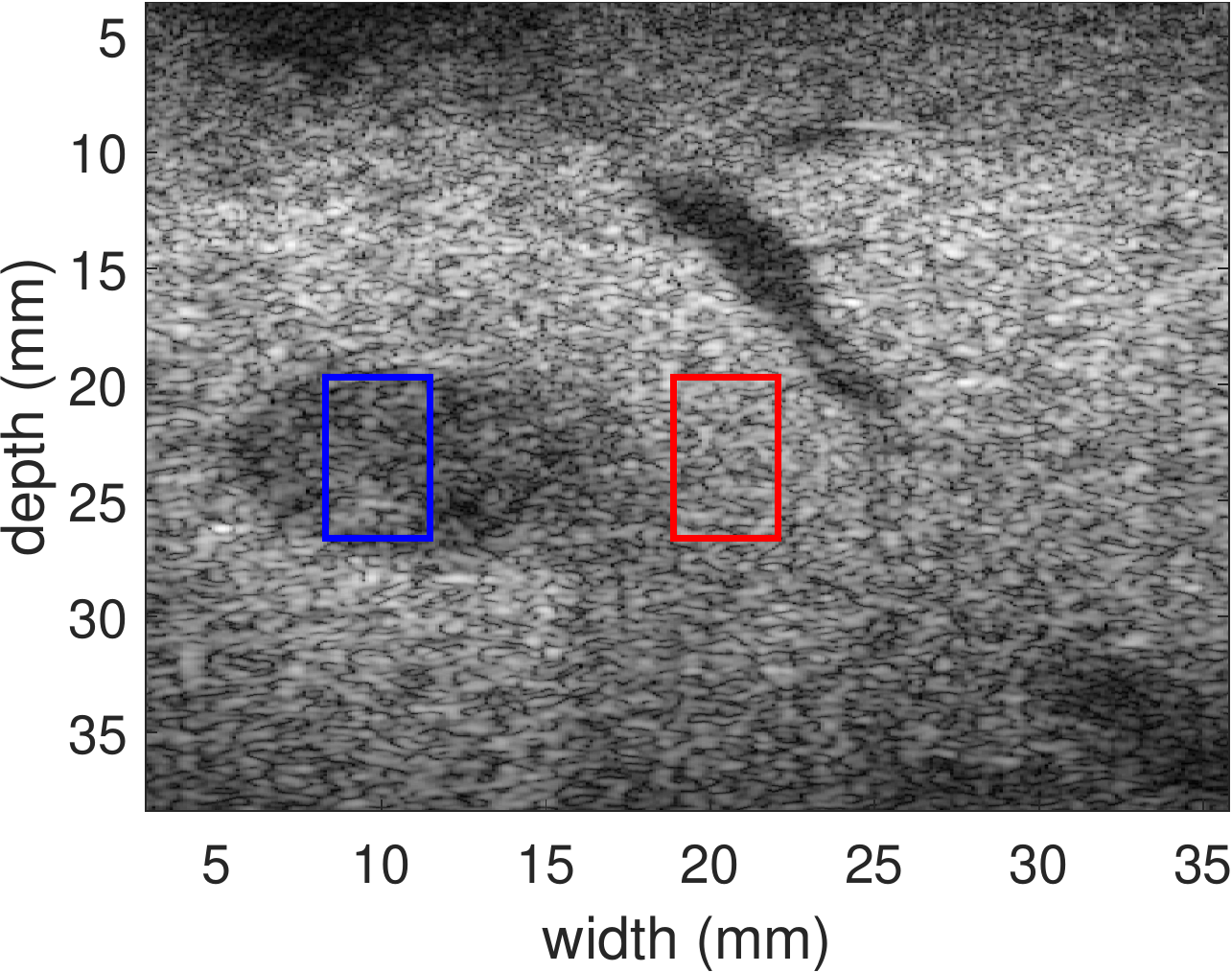} }}%
	\subfigure[Hybrid]{{\includegraphics[width=0.25\textwidth,height=0.28\textwidth]{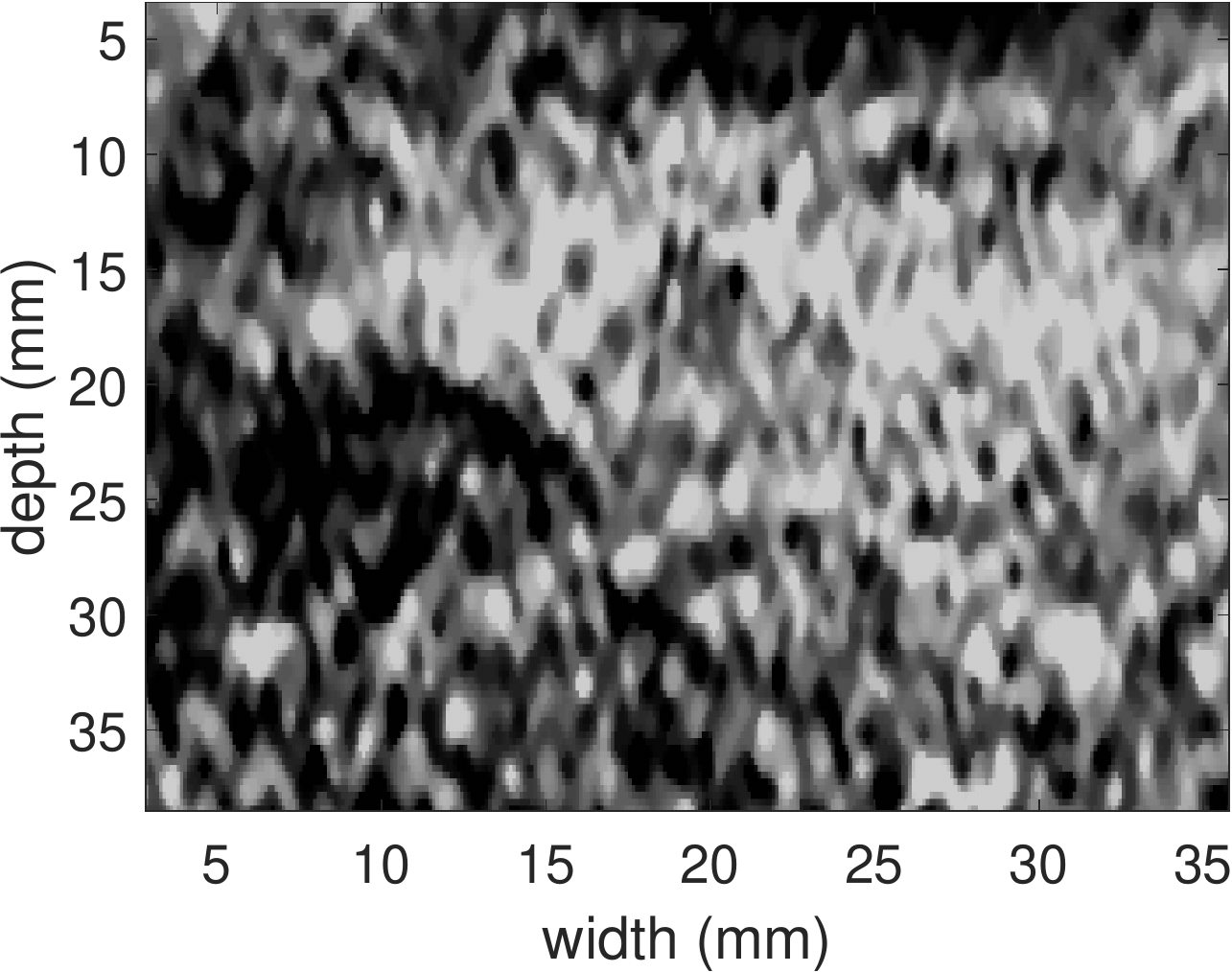} }}%
	\subfigure[GLUE]{{\includegraphics[width=0.25\textwidth,height=0.28\textwidth]{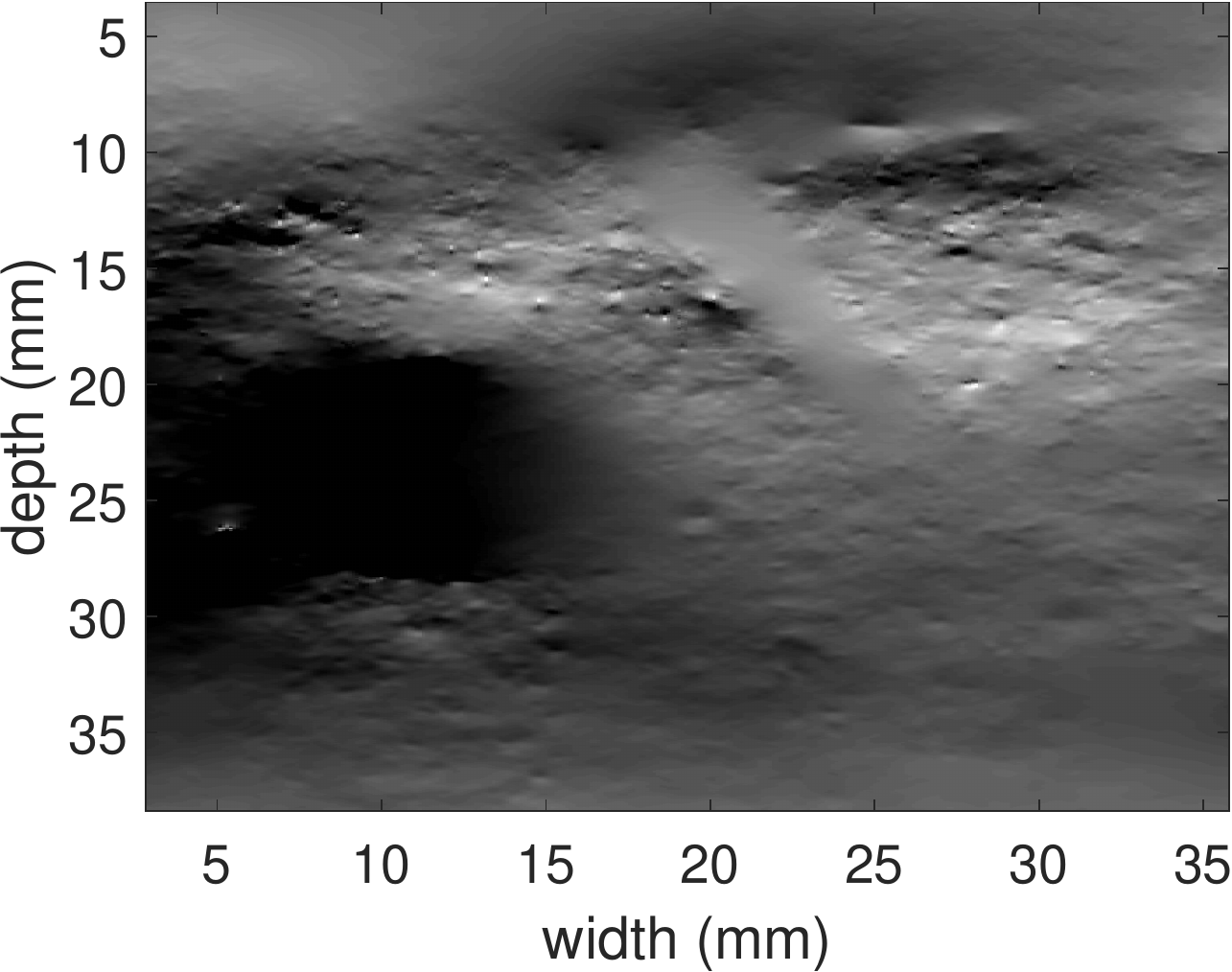} }}%
	\subfigure[rGLUE]{{\includegraphics[width=0.25\textwidth,height=0.28\textwidth]{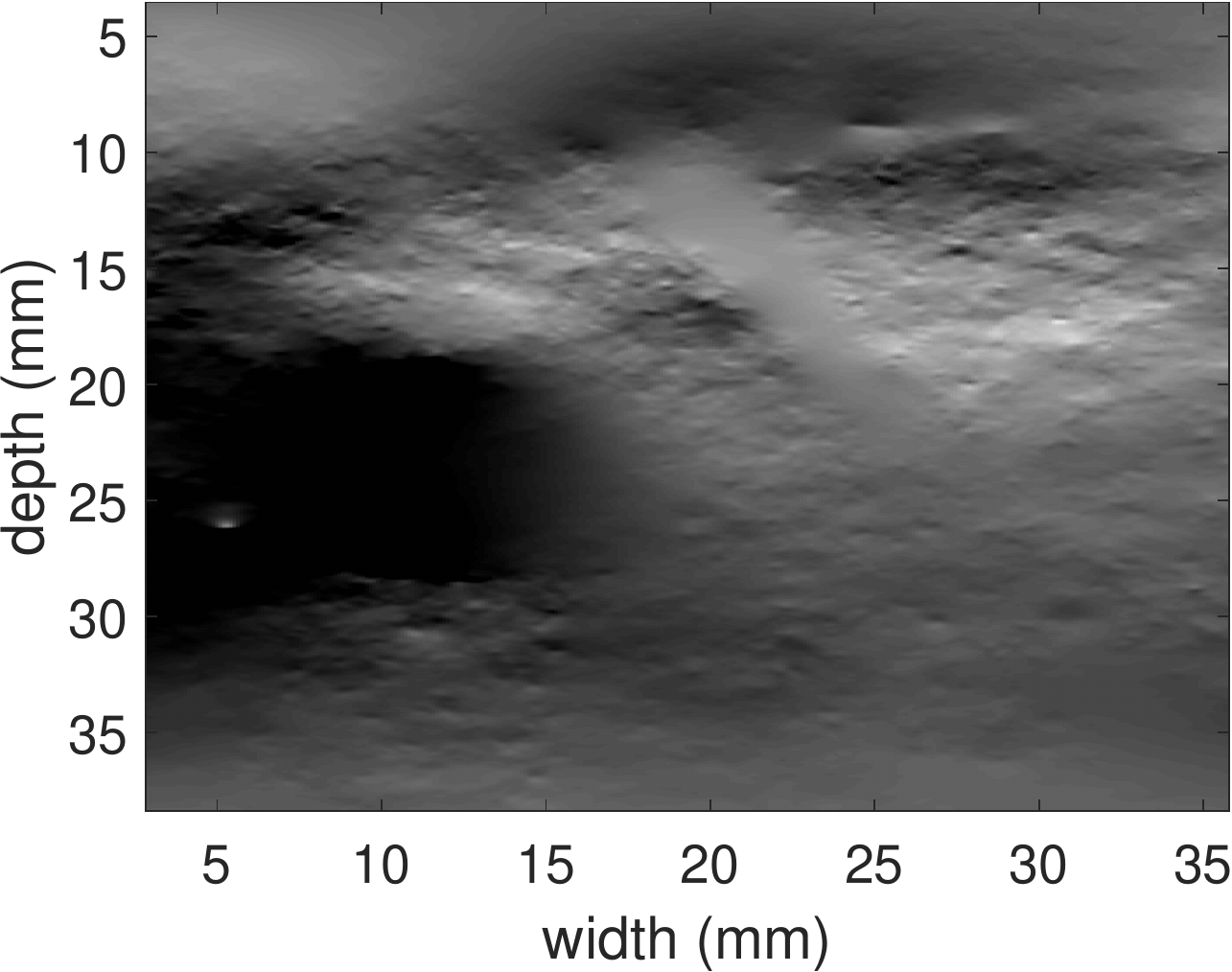} }}
	\subfigure[Strain]{{\includegraphics[width=0.3\textwidth]{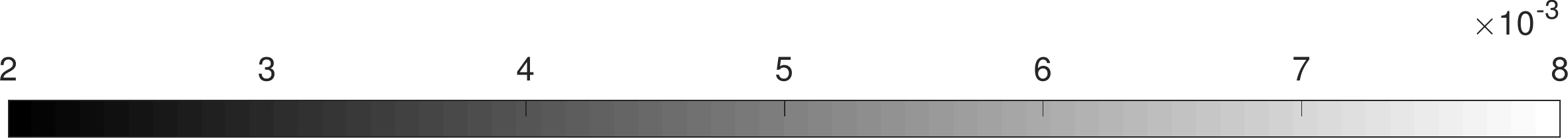}}}
	\caption{Results for the liver patient 1. (a) represents the B-mode image. (b)-(d) show the strain images obtained from Hybrid, GLUE, and rGLUE, respectively. (e) shows the color bar. The blue and red colored windows show foreground and background windows, respectively. The least-squares window length is set to 3 for calculating strain from displacement.}
	\label{liver_p2}
\end{figure*}

\begin{table}[tb]  
	\centering
	\caption{SNR and CNR of the strain images for the \textit{in vivo} liver patient 1. CNR is calculated from blue colored target and red colored background windows depicted in Fig.~\ref{liver_p2}(a). SNR is calculated on the red colored background window. rGLUE's improvement over GLUE is also reported.} 
	\label{table_liver}
	\begin{tabular}{c c c c c c c} 
		\hline
		$ $  $ $&    SNR & CNR\\
		\hline
		Hybrid & 4.75 & 2.85\\
		GLUE & 10.77 &  8.87\\
		rGLUE & \textbf{12.26}  & \textbf{9.75}\\
		Improvement & 13.83\% &  9.92\%\\
		\hline
	\end{tabular}
\end{table}

\begin{table}[tb]  
	\centering
	\caption{SNR and CNR for the \textit{in vivo} liver patient 2. CNR is calculated using the blue colored target and red colored background windows depicted in Fig.~\ref{liver_p3}(d). SNR is calculated on the background window only. The improvement of rGLUE over GLUE is also reported.} 
	\label{table_liver_p3}
	\begin{tabular}{c c c c c c c} 
		\hline
		$ $  $ $&    SNR & CNR\\
		\hline
		Hybrid & 10.93 & 5.79\\
		GLUE & 17.51 &  5.67\\
		rGLUE & \textbf{17.97}  & \textbf{6.36}\\
		Improvement & 2.63\% &  12.17\%\\
		\hline
	\end{tabular}
\end{table}

\begin{table}[tb]  
	\centering
	\caption{SNR and CNR of the strain images for the \textit{in vivo} breast dataset. CNR values are calculated from blue colored target window and red colored background window depicted in Fig.~\ref{breast_p1}(a). SNR values are calculated on the red colored background window. rGLUE's improvement over GLUE is also calculated.}
	\label{table_breast}
	\begin{tabular}{c c c c c c c} 
		\hline
		$ $  $ $&    SNR & CNR\\
		\hline
		Hybrid & \textbf{11.28} & 6.56\\
		GLUE & 7.23 &  8.40\\
		rGLUE & 8.87  & \textbf{10.49}\\
		Improvement & 22.68\% &  24.88\%\\
		\hline
	\end{tabular}
\end{table}

\subsubsection{Layer phantoms}
Fig. 4 of the Supplementary Material demonstrates that Hybrid suffers from extensive strain variability while GLUE exhibits background artifacts in case of the four-layer phantom. rGLUE resolves the issues associated with Hybrid and GLUE which is substantiated by the RMSE values reported in Table I of the Supplementary Material.

The axial strain estimates for the thin-layer and low-contrast phantoms (Figs. 5 and 6 of the Supplementary Material) indicate that Hybrid fails to detect the hard layers in both cases and the outlier regions are visible in the GLUE strain images. rGLUE obtains high-quality strain maps in both phantoms which is corroborated by the RMSE values presented in Table I of the Supplementary Material.      

\subsection{Phantom Results}
The B-mode image and the axial strain images obtained from the CIRS breast elastography phantom have been reported in Fig.~\ref{phan_1}. It has been observed that all three algorithms show good contrast between the low-strain region corresponding to the hard inclusion and the uniform background. Both GLUE and rGLUE obtain spatially smooth strain images, whereas Hybrid yields undesired strain fluctuations in both background and target tissue regions. rGLUE generates a substantially better strain map than Hybrid and GLUE by removing the unexpected nonuniformities induced by the outlier samples. It is worth mentioning that the SNR values calculated on the red colored background window~(Fig.~\ref{phan_1}(a)) and the CNR values computed between the blue colored target and red colored background windows~(Fig.~\ref{phan_1}(a)) reported in Table~\ref{table_phan} substantiate our visual judgement. To further demonstrate the techniques' quantitative performance at different spatial locations, we calculate the histogram of 120 CNR values (see Fig.~\ref{cnr_histograms}(b)) between 6 target and 20 background windows. It is evident that rGLUE dominates in occupying the high CNR values. The average of the aforementioned 120 CNR values obtained by Hybrid, GLUE, and rGLUE are 4.70, 4.98, and 5.85, respectively. Fig. 7 of the Supplementary Material demonstrates that Hybrid and GLUE, respectively, exhibit global and local strain alterations in response to locally-inserted outliers, whereas the rGLUE strain remains unaffected.

\begin{figure*}
	\centering
	\subfigure[B-mode]{{\includegraphics[width=0.24\textwidth,height=0.33\textwidth]{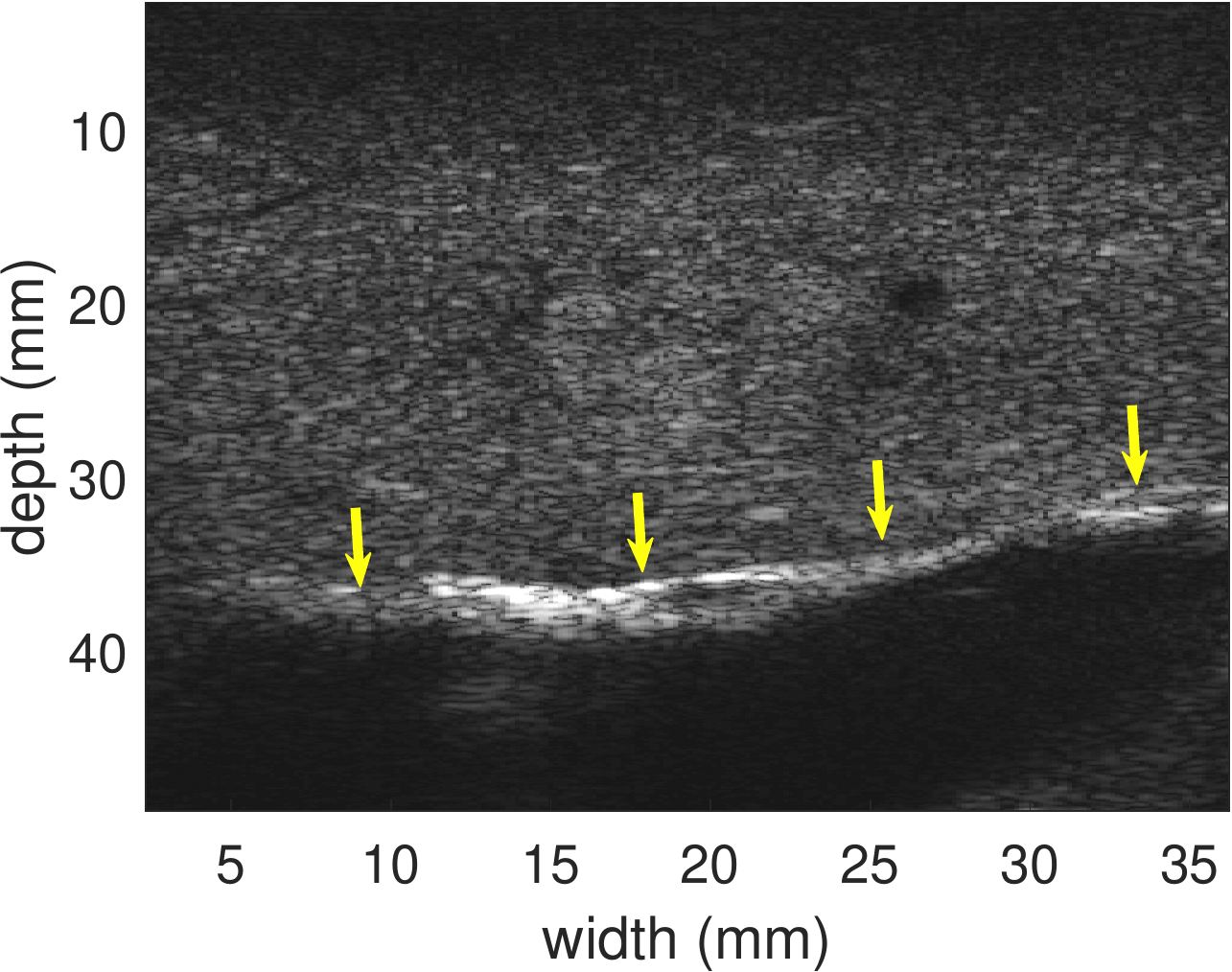} }}%
	\subfigure[Hybrid]{{\includegraphics[width=0.24\textwidth,height=0.33\textwidth]{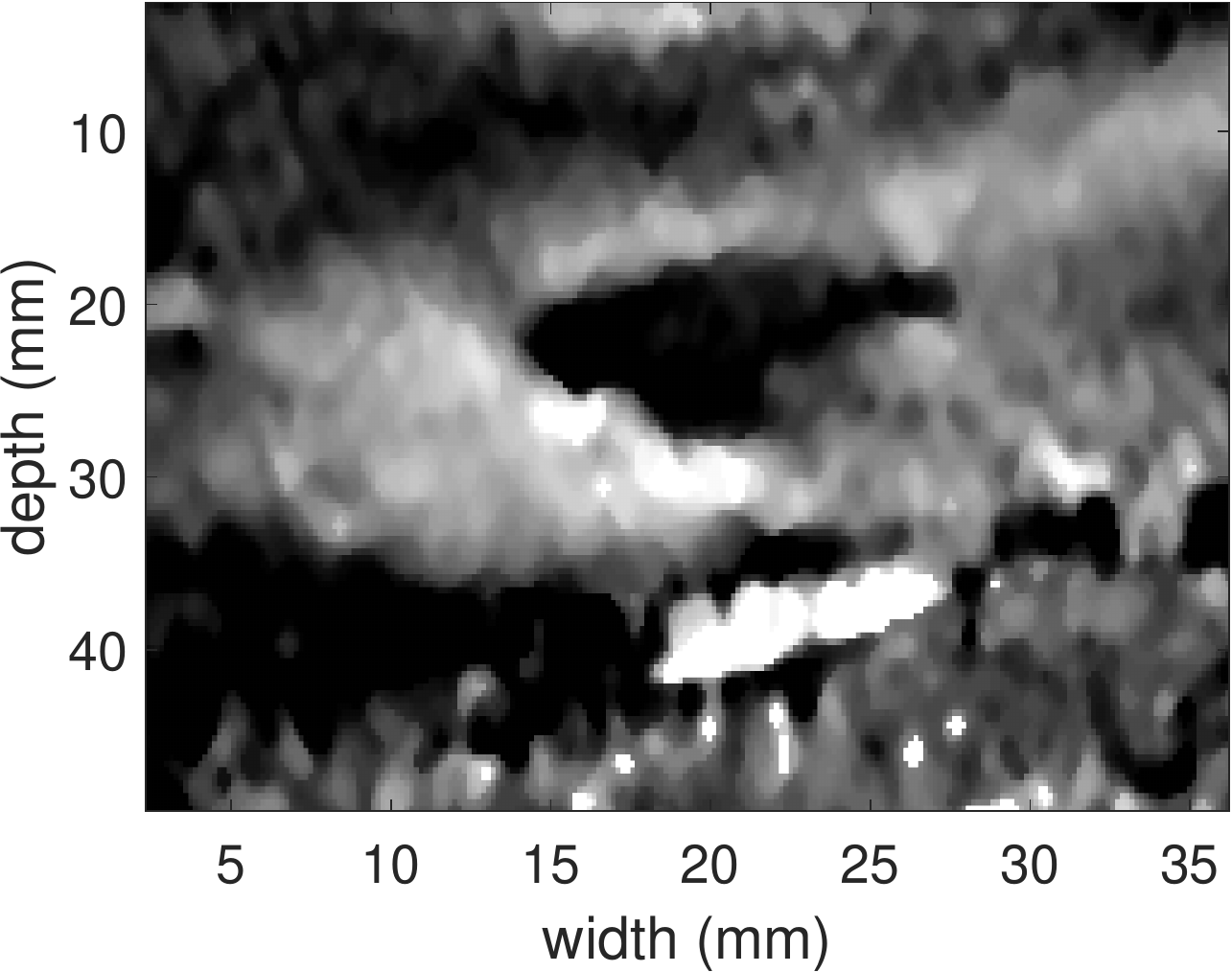} }}%
	\subfigure[GLUE]{{\includegraphics[width=0.24\textwidth,height=0.33\textwidth]{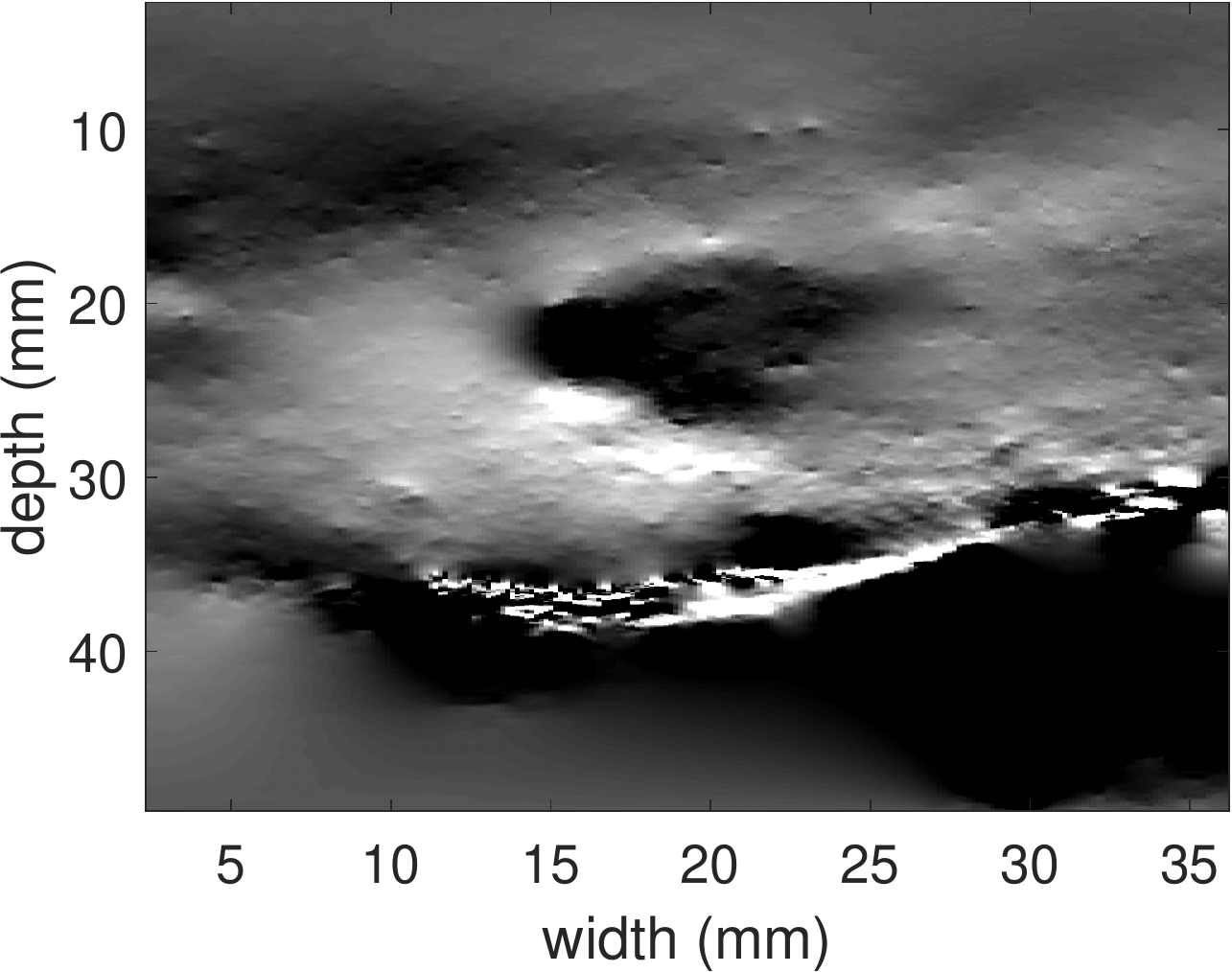} }}%
	\subfigure[rGLUE]{{\includegraphics[width=0.24\textwidth,height=0.33\textwidth]{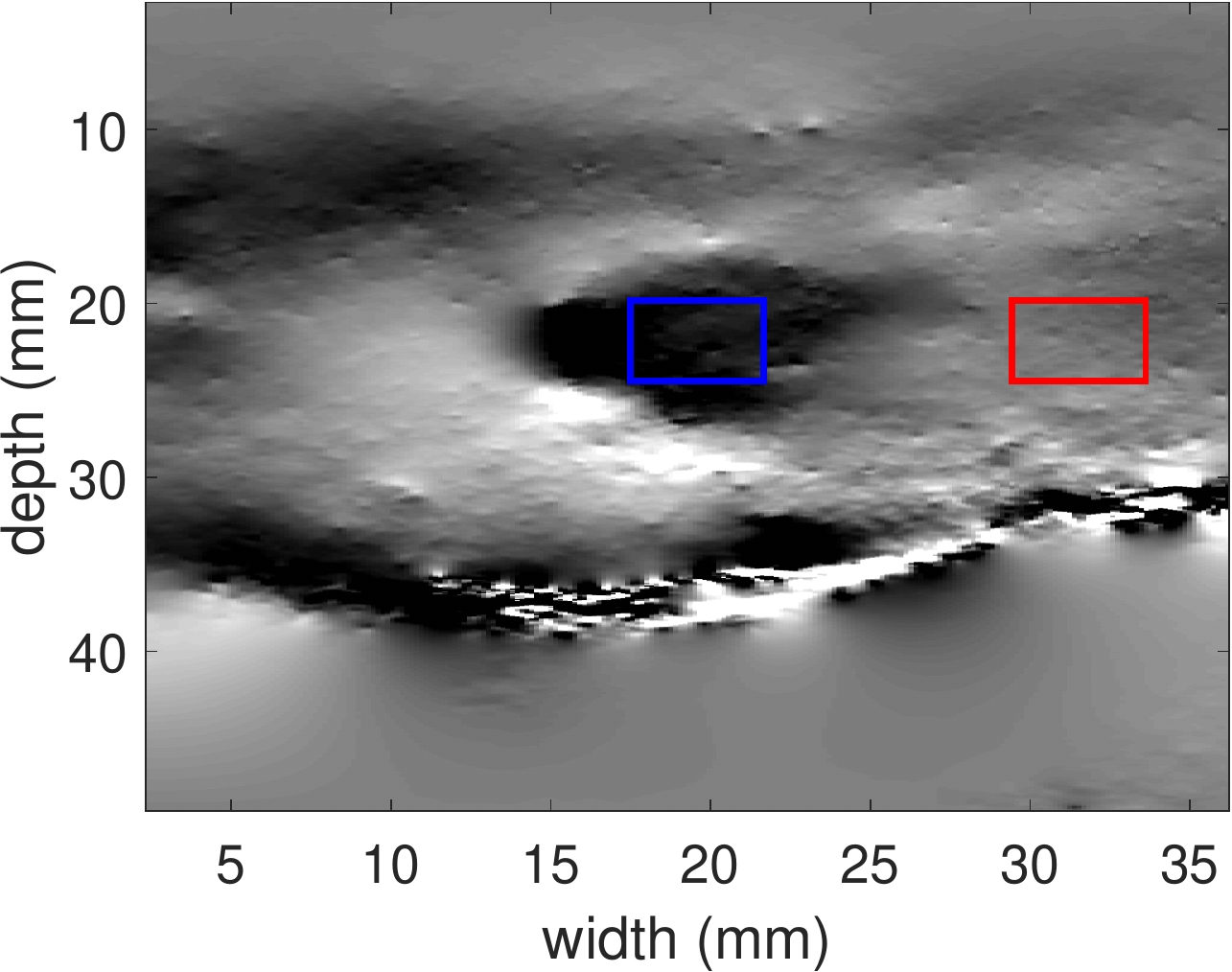} }}
	\subfigure[Strain]{{\includegraphics[width=0.3\textwidth]{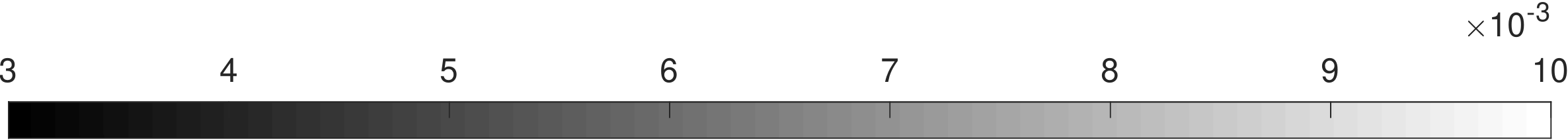}}}
	\caption{Results for the liver patient 2. Columns 1-4 show the B-mode image and the strain images obtained from Hybrid, GLUE, and rGLUE, respectively. The blue and red colored windows show foreground and background windows, respectively. The yellow arrows indicate the rib bones. The least-squares window length is set to 3 for calculating strain from displacement.}
	\label{liver_p3}
\end{figure*}

\begin{figure*}
	\centering
	\subfigure[B-mode]{{\includegraphics[width=0.25\textwidth]{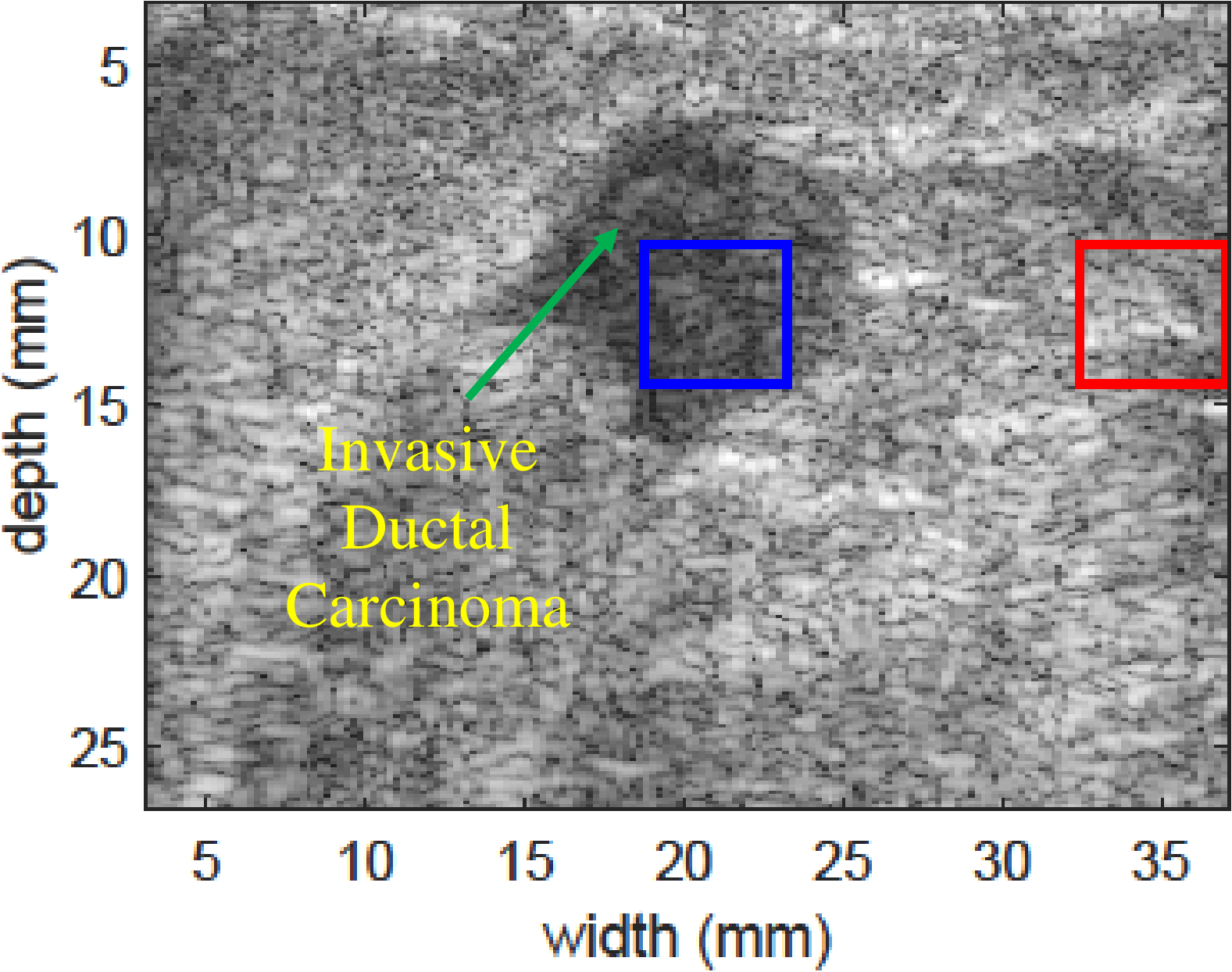} }}%
	\subfigure[Hybrid]{{\includegraphics[width=0.25\textwidth]{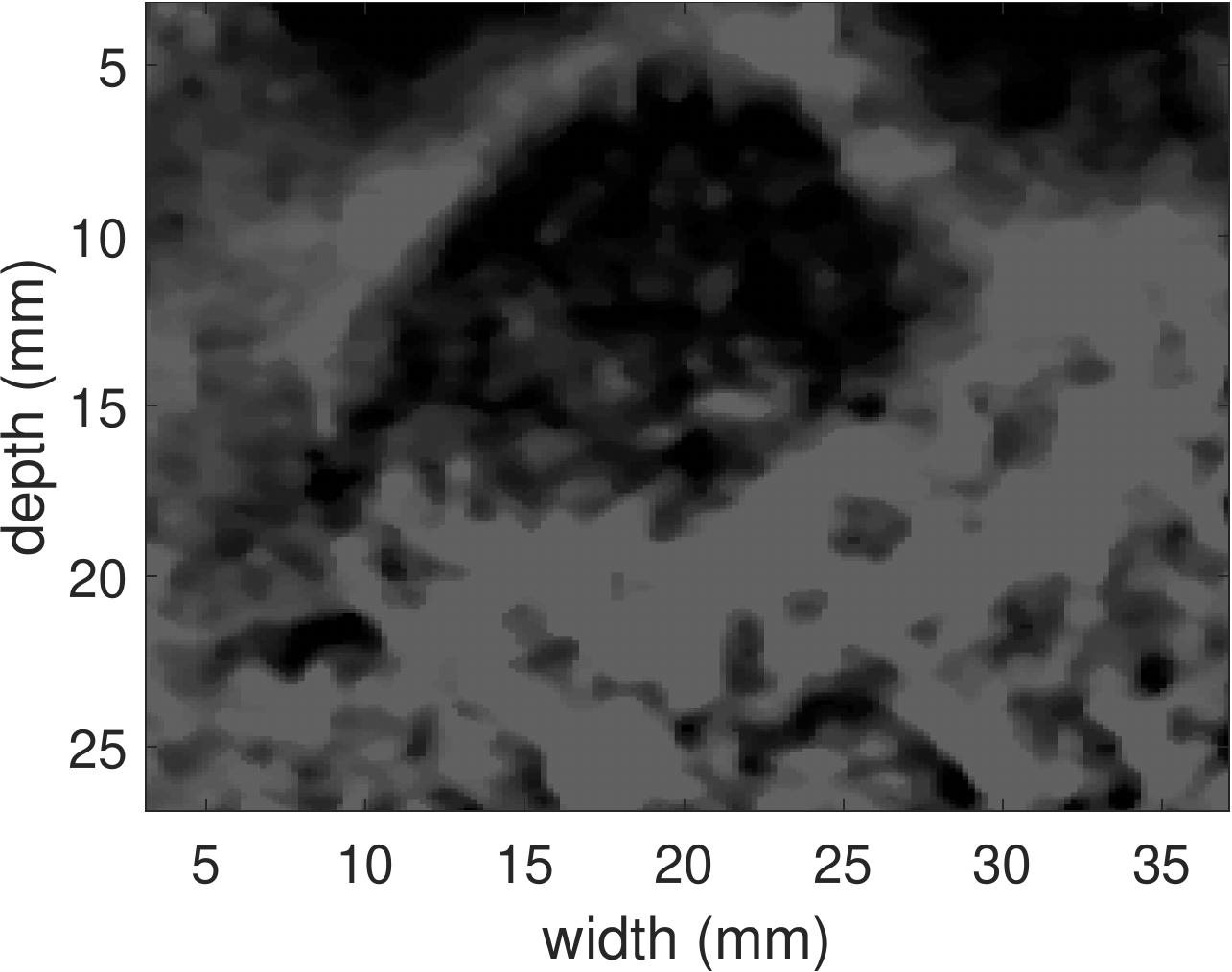} }}%
	\subfigure[GLUE]{{\includegraphics[width=0.25\textwidth]{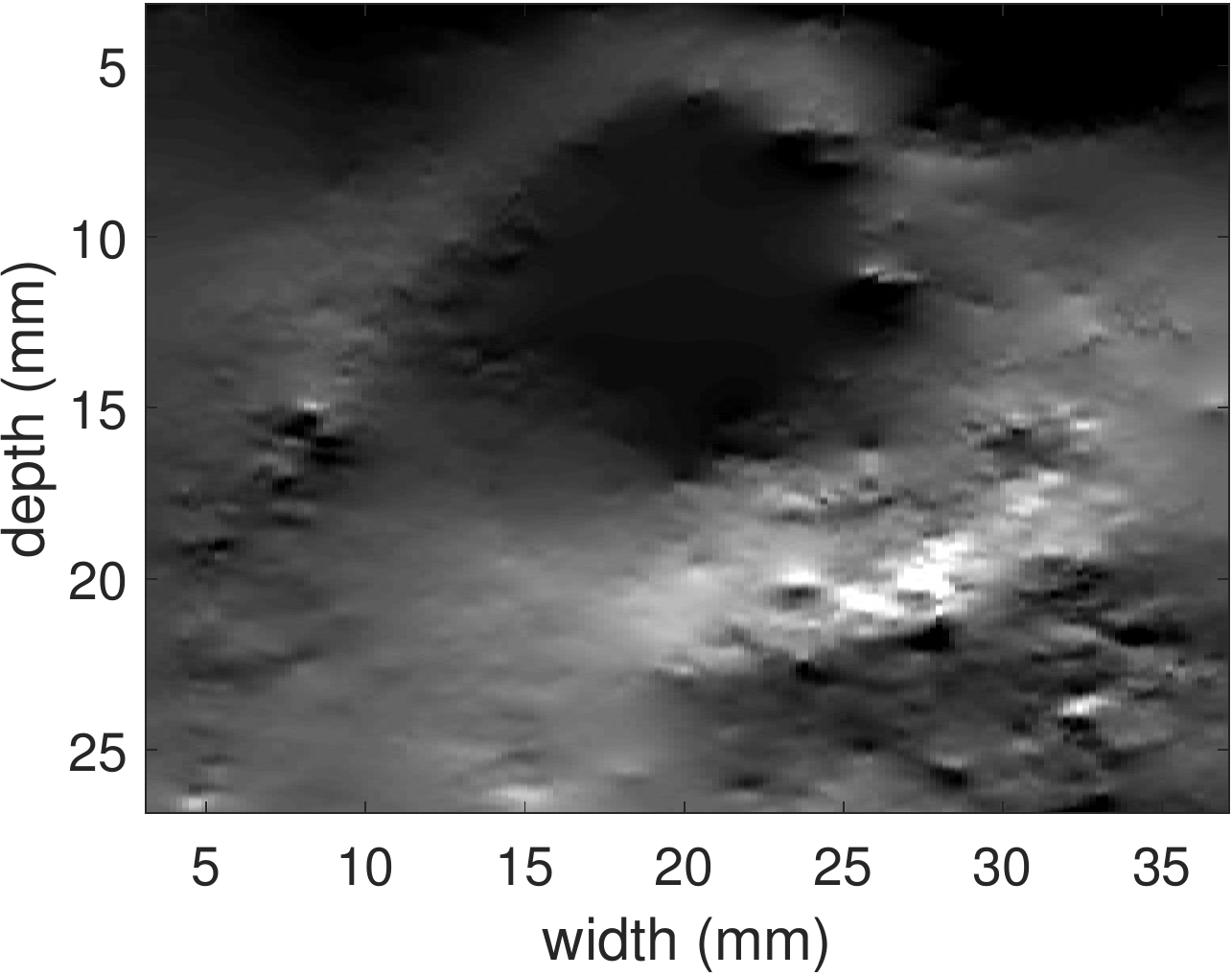} }}%
	\subfigure[rGLUE]{{\includegraphics[width=0.25\textwidth]{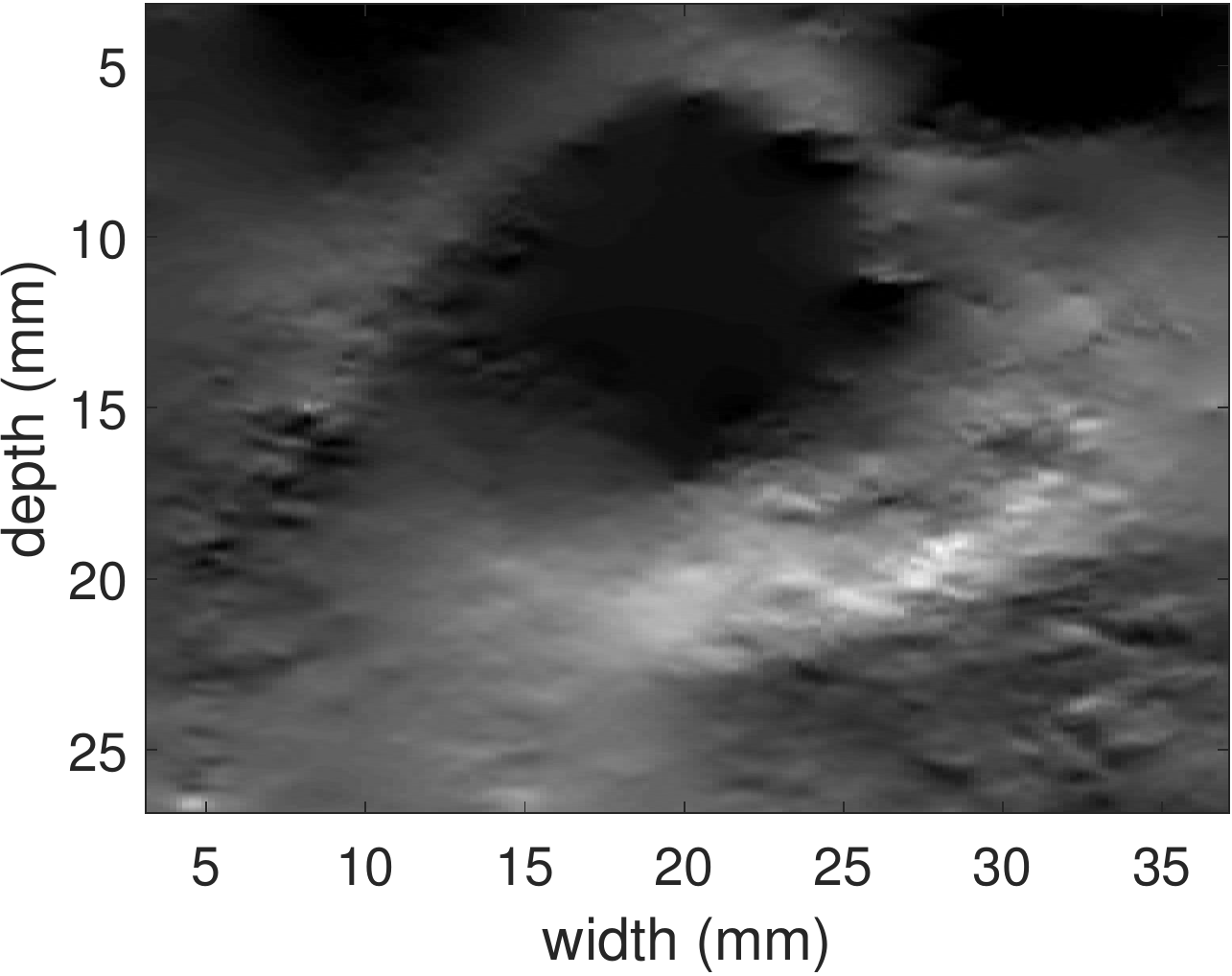} }}
	\subfigure[Strain]{{\includegraphics[width=0.3\textwidth]{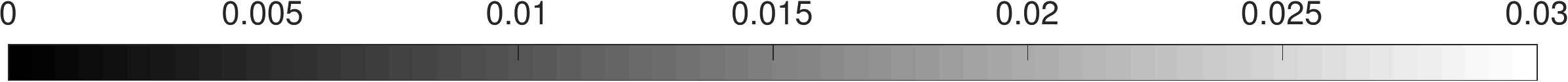} }}
	\caption{Results of \textit{in vivo} breast data. Columns 1-4 represent the B-mode image and the strain images obtained from Hybrid, GLUE, and rGLUE, respectively. (e) shows the color bar. The blue and red colored windows show foreground and background windows, respectively.}
	\label{breast_p1}
\end{figure*}

\begin{figure*}
	\centering
	\subfigure[Simulated phantom]{{\includegraphics[width=.2\textwidth]{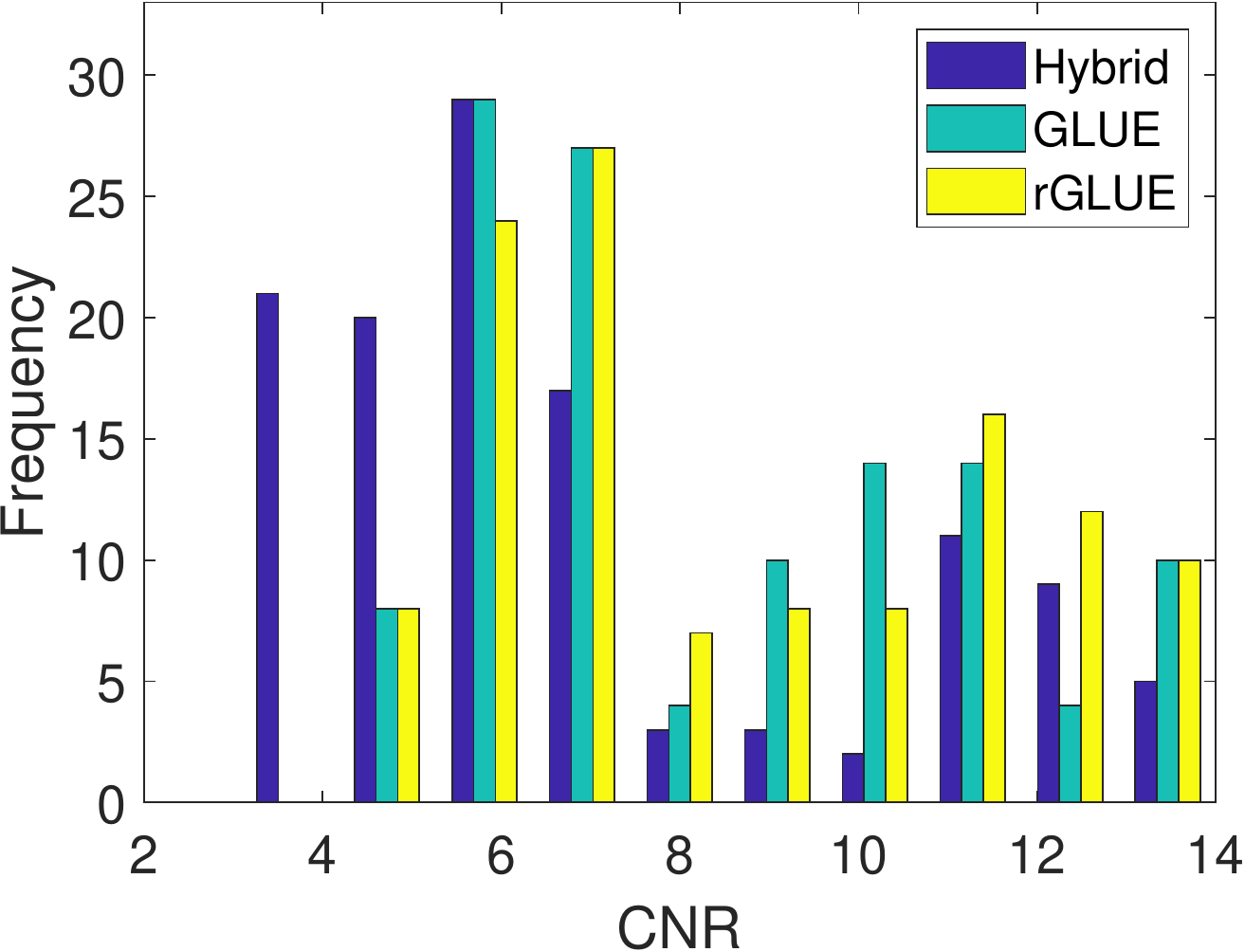}}}%
	\subfigure[Breast phantom]{{\includegraphics[width=.2\textwidth]{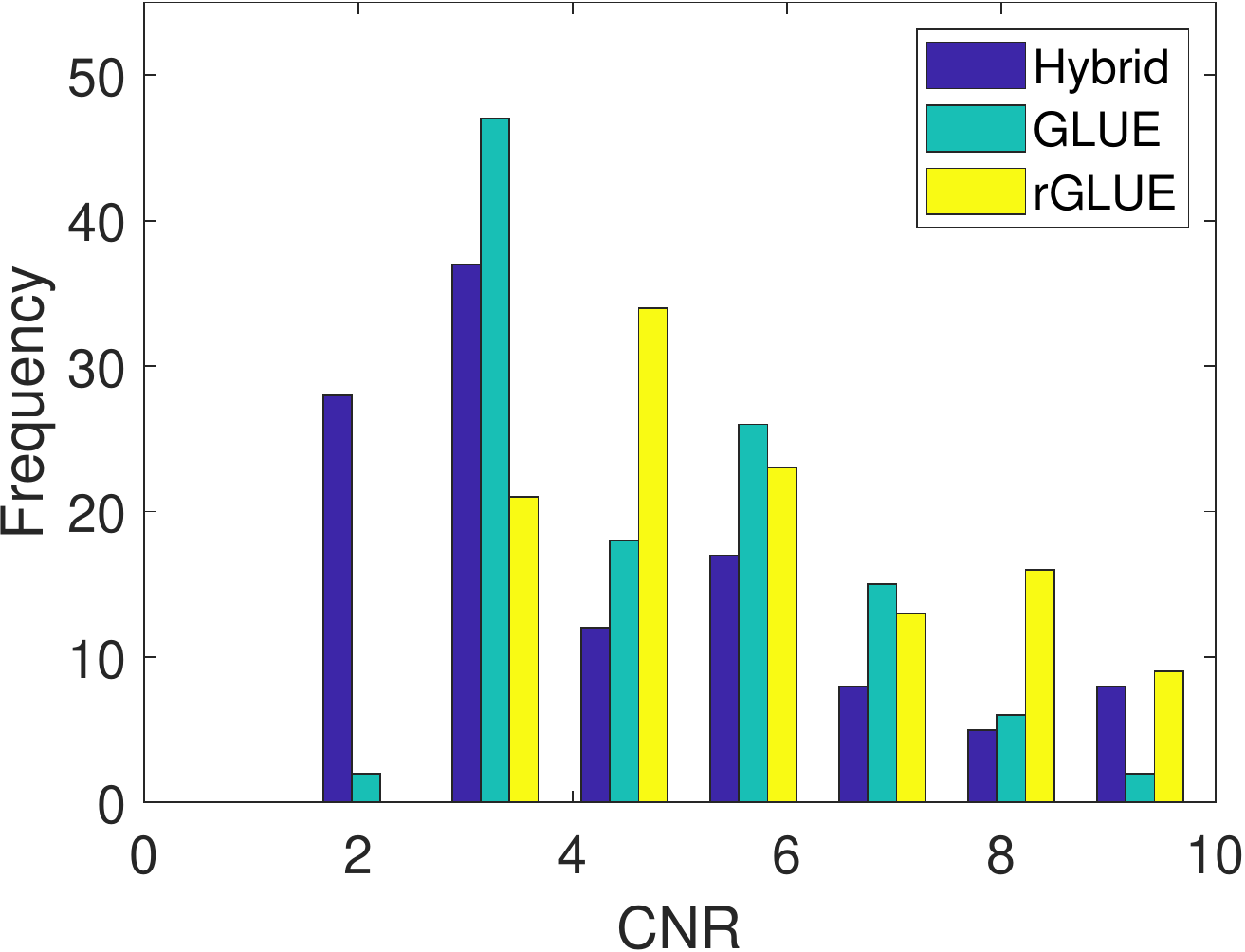}}}%
	\subfigure[\textit{In vivo} liver data 1]{{\includegraphics[width=.2\textwidth]{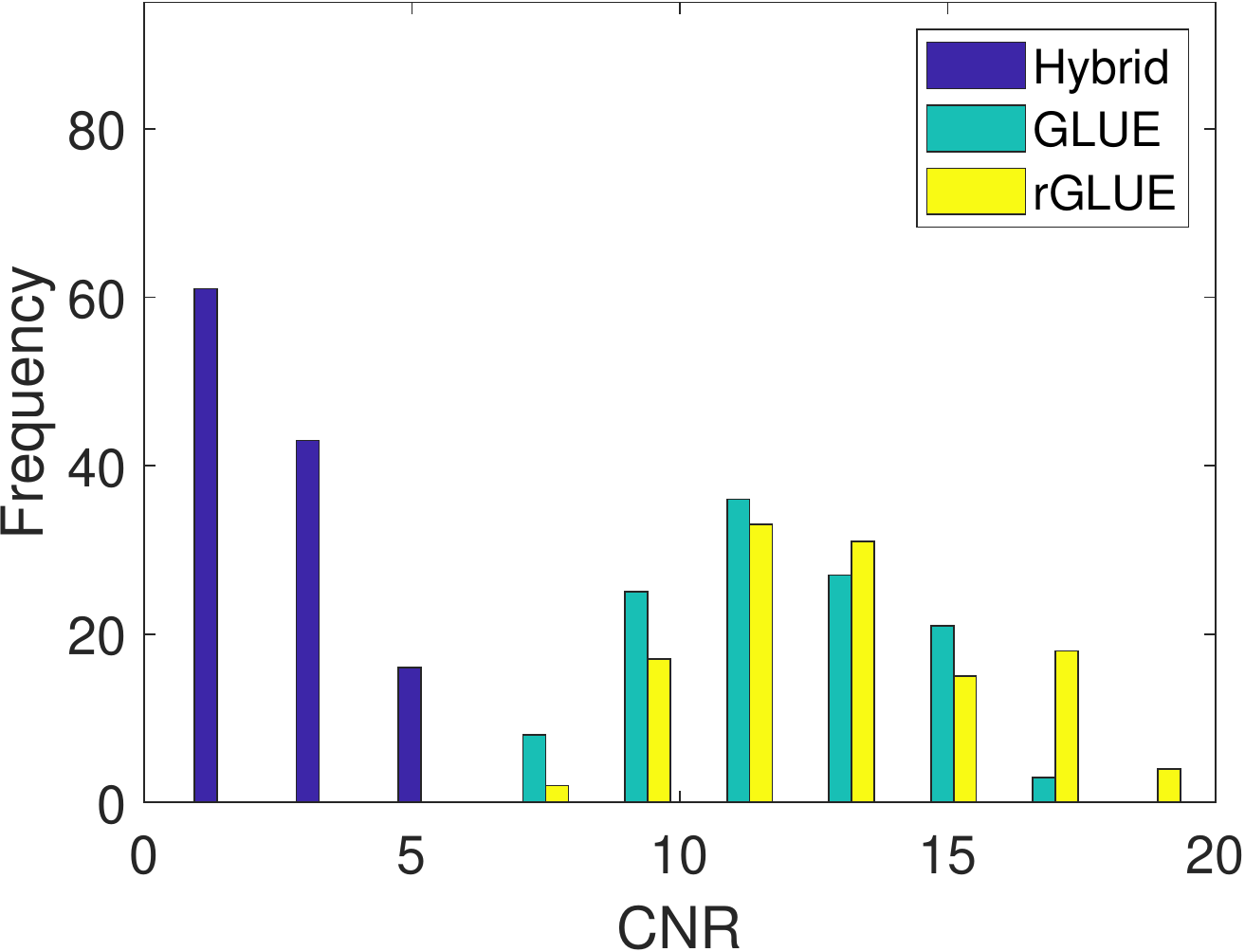} }}%
	\subfigure[\textit{In vivo} liver data 2]{{\includegraphics[width=.2\textwidth]{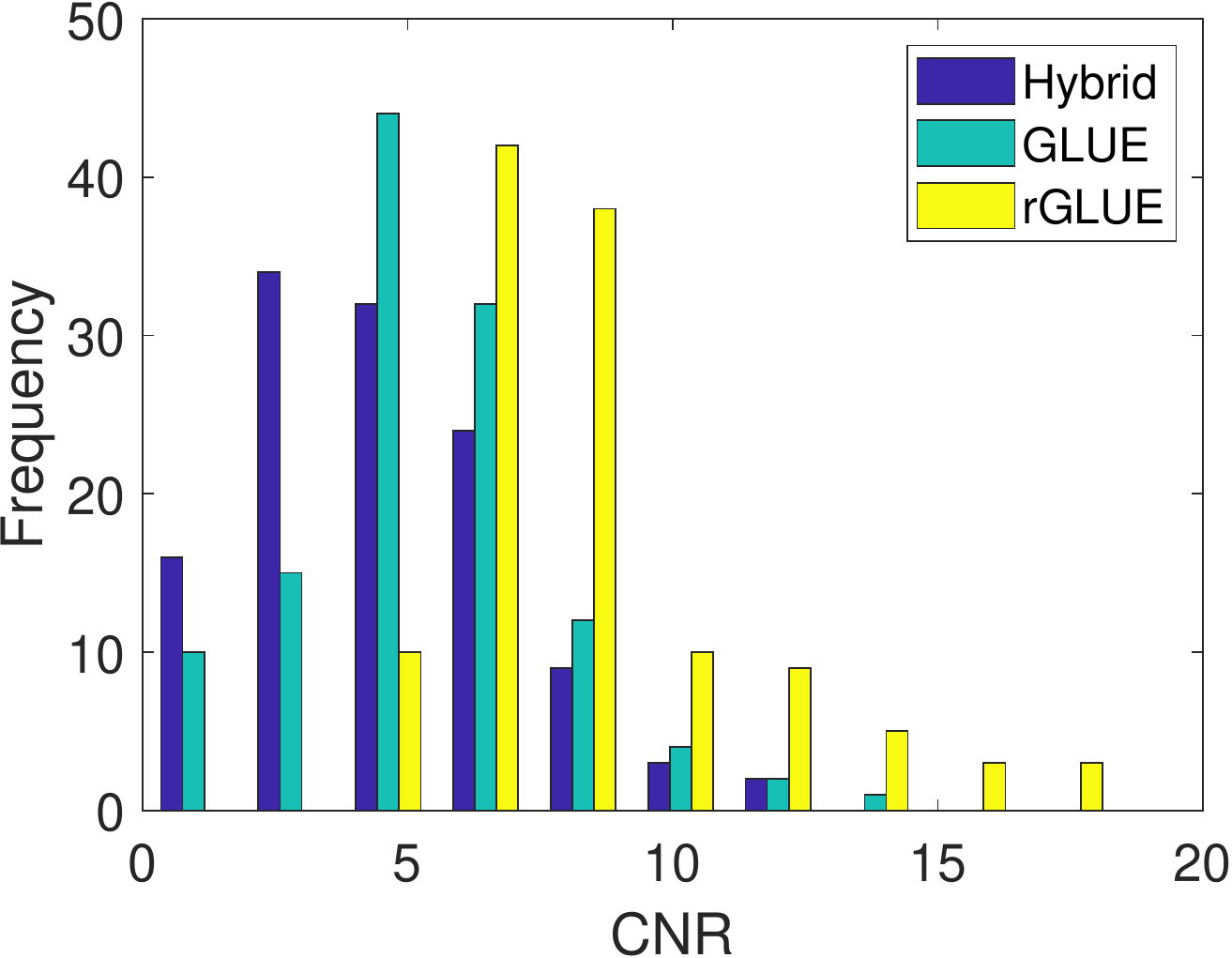} }}%
	\subfigure[\textit{In vivo} breast data]{{\includegraphics[width=.2\textwidth]{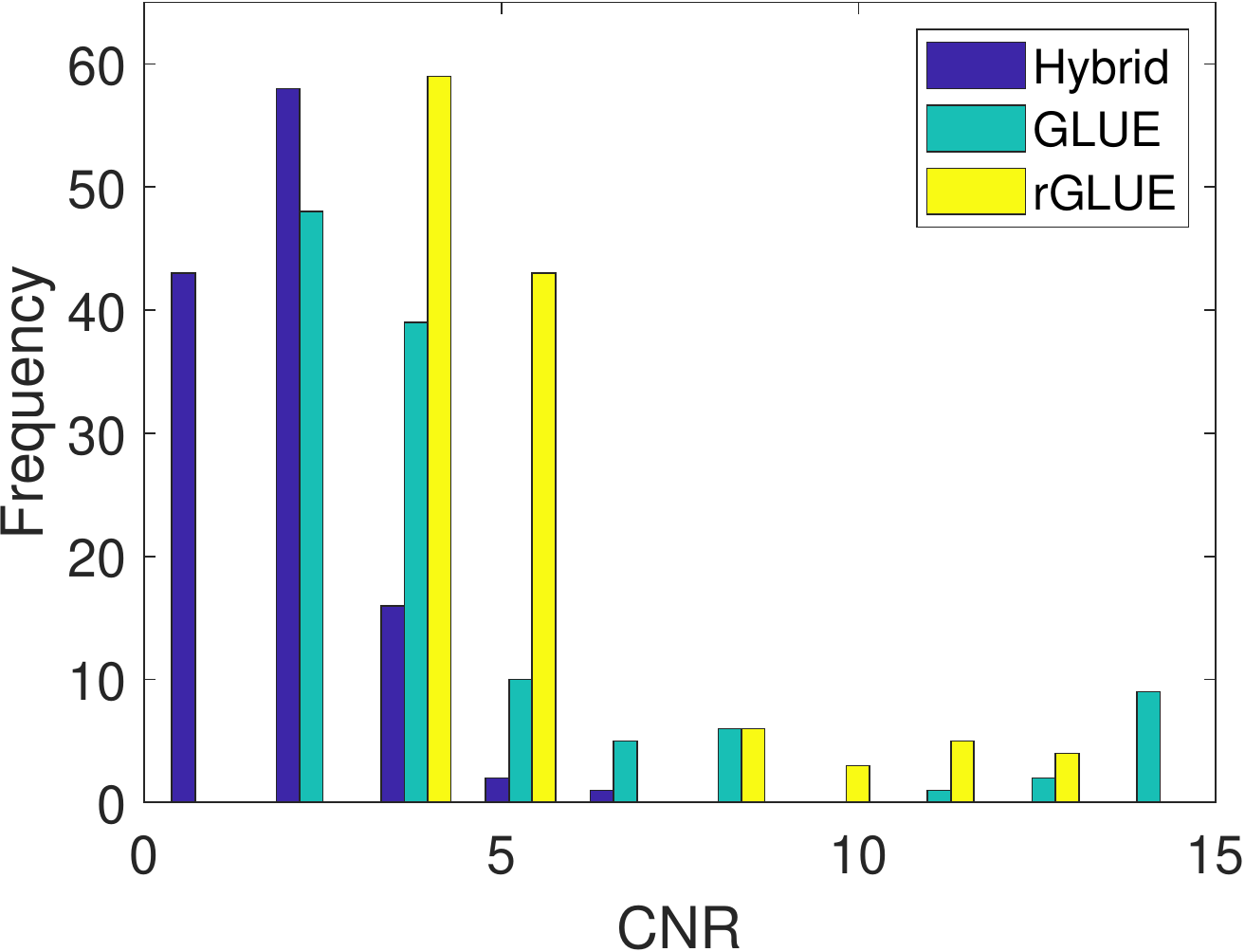} }}%
	\caption{CNR histograms obtained utilizing 120 target-background window pairs. Columns 1-5 correspond to soft-inclusion simulated phantom, breast phantom, \textit{in vivo} liver datasets 1 and 2, and the breast dataset, respectively.}
	\label{cnr_histograms}
\end{figure*}



\subsection{\textit{in vivo} Results}
\subsubsection{Liver Patient 1}
Fig.~\ref{liver_p2} shows the B-mode image and the strain images for the liver patient 1. Due to the echogenic difference between the tumor and the normal tissue, the tumor is visible in the B-mode image. However, in the B-mode image, the view of the tumor is corrupted with speckle and clutter echo signal. The strain images obtained from GLUE and rGLUE provide clearer realization of the tumor. The Hybrid technique obtains the noisiest strain image. The GLUE strain image suffers from large artifacts. rGLUE removes the artifacts and provides a higher quality strain map with a better contrast between the tumor and the background. The quantitative values reported in Table~\ref{table_liver} also corroborate our visual evaluation. Similar to simulation and phantom experiments, we report the histogram of 120 CNR values in Fig.~\ref{cnr_histograms}(c) which demonstrates the high quantitative performance of the proposed technique throughout the image. Hybrid obtains an average CNR value of 2.72 whereas the average CNR values corresponding to GLUE and rGLUE are 11.70 and 12.86, respectively. One interesting observation about GLUE and rGLUE is that both techniques slightly overestimate the tumor size. This issue arises from the dynamic range of the frames under consideration. Due to low sample amplitudes, the regularization term dominates the data term in the tumor region resulting in an underestimation of strain in the surrounding healthy regions. This propagation of low strain values leads to the overestimation of tumor size.

\subsubsection{Liver Patient 2}
Fig.~\ref{liver_p3} shows the axial strain estimates for the liver patient 2. The echogenic difference between the healthy tissue and the tumor is negligible in the B-mode image. In contrast, the strain images clearly delineate the pathologic tissue. The yellow arrows in the B-mode image indicate the rib bones. The specular bone surface introduces large outliers to the RF data, creating noticeable artifacts in Hybrid and GLUE strain images. Hybrid produces the noisiest strain estimate. GLUE exhibits a better noise suppression performance than Hybrid. In addition, like Hybrid, GLUE manifests strain underestimation in the shallow regions. rGLUE adaptively handles the RF samples and provides a brighter shallow tissue region. In addition, rGLUE reduces the strain artifacts around the bone surface. Furthermore, the rGLUE strain image presents a smoother background than Hybrid and GLUE, while preserving the target-background contrast. The SNR and CNR values reported in Table~\ref{table_liver_p3} support our visual inference. In addition, the histogram of 120 CNR values shown in Fig.~\ref{cnr_histograms}(d) demonstrates the superiority of rGLUE all over the strain image. The average CNR values obtained by Hybrid, GLUE, and rGLUE, respectively, are 4.47, 5.30, and 8.40.         

\subsubsection{Breast Data}
Fig.~\ref{breast_p1} reports the B-mode and the strain images for the breast dataset. The IDC observed in the B-mode image is much better visualized in the strain images obtained by GLUE and rGLUE. Although Hybrid is able to show contrast between IDC and the healthy tissue, the IDC's bottom is highly corrupted with noise. GLUE outperforms Hybrid in the pathologic region. However, the strain image obtained by GLUE is still noisy, which is resolved by the proposed rGLUE technique. The quantitative metrics (Table~\ref{table_breast}) reassure that rGLUE substantially outperforms GLUE and Hybrid. Besides single CNR value, we report the histogram of 120 CNR values (see Fig.~\ref{cnr_histograms}(e)) calculated utilizing 6 target and 20 background windows. Hybrid and rGLUE, respectively, show the lowest and highest frequencies of high CNR values. Hybrid, GLUE, and rGLUE obtain average CNR values of 2.01, 4.66, and 5.25, respectively. Similar to the first liver data, GLUE and rGLUE slightly overestimate the tumor size due to high echogenic contrast between tumor and healthy tissue. It is worth noting that Hybrid substantially overestimate the tumor size in this case. This might be caused by the erroneous propagation of low strain values due to incorporation of neighboring samples in displacement calculation.



\section{Discussion}
Sample amplitude similarity constraint, in conjunction with spatial and temporal regularization, often leads to good quality strain map. However, ultrasound RF data contain outlier samples originating from numerous sources. To obtain high quality strain image in the presence of notable amount of outlier samples, sample amplitude alone is not sufficient and therefore the condition of image gradient similarity needs to be imposed. The strain imaging performance reported in this paper show the promise of the proposed technique in robust TDE.

Since lesions can be both solid and fluid-filled, a pathologic tissue region is either softer or harder than surrounding healthy tissue. Therefore, to obtain comprehensive knowledge about the proposed technique's strength and weakness, we have conducted several experiments with both soft- and hard-inclusion datasets. It is worth noting that the proposed rGLUE algorithm has been compared against two recently published techniques: Hybrid, which is especially designed to handle both solid and fluid-filled lesions, and GLUE, which is being used by many research labs as the state-of-the-art. rGLUE proves its potential by substantially outperforming both of the aforementioned techniques in all of our validation experiments.     

In this work, the spatial regularization weight for each sample has been considered to be constant. However, different levels of continuity in different regions of the image might be of immense importance in certain applications. Therefore, an adaptive regularization scheme can be developed to use distinct regularization weight for each sample. Echogenic properties of the B-mode image can be investigated to ascertain the proper values of regularization. However, the echogenic properties do not always correlate with the elastic properties of the tissue. Therefore, an alternative strategy can generate the strain map with a gold standard technique first and then use its noise statistics as a reference to re-estimate the displacement field with an adaptive distribution of regularization parameter values. However, this framework aiming at maintaining a proper balance between motion continuity and discontinuity is beyond the scope of this work and calls for further research.

A manual selection scheme was taken into account to obtain the optimal regularization parameters in this work. The strain images corresponding to different sets of continuity weights were assessed visually to select the best ones. This procedure can be semi-automatized by choosing the best parameter set based on the peak value of some quality metric such as SNR or CNR. The renowned L-curve technique~\cite{hansen1999curve} can also be taken into account to fully automate the parameter selection process. Another approach for accomplishing this crucial task is to develop an automatic machine learning-based model which would be trained on a large dataset obtained from a semiautomatic technique. The aforementioned techniques are beyond the scope of this paper and interesting avenues of future work. It is worth noting that the optimal continuity weights for different organs such as liver, breast, etc., can be saved as elasticity imaging presets of a commercial ultrasound machine. These weights need to be adjusted only when imaging a new organ, or when changing the imaging settings.

Simulation studies show that rGLUE efficiently distinguishes outliers from small targets. The adaptive weighting technique embedded in rGLUE estimates the weight of each sample taking the data and gradient residuals of the previous iteration into account. If the data residual for a particular sample is too high in comparison to the gradient residual, rGLUE detects it as an outlier and reduces its data weight. Therefore, its displacement is estimated depending mostly on the regularization term. However, this is not the case for small targets since there is correspondence between the pre- and post-deformed samples, even in the boundary region of the target. Due to this correspondence, the data residual in the target boundary is not high. Hence, rGLUE assigns usual weights to the target samples, correctly classifying them as actual data.

The proposed technique follows an iterative approach to optimally select the weights of data and gradient constancy terms. In our experience, five iterations are often sufficient for the optimality of the aforementioned weights. In addition, the weight balancing parameter $\lambda$ plays a crucial role in proper selection of the weights. It is clear from the formation of $\theta (i,j)$ that a very low value of $\lambda$ is equivalent to putting the same weight on both of the constancy terms. On the other hand, a very high value to $\lambda$ leads to an oversmooth displacement map due to the negligible weight of the data similarity term. Hence a careful tuning of the balancing parameter is required for a robust estimation of the displacement field. The optimal value of $\lambda$ is attained by manual intervention in this work. An automatic selection of $\lambda$ would certainly introduce ease to our technique which we postpone as a potential area of future work.                                                   
        
\section{Conclusion}	
A robust regularized optimization-based TDE scheme called rGLUE for quasi-static ultrasound elastography has been proposed. In this novel approach, total data cost has been formulated considering the adaptively weighted effects of both amplitude and gradient mismatches. An iterative approach has also been introduced for the selection of optimal weights for each dissimilarity term. The non-linear cost function devised herein has been efficiently optimized to derive a linear system of equations which is solved for millions of variables within a few seconds on a standard personal computer. Extensive validation against simulated, phantom, \textit{in vivo} liver, and breast datasets demonstrate the superiority of the proposed algorithm over state-of-the-art elastography algorithms.

\section*{Acknowledgment}
This work was supported by Natural Sciences and Engineering Research Council of Canada (NSERC) RGPIN-2020-04612 and by the Quebec Bio-Imaging Network of the Fonds de Recherche du Québec. We thank Drs. E. Boctor, M. Choti, and G. Hager for allowing us to use the \textit{in vivo} liver data. We also thank Dr. A. Nahiyan and Dr. M. K. Hasan for sharing the Hybrid code with us and for helping us to tune its parameters. Authors thank the anonymous reviewers for their constructive feedback.

\FloatBarrier
\balance
\bibliographystyle{IEEEtran}
\bibliography{ref}

\end{document}


\title{Incorporating Gradient Similarity for\\ Robust Time Delay Estimation in\\ Ultrasound Elastography}
%

%
\author{Md~Ashikuzzaman, \textit{Graduate Student Member}, \textit{IEEE},
	Timothy~J.~Hall, \textit{Member}, \textit{IEEE},
	and~Hassan~Rivaz, \textit{Senior Member}, \textit{IEEE}}

\maketitle
In this Supplementary Material, we will define matrices $H_{I}$, $H_{\nabla I_{y}}$, $H_{\nabla I_{x}}$, $P_{I}$, $P_{\nabla I_{y}}$, and $P_{\nabla I_{x}}$ and vectors $\mu_{1}$, $\mu_{2}$, and $\mu_{3}$. We will compare Hybrid, GLUE, and rGLUE for soft-inclusion simulated datasets with and without outlier. Besides, we will show the strain images corresponding to the four-layer, thin-layer, and low-contrast simulated phantoms. In addition, we will present the results for real breast phantom in the presence of artificial outlier region. Furthermore, We will present the lateral strain images for the breast elastography phantom. Finally, we will assess the effect of median filtering on GLUE strain estimate.

\section{Method}
$H_{I}=diag({h_{1}^{'}}^{2}(1,1),{h_{1}^{'}}^{2}(1,2),{h_{1}^{'}}^{2}(1,3),...,{h_{1}^{'}}^{2}(m,n))$, $H_{\nabla I_{y}}=diag({h_{2}^{'}}^{2}(1,1),{h_{2}^{'}}^{2}(1,2),{h_{2}^{'}}^{2}(1,3),...,{h_{2}^{'}}^{2}(m,n))$, and $H_{\nabla I_{x}}=diag({h_{3}^{'}}^{2}(1,1),{h_{3}^{'}}^{2}(1,2),{h_{3}^{'}}^{2}(1,3),...,{h_{3}^{'}}^{2}(m,n))$ are symmetric tri-diagonal matrices of size $2mn \times 2mn$ where ${h_{1}^{'}}^{2}(i,j)$, ${h_{2}^{'}}^{2}(i,j)$, and ${h_{3}^{'}}^{2}(i,j)$ are defined as follows:

	\begin{equation}
	{h_{1}^{'}}^{2}(i,j)=\theta_{i,j}
	\begin{bmatrix}
	{I_{2,a}^{'}}^{2}(i,j) & {I_{2,a}^{'}}(i,j){I_{2,l}^{'}}(i,j) \\
	{I_{2,a}^{'}}(i,j){I_{2,l}^{'}}(i,j) & {I_{2,l}^{'}}^{2}(i,j)
	\end{bmatrix}
	\label{eq:hsquare_i}
	\end{equation}

	\begin{equation}
	\begin{aligned}
	&{h_{2}^{'}}^{2}(i,j)=\\ 
	&\Gamma_{i,j}
	\begin{bmatrix}
	{\nabla I_{2,y,a}^{'}}^{2}(i,j) & {\nabla I_{2,y,a}^{'}}(i,j){\nabla I_{2,y,l}^{'}}(i,j) \\
	{\nabla I_{2,y,a}^{'}}(i,j){\nabla I_{2,y,l}^{'}}(i,j) & {\nabla I_{2,y,l}^{'}}^{2}(i,j)
	\end{bmatrix}
	\label{eq:hsquare_y}
	\end{aligned}
	\end{equation}

	\begin{equation}
	\begin{aligned}
	&{h_{3}^{'}}^{2}(i,j)=\\
	&\Gamma_{i,j}
	\begin{bmatrix}
	{\nabla I_{2,x,a}^{'}}^{2}(i,j) & {\nabla I_{2,x,a}^{'}}(i,j){\nabla I_{2,x,l}^{'}}(i,j) \\
	{\nabla I_{2,x,a}^{'}}(i,j){\nabla I_{2,x,l}^{'}}(i,j) & {\nabla I_{2,x,l}^{'}}^{2}(i,j)
	\end{bmatrix}
	\label{eq:hsquare_x}
	\end{aligned}
	\end{equation}

\noindent

where $\Gamma_{i,j}=\gamma(1-\theta_{i,j})$. ${I_{2,a}^{'}}(i,j)$ and ${I_{2,l}^{'}}(i,j)$ denote the axial and lateral derivatives of $I_{2}$ at $(i+a_{i,j},j+l_{i,j})$. ${\nabla I_{2,y,a}^{'}}(i,j)$, ${\nabla I_{2,y,l}^{'}}(i,j)$ and ${\nabla I_{2,x,a}^{'}}(i,j)$, ${\nabla I_{2,x,l}^{'}}(i,j)$ refer to the axial and lateral derivatives of $\nabla I_{2,y}$ and $\nabla I_{2,x}$ at $(i+a_{i,j},j+l_{i,j})$, respectively. $P_{I}$, $P_{\nabla I_{y}}$, and $P_{\nabla I_{x}}$ are diagonal matrices of size $2mn \times 2mn$ and given by:\\

$P_{I}=diag({I_{2,a}^{'}}(1,1),{I_{2,l}^{'}}(1,1),{I_{2,a}^{'}}(1,2),{I_{2,l}^{'}}(1,2),\dots,\\ \qquad {I_{2,a}^{'}}(m,n),{I_{2,l}^{'}}(m,n))$\\

$P_{\nabla I_{y}}=diag({\nabla I_{2,y,a}^{'}}(1,1),{\nabla I_{2,y,l}^{'}}(1,1),{\nabla I_{2,y,a}^{'}}(1,2),\\{\nabla I_{2,y,l}^{'}}(1,2),\dots,{\nabla I_{2,y,a}^{'}}(m,n),{\nabla I_{2,y,l}^{'}}(m,n))$\\

$P_{\nabla I_{x}}=diag({\nabla I_{2,x,a}^{'}}(1,1),{\nabla I_{2,x,l}^{'}}(1,1),{\nabla I_{2,x,a}^{'}}(1,2),\\{\nabla I_{2,x,l}^{'}}(1,2),\dots,{\nabla I_{2,x,a}^{'}}(m,n),{\nabla I_{2,x,l}^{'}}(m,n))$\\

$\mu_{t}$, $t \in \{1,2,3\}$ is a vector of length $2mn$ which is defined as:
\begin{equation}
\mu_{t}=\begin{bmatrix}
g_{1,1}^{t},g_{1,1}^{t},g_{1,2}^{t},g_{1,2}^{t},\dots,g_{m,n}^{t} 
\end{bmatrix}^T
\label{eq:mut}
\end{equation}

\noindent

where
\begin{equation}
g_{i,j}^{1}=\theta_{i,j}(I_{1}(i,j)-I_{2}(i+a_{i,j},j+l_{i,j})) 
\label{eq:g1}
\end{equation}

\begin{equation}
g_{i,j}^{2}=\Gamma_{i,j}(\nabla I_{1,y}(i,j)-\nabla I_{2,y}(i+a_{i,j},j+l_{i,j})) 
\label{eq:g2}
\end{equation}

\begin{equation}
g_{i,j}^{3}=\Gamma_{i,j}(\nabla I_{1,x}(i,j)-\nabla I_{2,x}(i+a_{i,j},j+l_{i,j})) 
\label{eq:g3}
\end{equation}

\subsection{Simulation and data acquisition}
\subsubsection{Simulated data with soft inclusion}
To analyze the TDE techniques’ robustness to outliers, we introduced two types of outliers to the soft-inclusion simulated phantom. First, we artificially created an outlier region in the RF data by making the post-compressed sample amplitudes of that particular region three times the original values. Second, we increased the sample amplitudes of 10 RF lines of the post-compressed frame by 30\% of the maximum sample amplitude to demonstrate the techniques’ robustness to outlier RF lines. These outliers concentrated at a particular tissue region were intended to represent the acquisition scenario in certain applications such as thermal ablation, or when faulty piezoelectric elements or air bubbles in the coupling gel render some RF lines inaccurate. There are three reasons behind adding outlier samples to one of the frames under consideration. First, to simulate the case of ablation where outlier samples are present in the frame acquired after the thermal expansion, not in the frame which is collected before the ablation. Second, to investigate each technique’s strength to handle a challenging situation where there is no correspondence between the pre- and post-deformed samples belonging to a particular region. And third, adding outliers to one or both frames has a similar effect in that there will be non-corresponding samples in both cases. Therefore, adding outliers to both frames is expected to lead to similar results.

\subsubsection{Simulated datasets with layers}
Three homogeneous phantoms with background elastic modulus of $20$ kPa were simulated. The first phantom includes two hard layers with Young's moduli of $40$ and $80$ kPa. Random Gaussian noise with $20$ dB PSNR was added to the RF data to emulate the real acquisition scenario. The second phantom includes a thin layer with an elastic modulus of $40$ kPa, and the third phantom simulates a low target-background contrast scenario by including a hard layer with $22.86$ kPa Young's modulus. In both thin-layer and low-contrast phantoms, we create an outlier region by multiplying the sample amplitudes in a rectangular region of the post-deformed frames by 3.

All layer phantoms were compressed by $4\%$ utilizing closed form equations. RF datasets were generated using Field II, setting the center and sampling frequencies to $7.27$ and $40$ MHz, respectively.

\subsubsection{Phantom experiment}
The real phantom experiment was carried out with around $1.5\%$ frame-to-frame strain. Phantom data does not usually contain a notable amount of outlier samples. To better investigate the techniques' robustness to outliers, we increased the sample amplitude values of a few RF lines by $20\%$ of the maximum amplitude.

\section{Results}
\begin{figure*}[ht]
	\centering
	\subfigure[Hybrid, $1\%$ compression]{{\includegraphics[width=.2\textwidth,height=.24\textwidth]{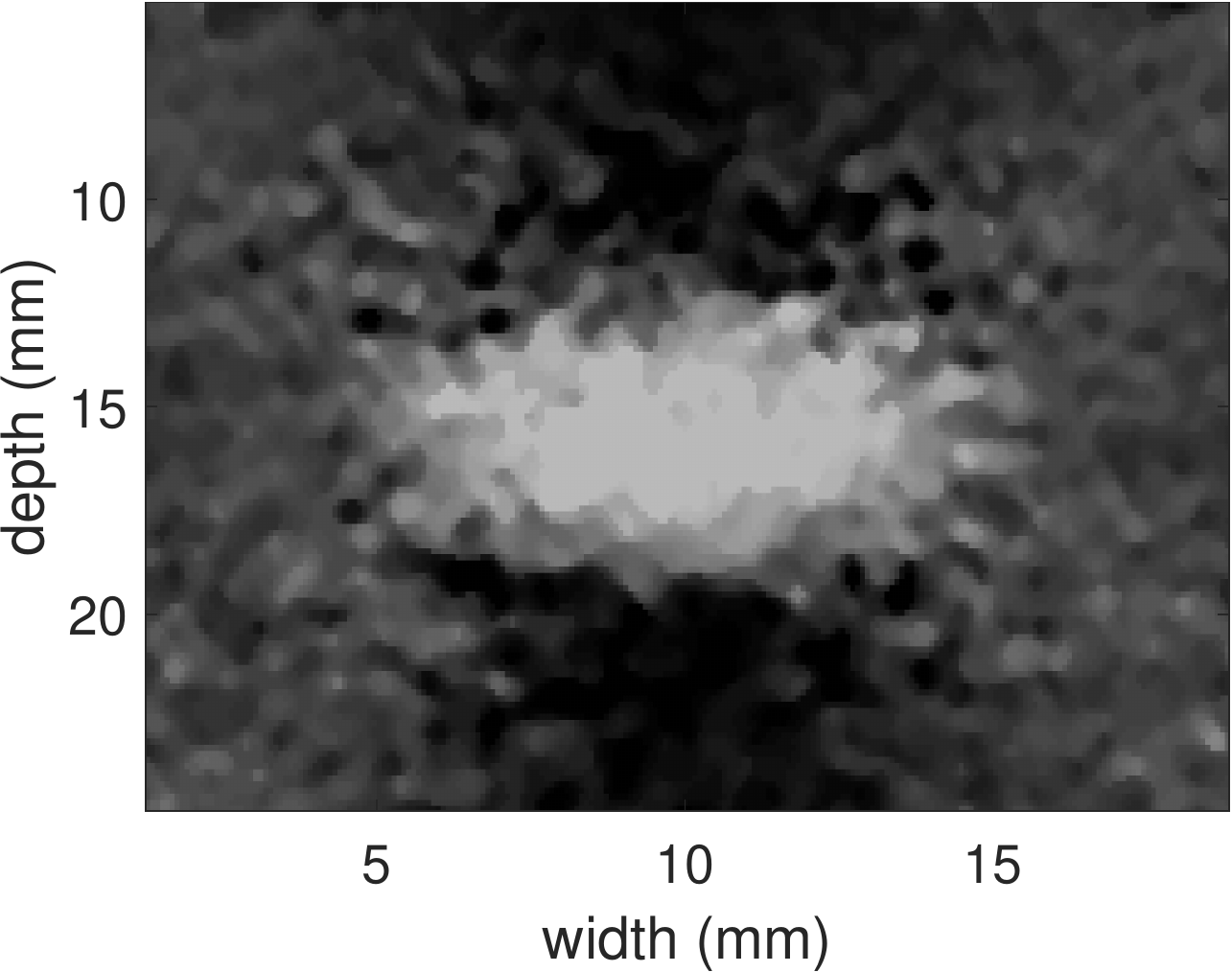} }}%
	\subfigure[Hybrid, $1.5\%$ compression]{{\includegraphics[width=.2\textwidth,height=.24\textwidth]{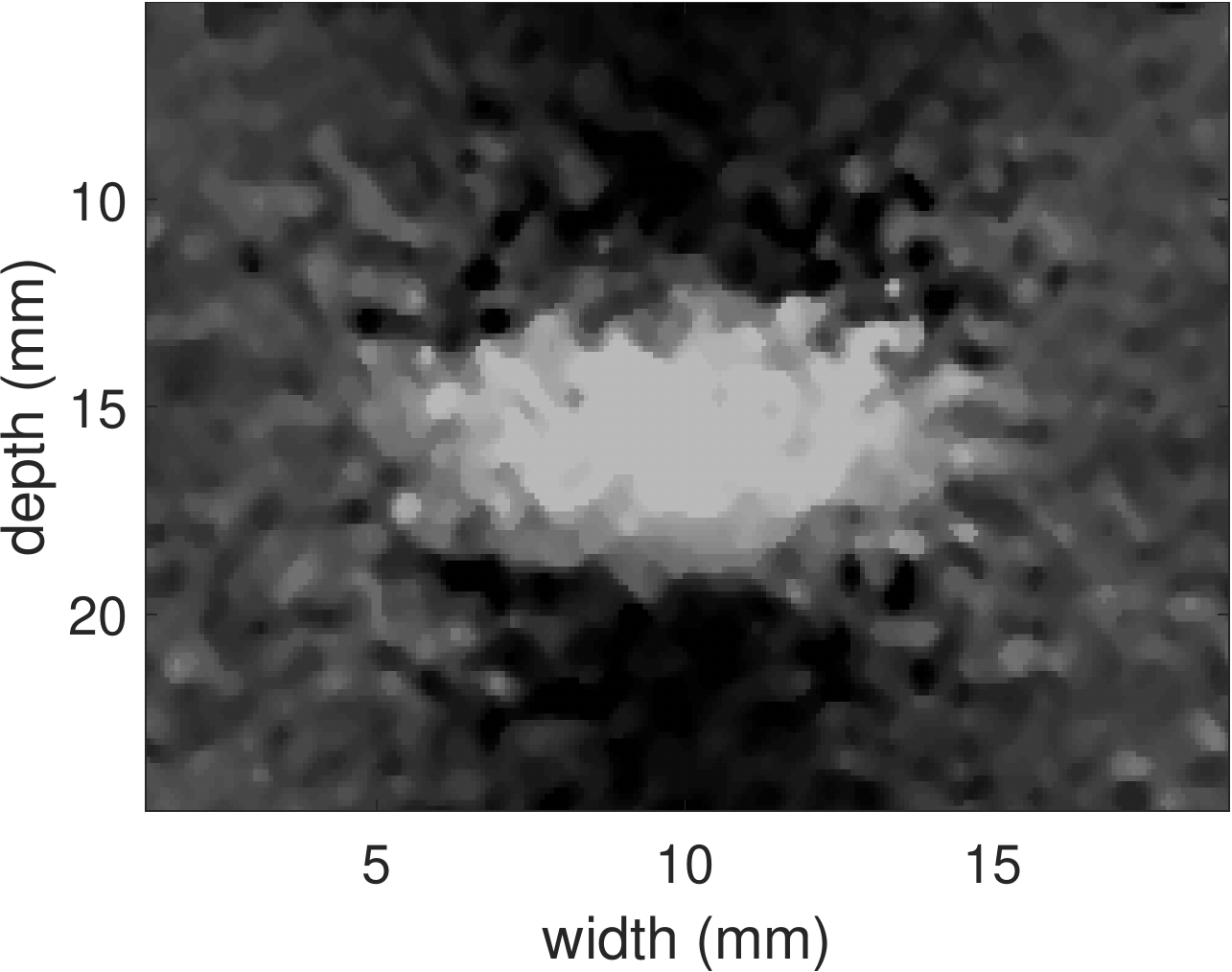} }}%
	\subfigure[Hybrid, $2\%$ compression]{{\includegraphics[width=.2\textwidth,height=.24\textwidth]{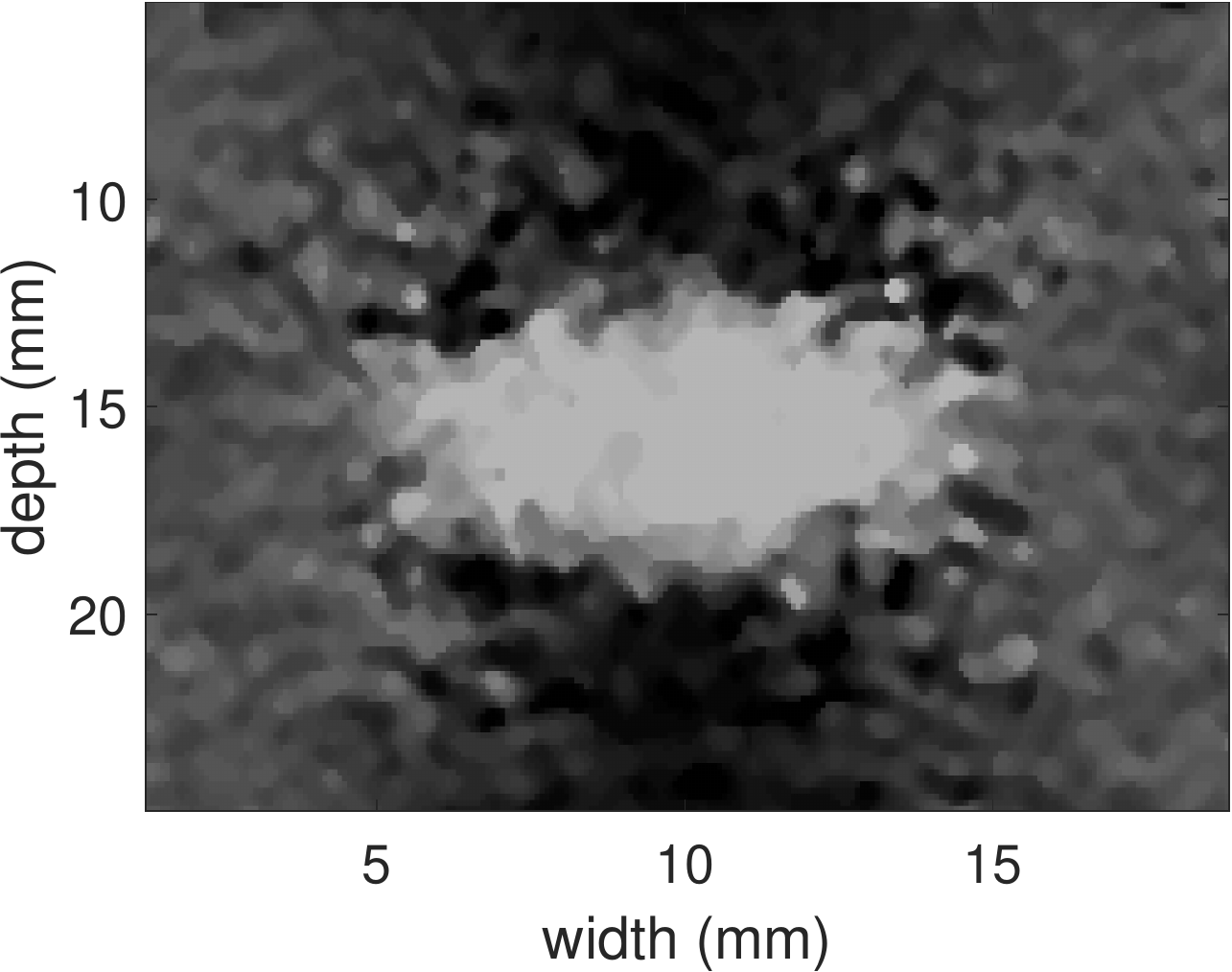} }}%
	\subfigure[Hybrid, $2.5\%$ compression]{{\includegraphics[width=.2\textwidth,height=.24\textwidth]{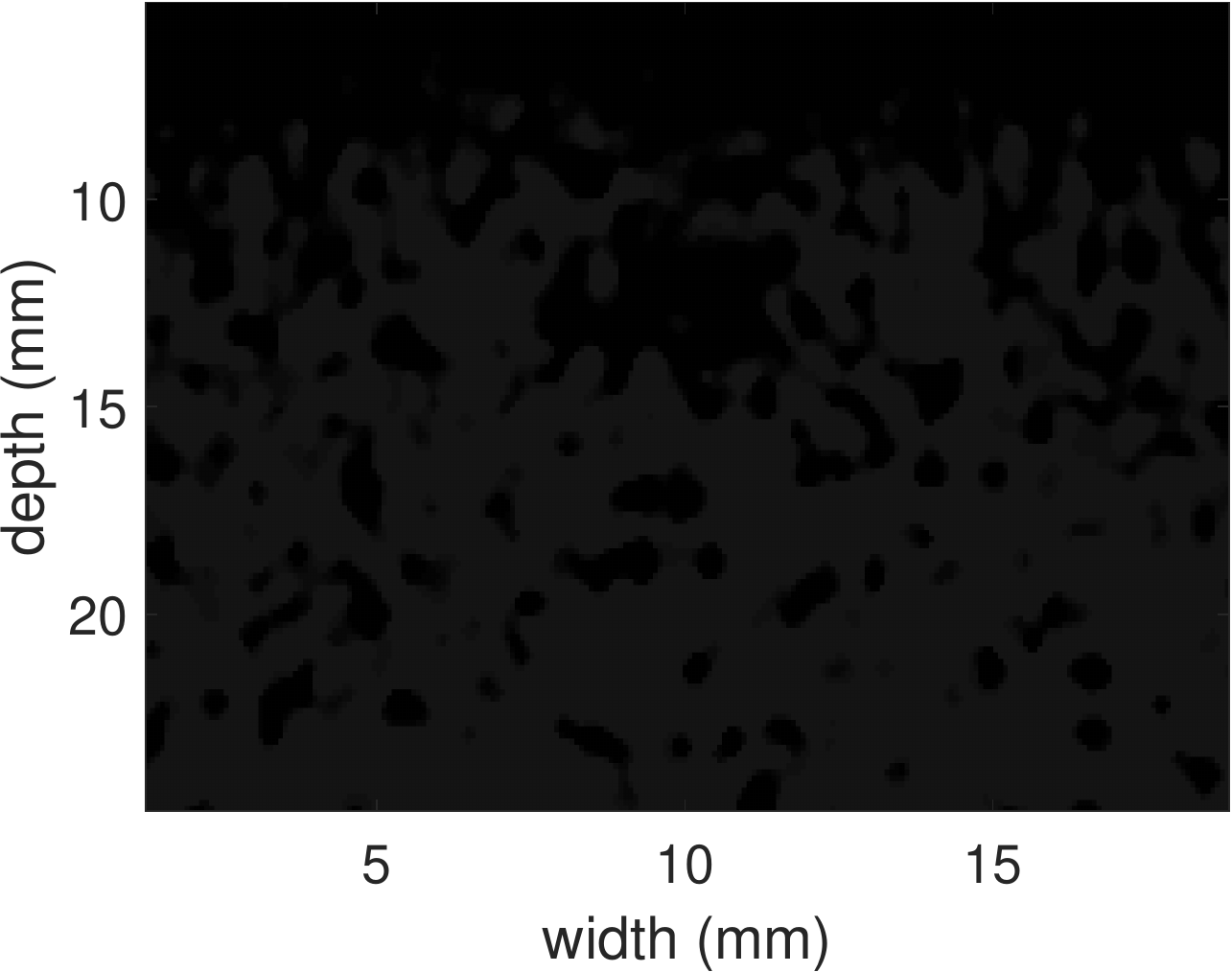} }}%
	\subfigure[Hybrid, $3\%$ compression]{{\includegraphics[width=.2\textwidth,height=.24\textwidth]{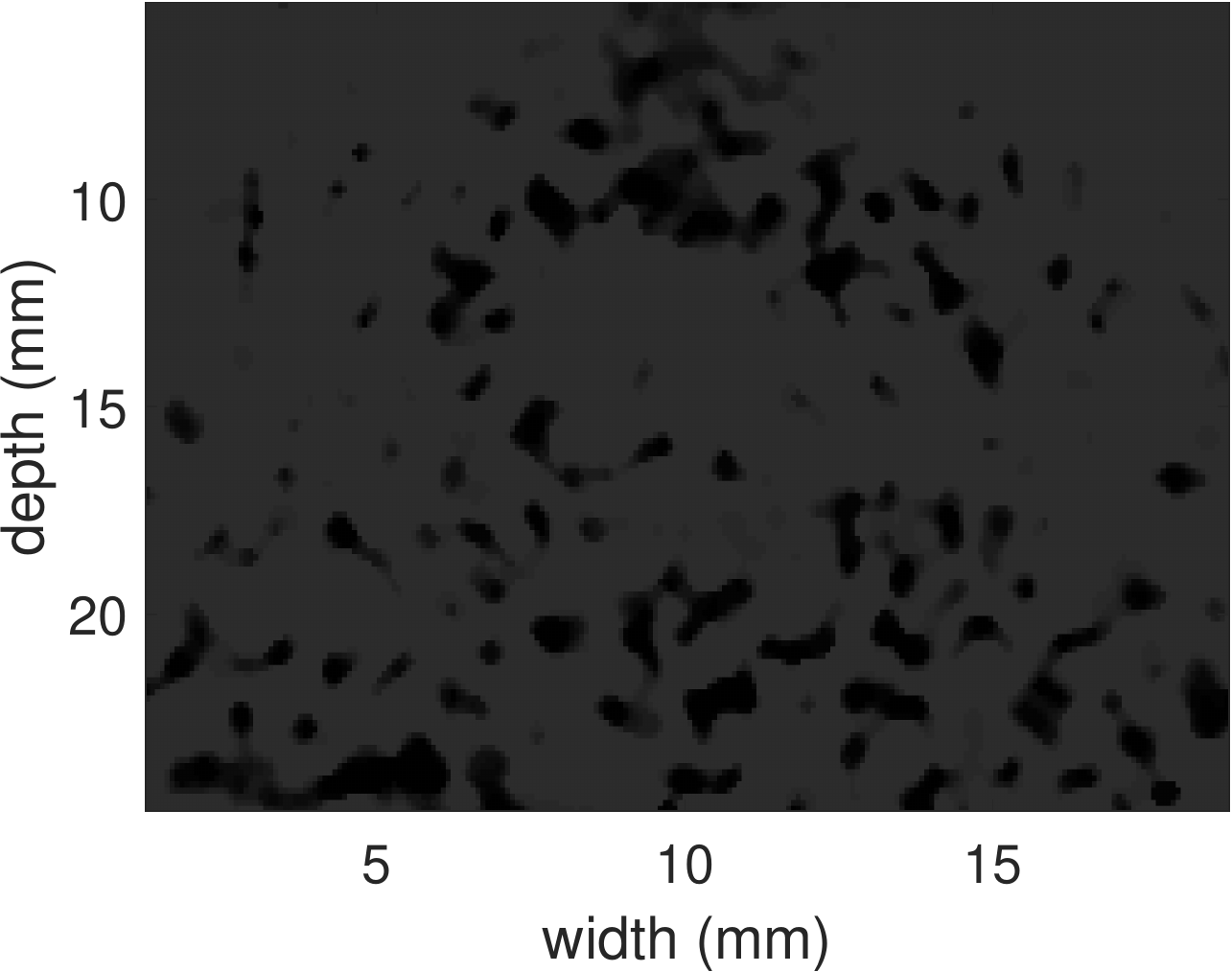} }}
	\subfigure[GLUE, $1\%$ compression]{{\includegraphics[width=.2\textwidth,height=.24\textwidth]{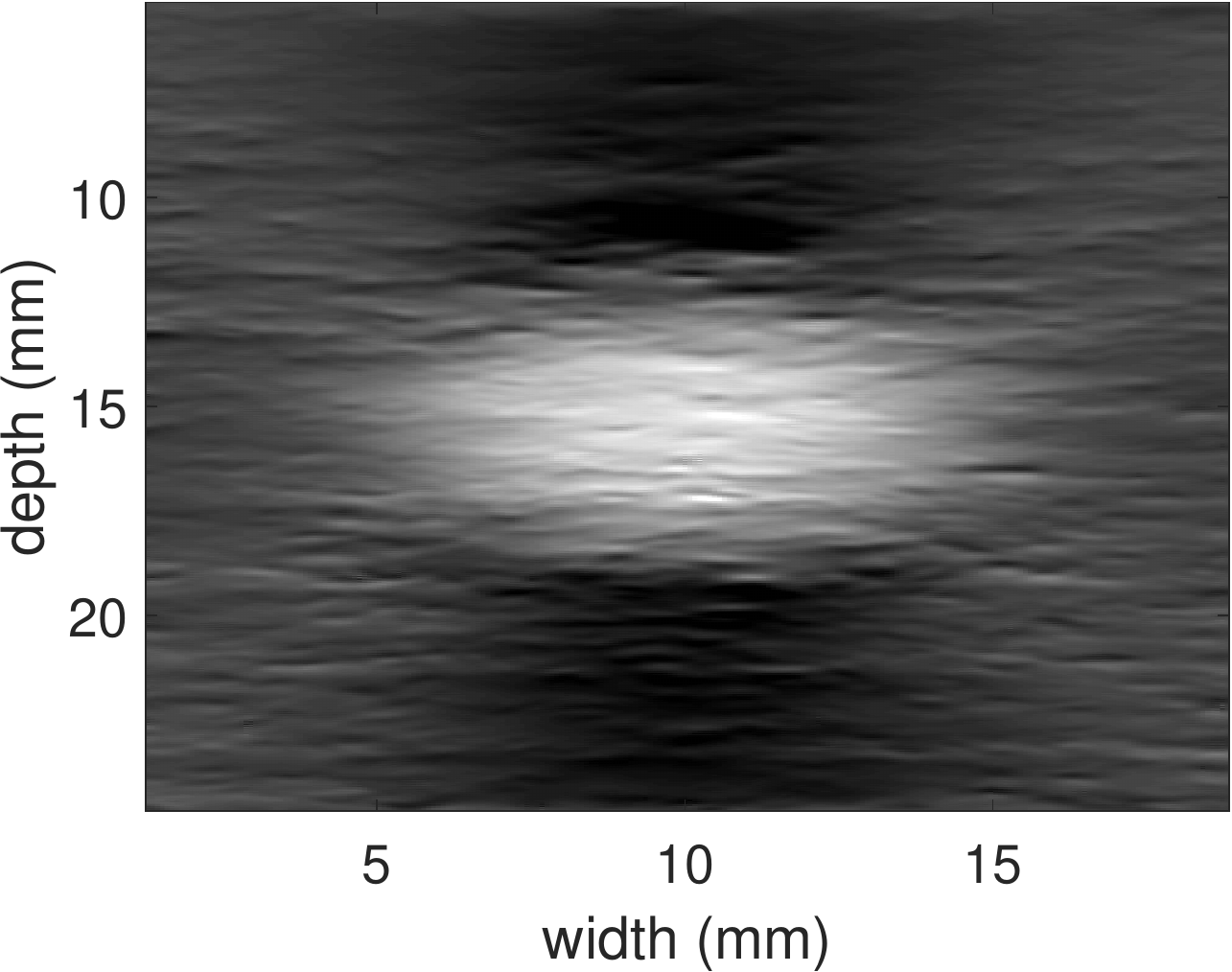} }}%
	\subfigure[GLUE, $1.5\%$ compression]{{\includegraphics[width=.2\textwidth,height=.24\textwidth]{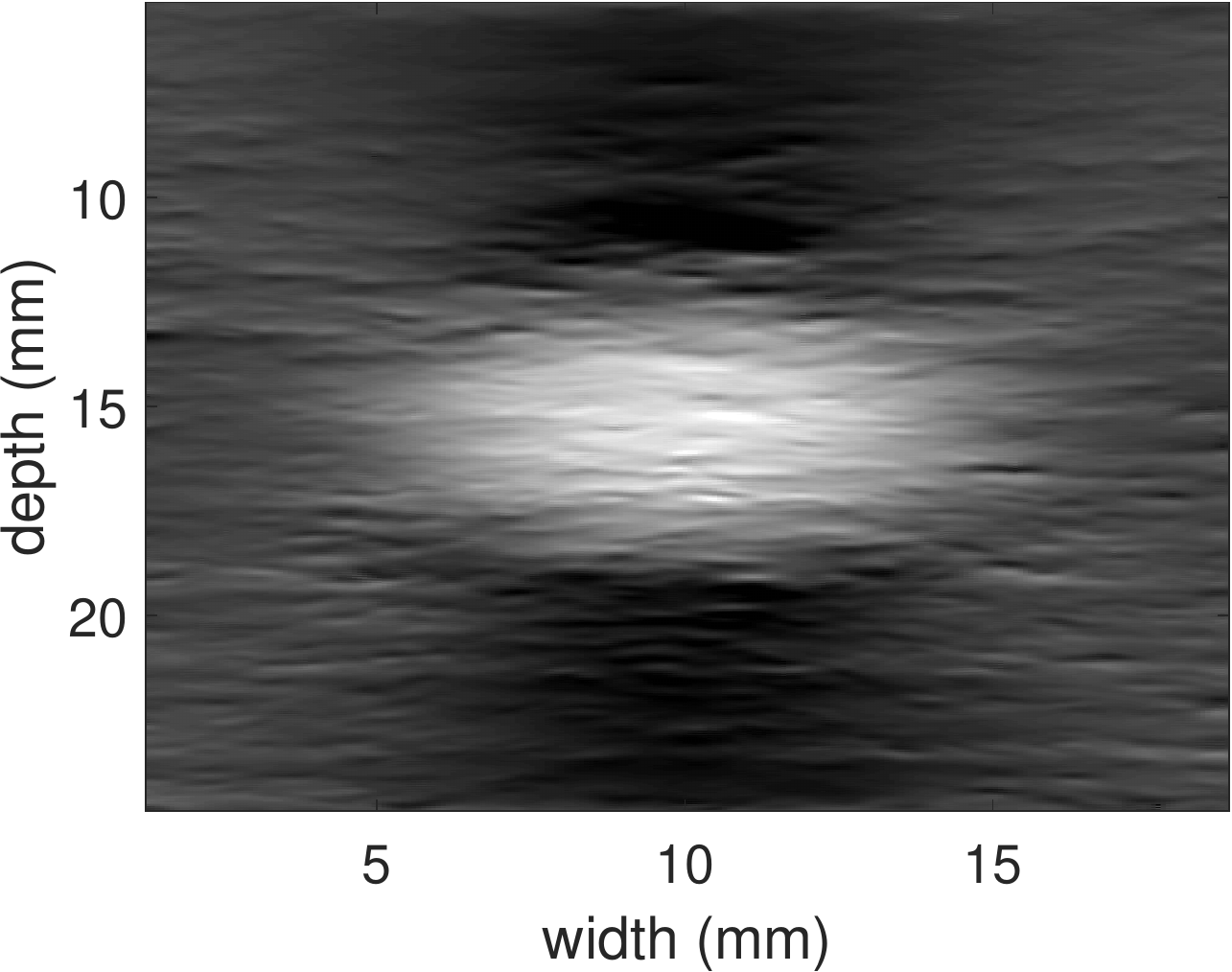} }}%
	\subfigure[GLUE, $2\%$ compression]{{\includegraphics[width=.2\textwidth,height=.24\textwidth]{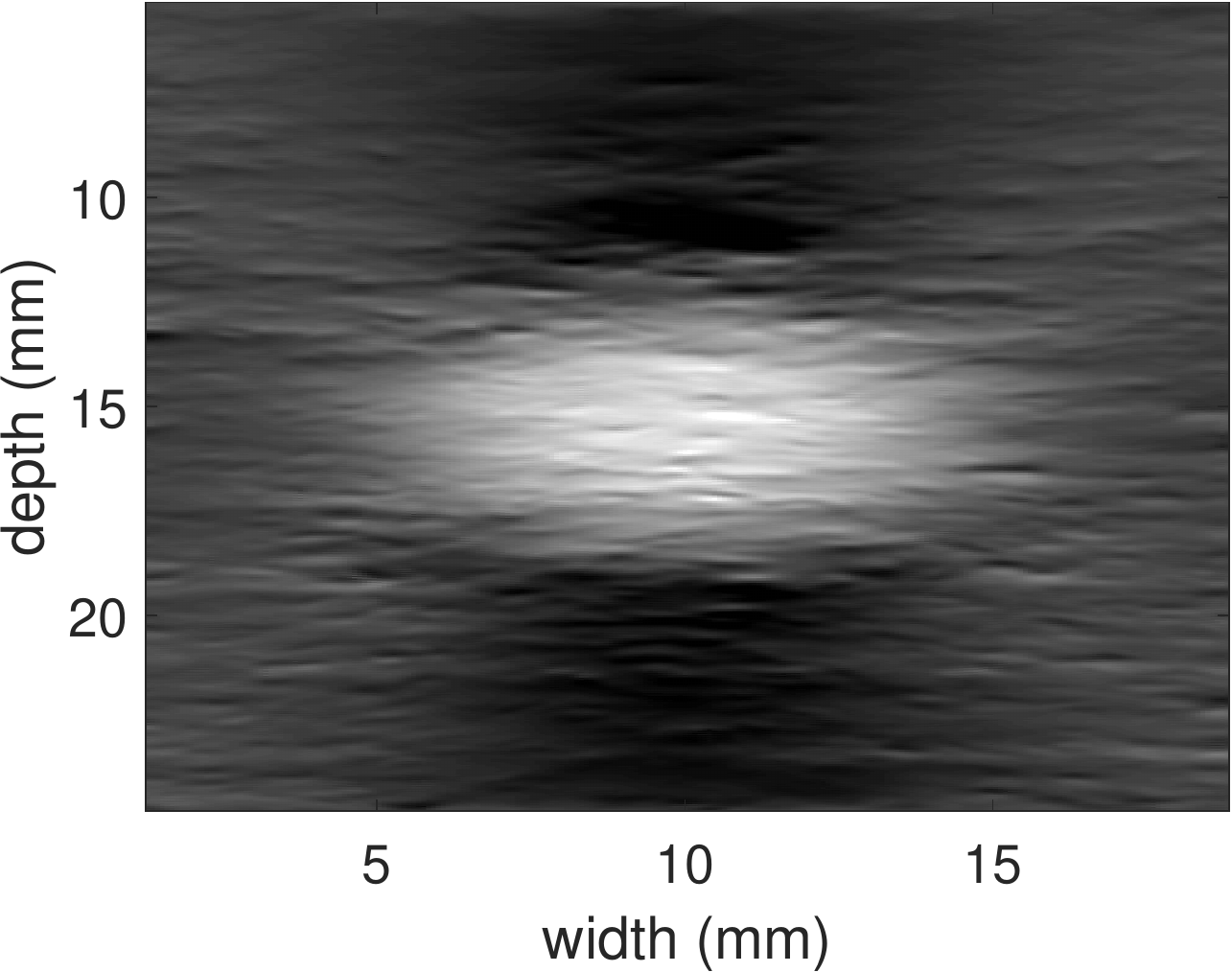} }}%
	\subfigure[GLUE, $2.5\%$ compression]{{\includegraphics[width=.2\textwidth,height=.24\textwidth]{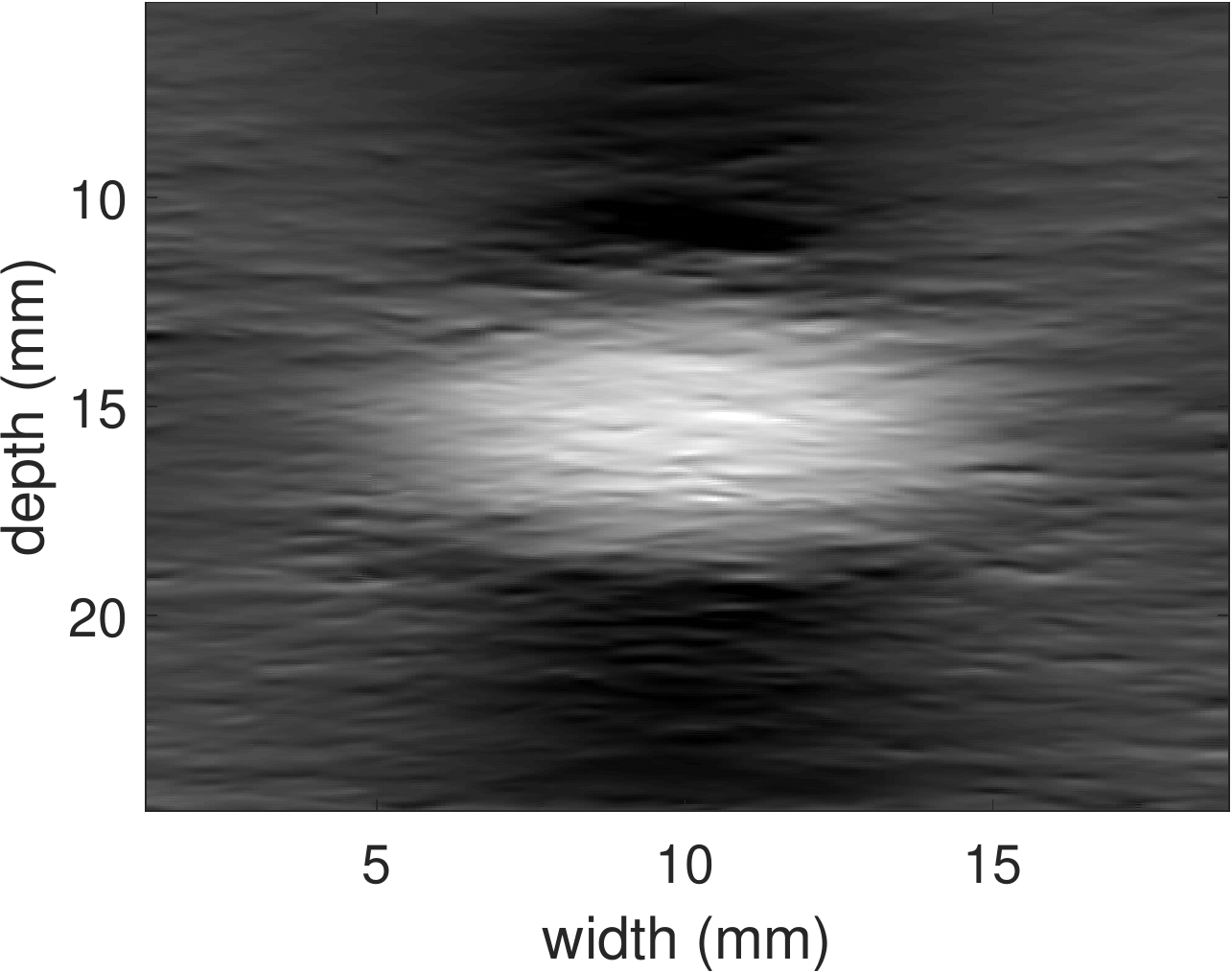} }}%
	\subfigure[GLUE, $3\%$ compression]{{\includegraphics[width=.2\textwidth,height=.24\textwidth]{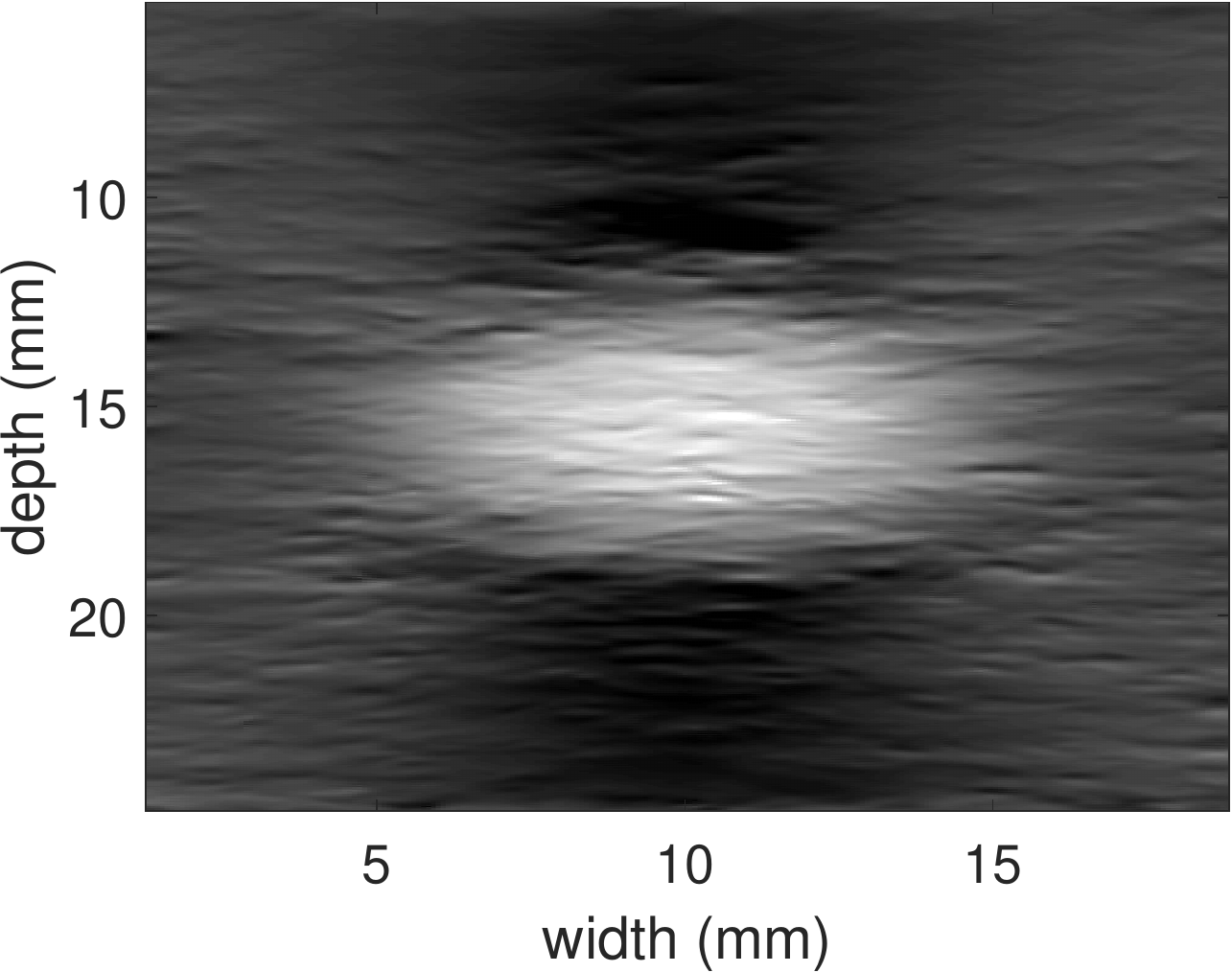} }}
	\subfigure[rGLUE, $1\%$ compression]{{\includegraphics[width=.2\textwidth,height=.24\textwidth]{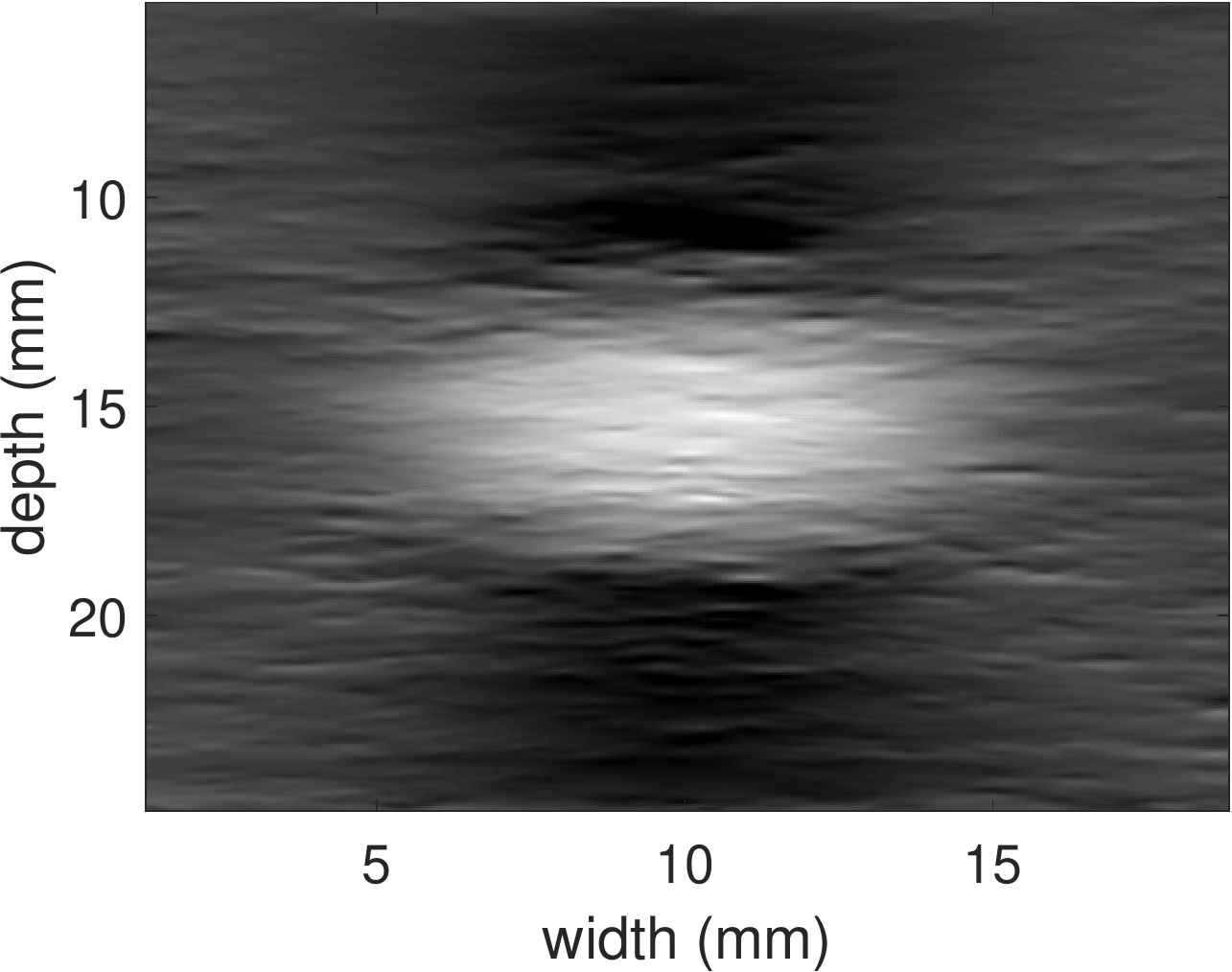} }}%
	\subfigure[rGLUE, $1.5\%$ compression]{{\includegraphics[width=.2\textwidth,height=.24\textwidth]{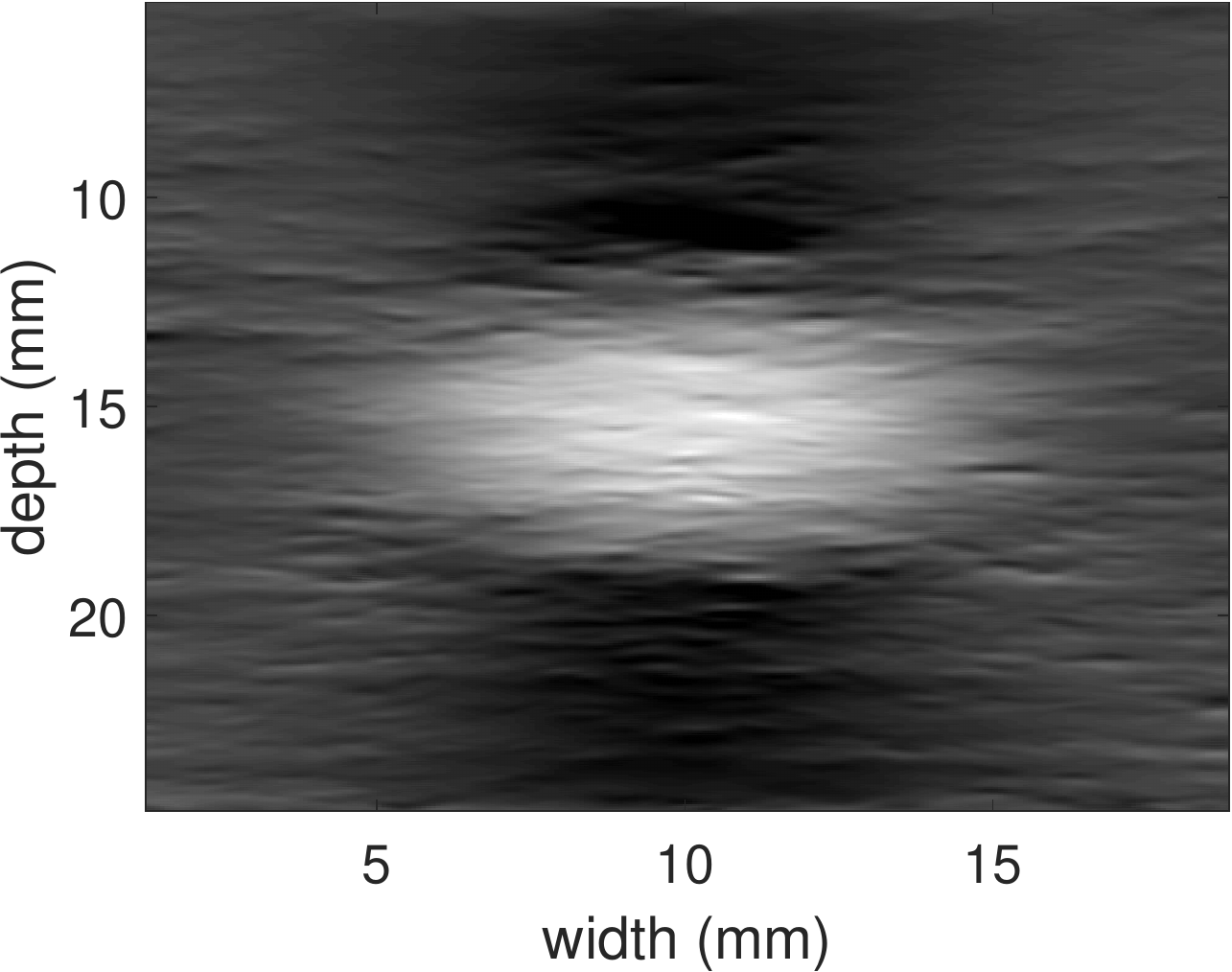} }}%
	\subfigure[rGLUE, $2\%$ compression]{{\includegraphics[width=.2\textwidth,height=.24\textwidth]{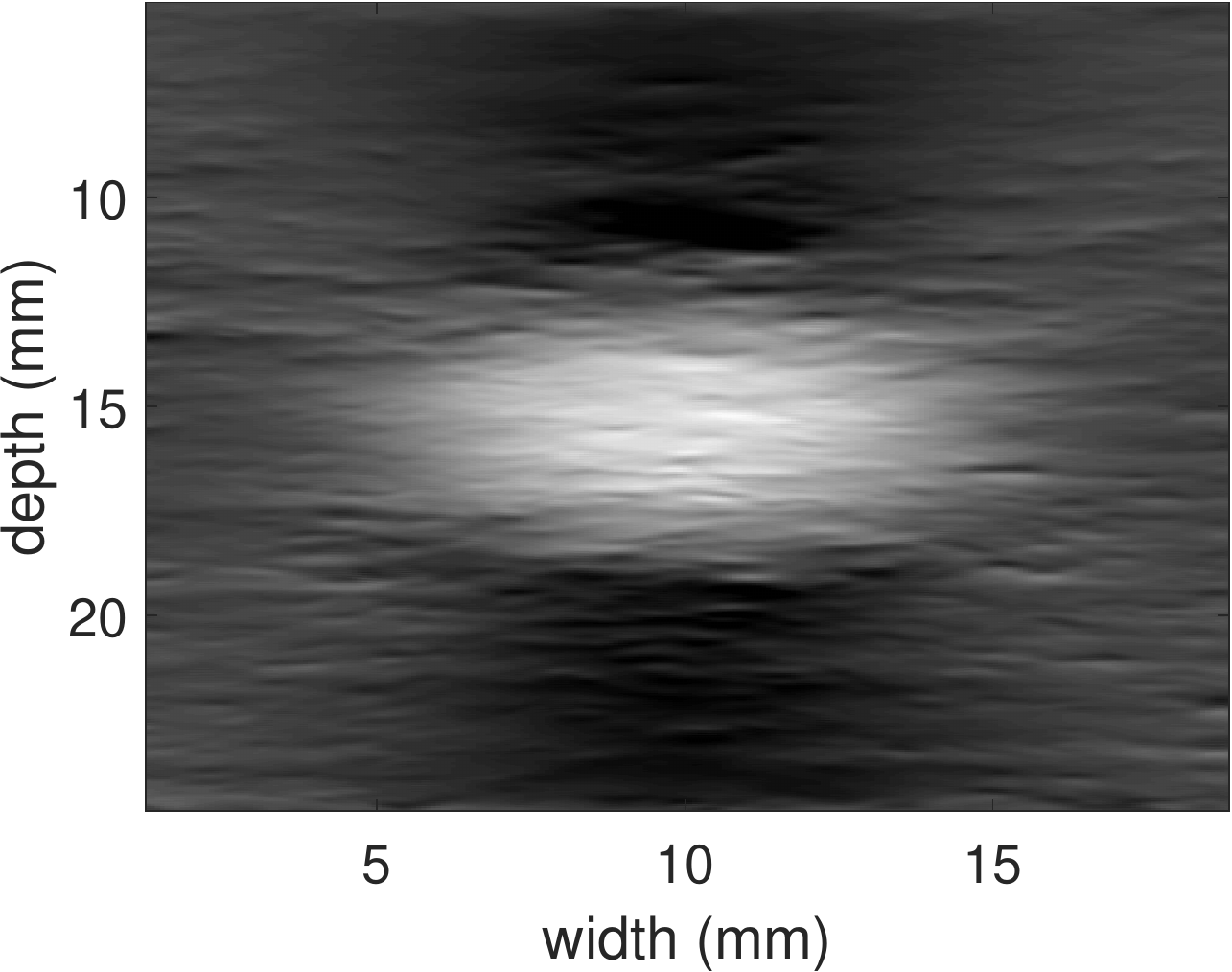} }}%
	\subfigure[rGLUE, $2.5\%$ compression]{{\includegraphics[width=.2\textwidth,height=.24\textwidth]{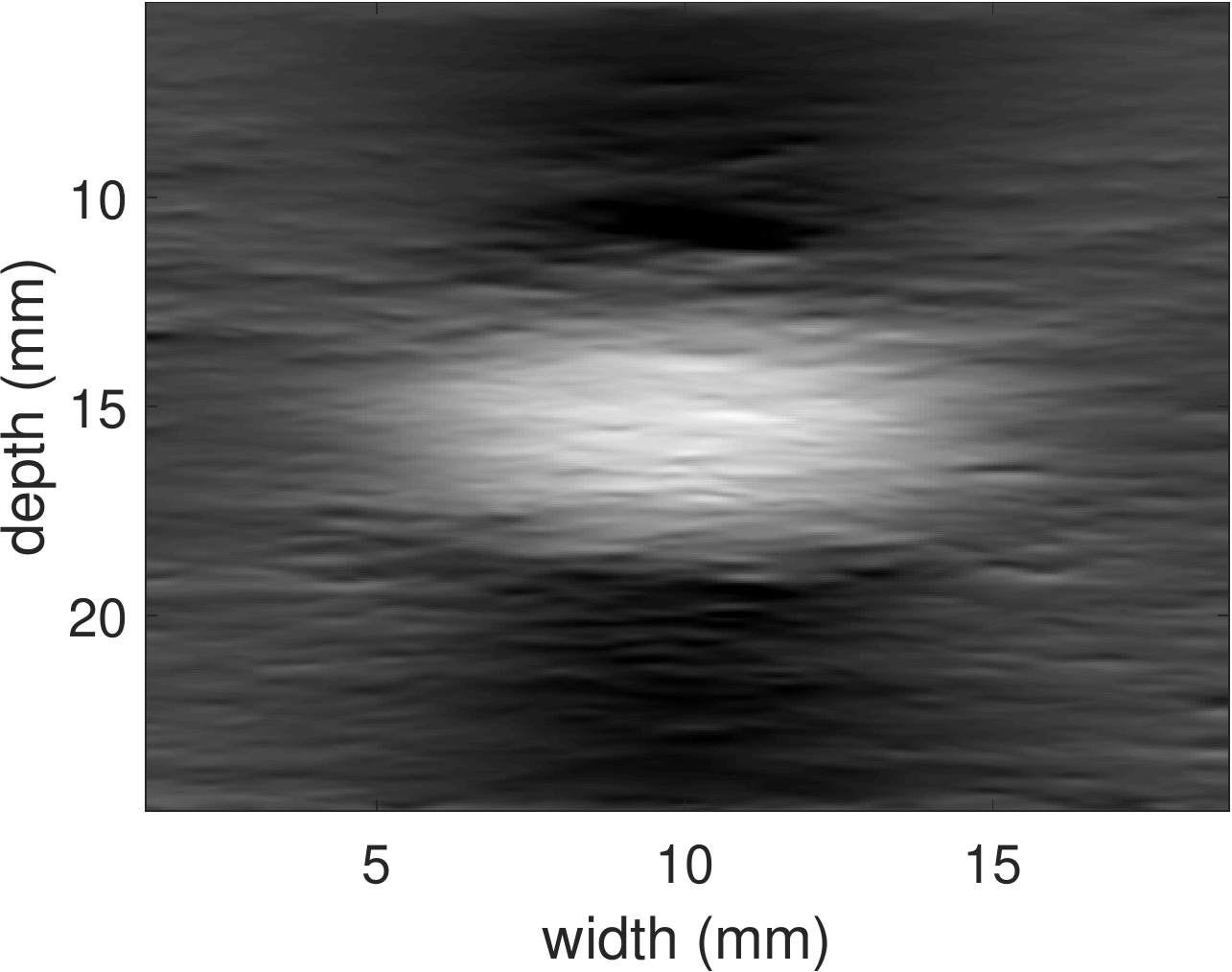} }}%
	\subfigure[rGLUE, $3\%$ compression]{{\includegraphics[width=.2\textwidth,height=.24\textwidth]{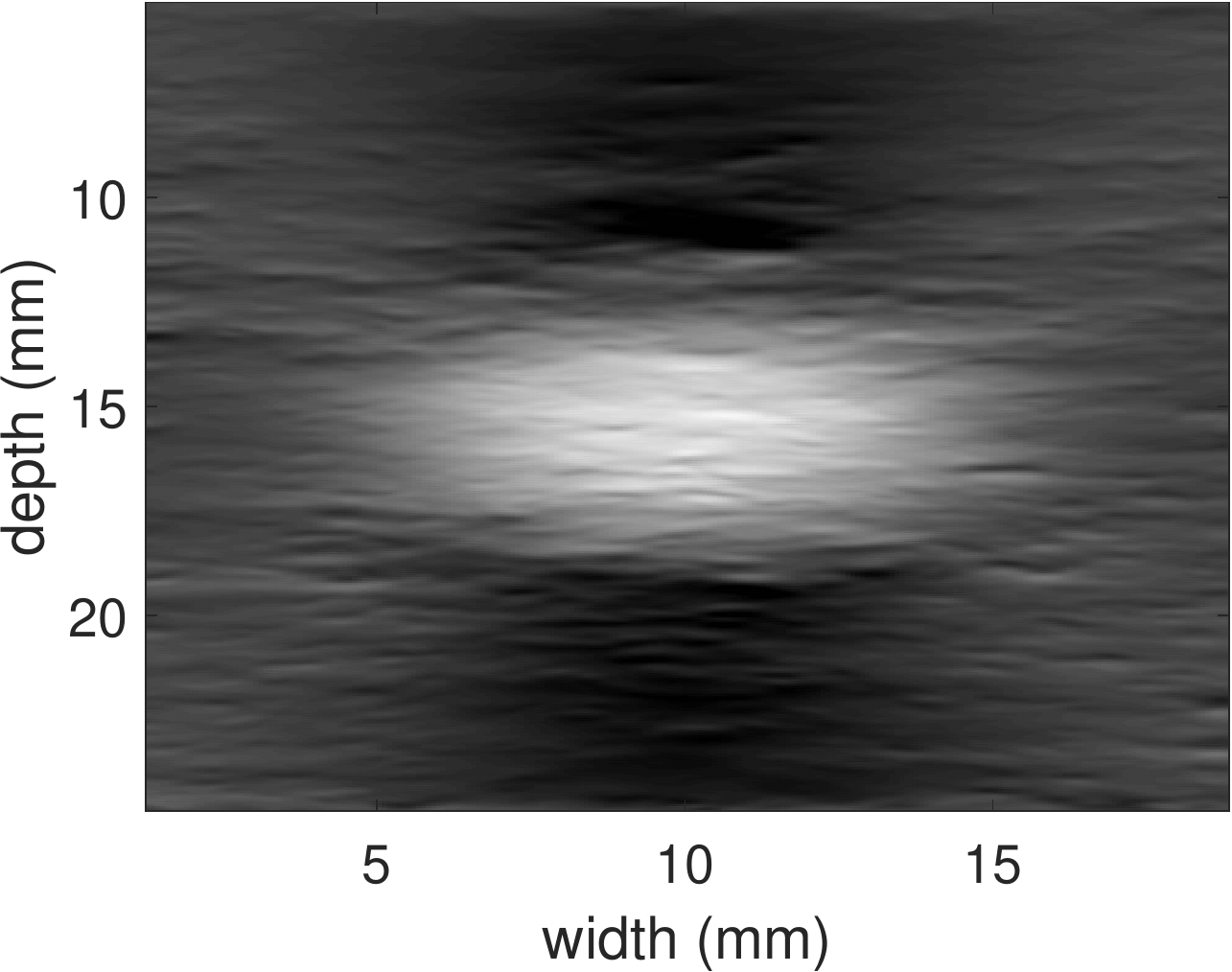} }}
	\subfigure[Strain, $1\%$ compression]{{\includegraphics[width=.2\textwidth]{results/simulation/color_1} }}%
	\subfigure[Strain, $1.5\%$ compression]{{\includegraphics[width=.2\textwidth]{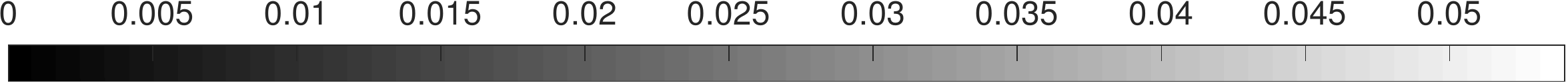} }}%
	\subfigure[Strain, $2\%$ compression]{{\includegraphics[width=.2\textwidth]{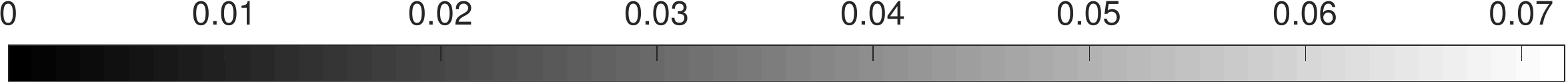} }}%
	\subfigure[Strain, $2.5\%$ compression]{{\includegraphics[width=0.2\textwidth]{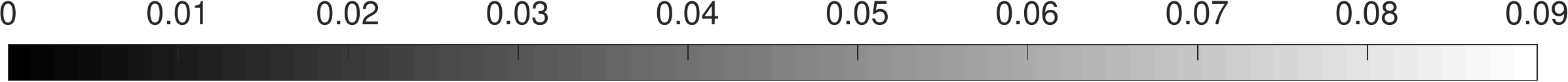} }}%
	\subfigure[Strain, $3\%$ compression]{{\includegraphics[width=0.2\textwidth]{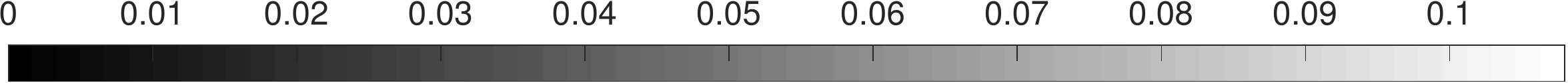} }}
	\caption{Axial strain images for the soft-inclusion simulation phantom. Rows 1-3 correspond to Hybrid, GLUE, and rGLUE, respectively. Columns 1-5 correspond to the compression levels of $1\%$, $1.5\%$, $2\%$, $2.5\%$, and $3\%$, respectively.}
	\label{simulation1}
\end{figure*}

Fig.~\ref{simulation1} shows the soft-inclusion simulation strain images for $1-3\%$ compression levels. GLUE and rGLUE exhibit similar performance in this range of deformation. Hybrid successfully distinguishes the soft inclusion from the background for $1-2\%$ deformations, whereas it fails to obtain acceptable strain image while the phantom undergoes more than $2\%$ deformation.

\begin{figure*}
	\centering
	\subfigure[B-mode, pre-compressed]{{\includegraphics[width=0.2\textwidth,height=0.224\textwidth]{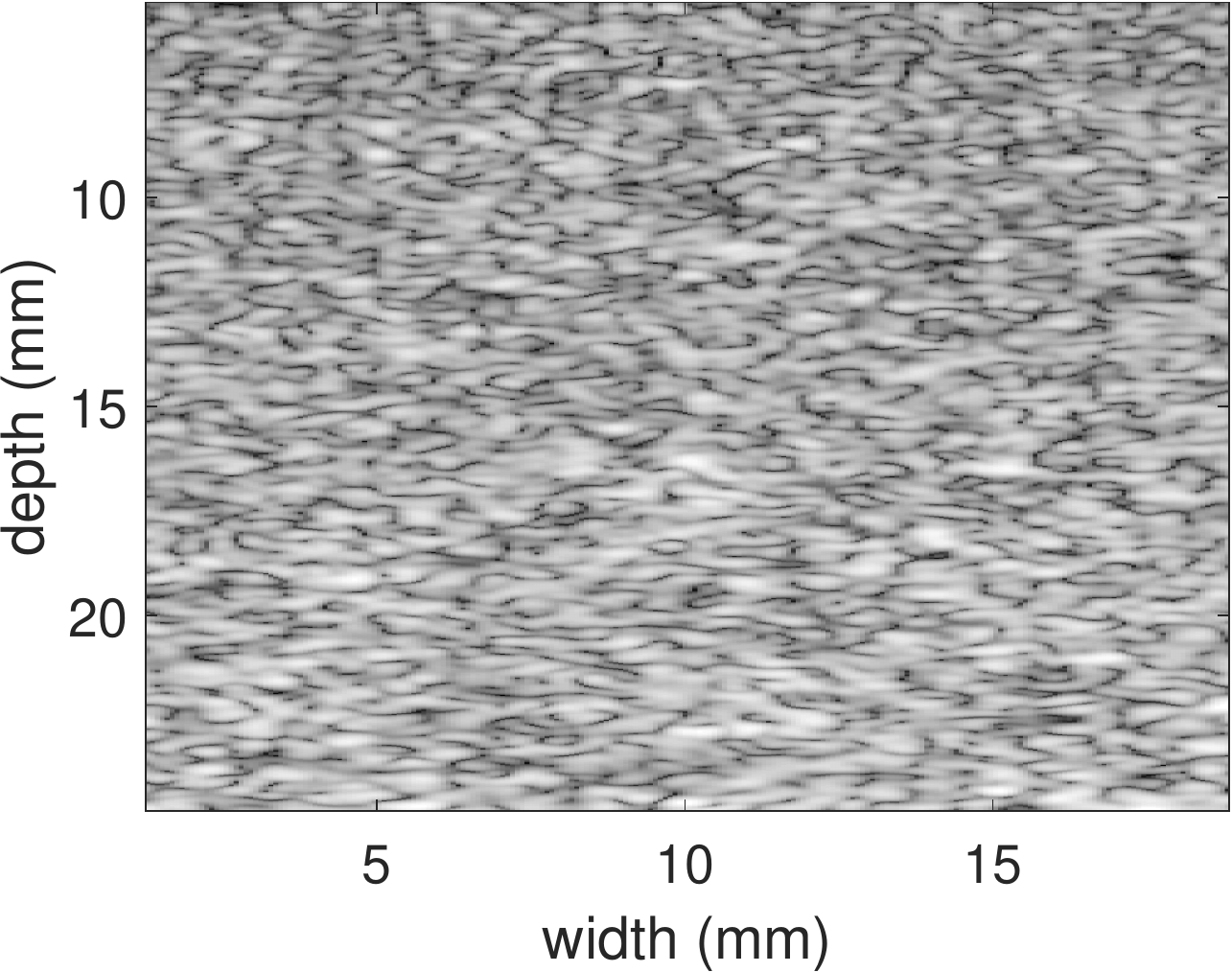} }}%
	\subfigure[B-mode, post-compressed]{{\includegraphics[width=0.2\textwidth,height=0.224\textwidth]{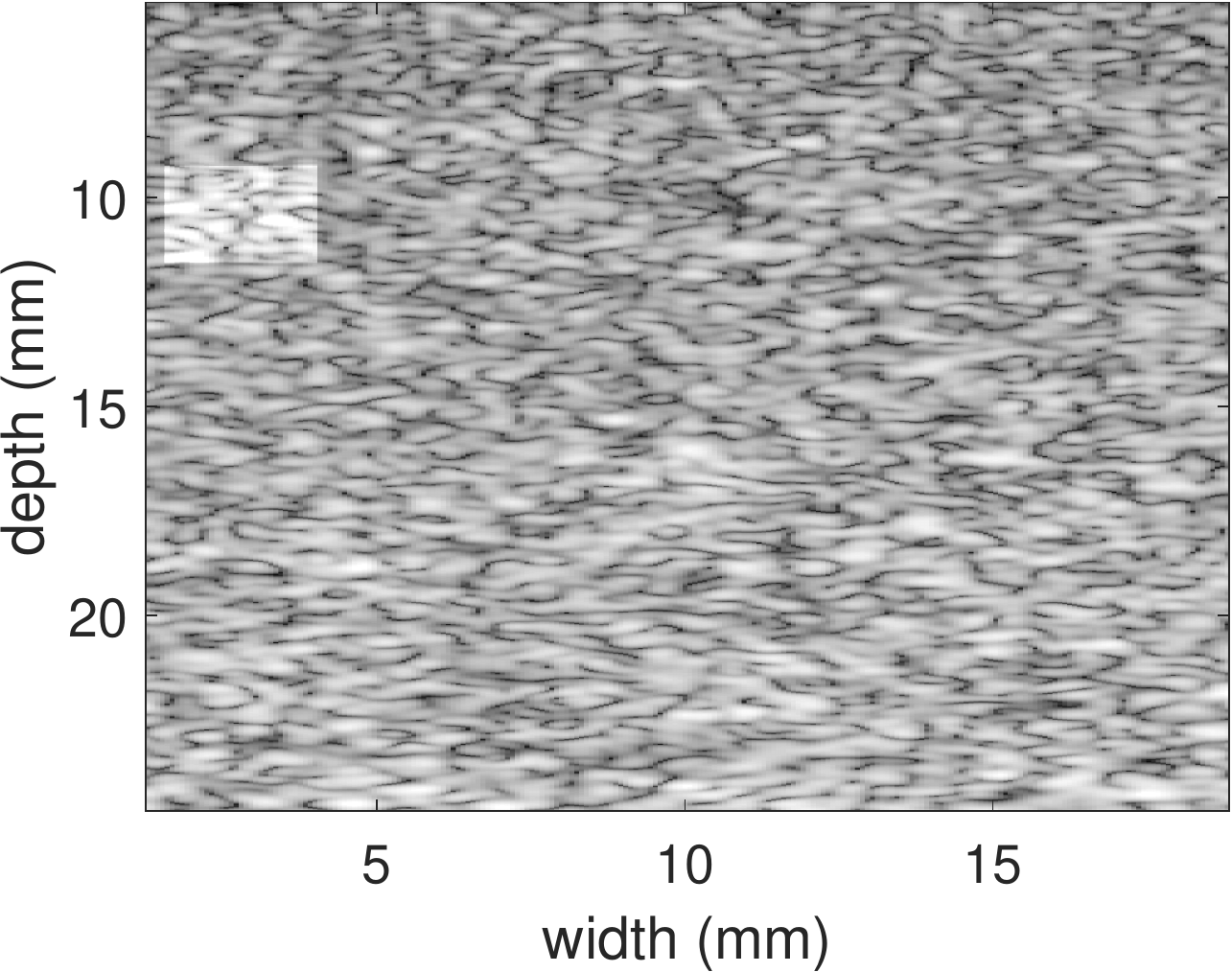} }}%
	\subfigure[GLUE for $\alpha_{1}=\beta_{1}=4$, $\alpha_{2}=\beta_{2}=0.4$]{{\includegraphics[width=0.2\textwidth,height=0.224\textwidth]{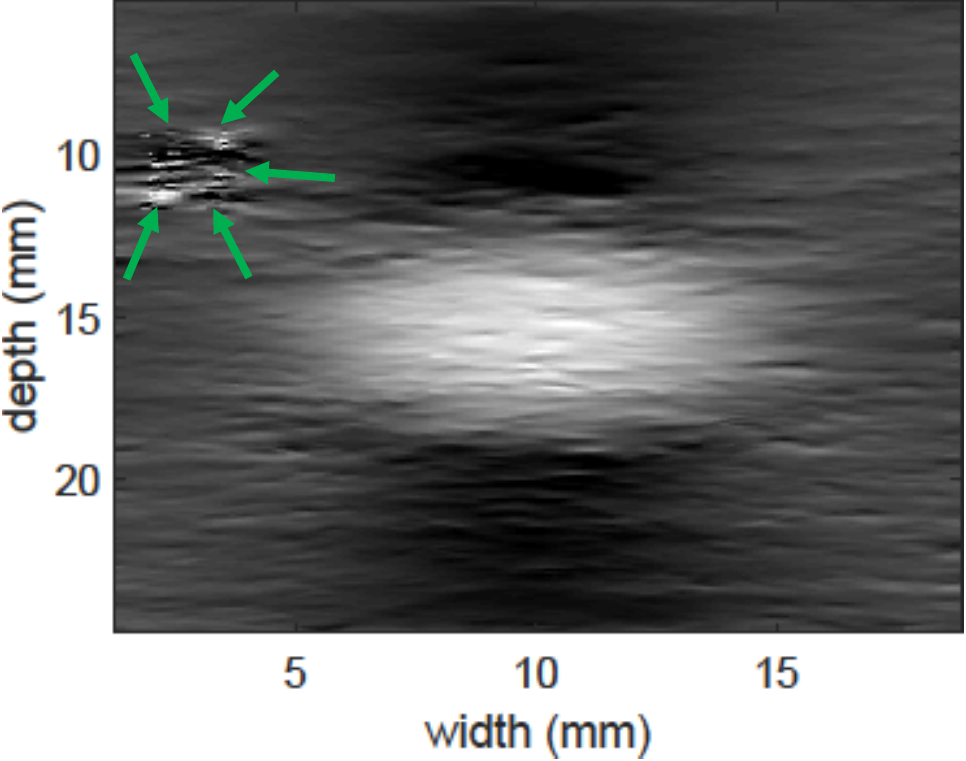} }}%
	\subfigure[GLUE for $\alpha_{1}=\beta_{1}=12$, $\alpha_{2}=\beta_{2}=1.2$]{{\includegraphics[width=0.2\textwidth,height=0.224\textwidth]{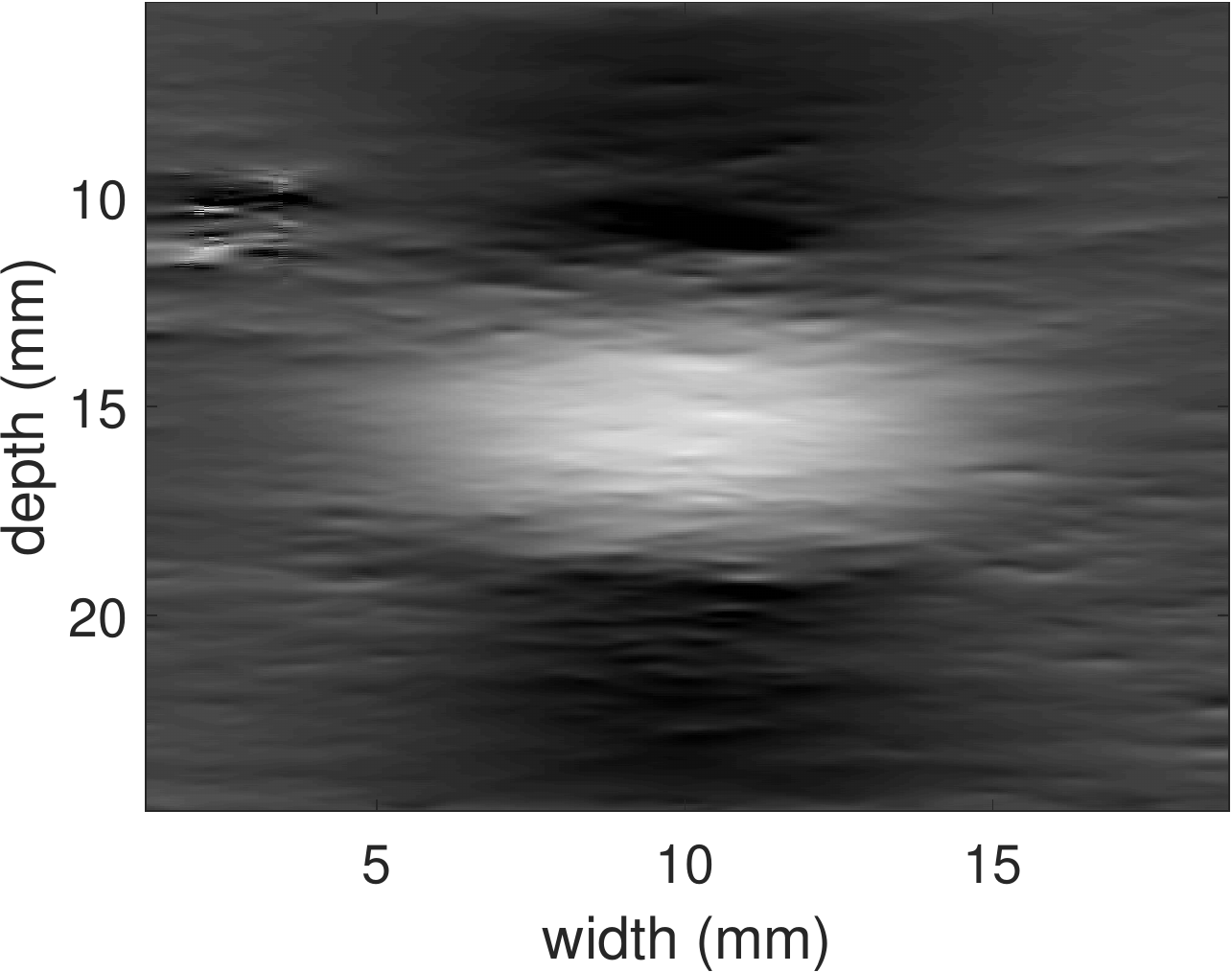} }}%
	\subfigure[rGLUE for $\alpha_{1}=\beta_{1}=3$, $\alpha_{2}=\beta_{2}=0.3$, $\lambda=20$]{{\includegraphics[width=0.2\textwidth,height=0.224\textwidth]{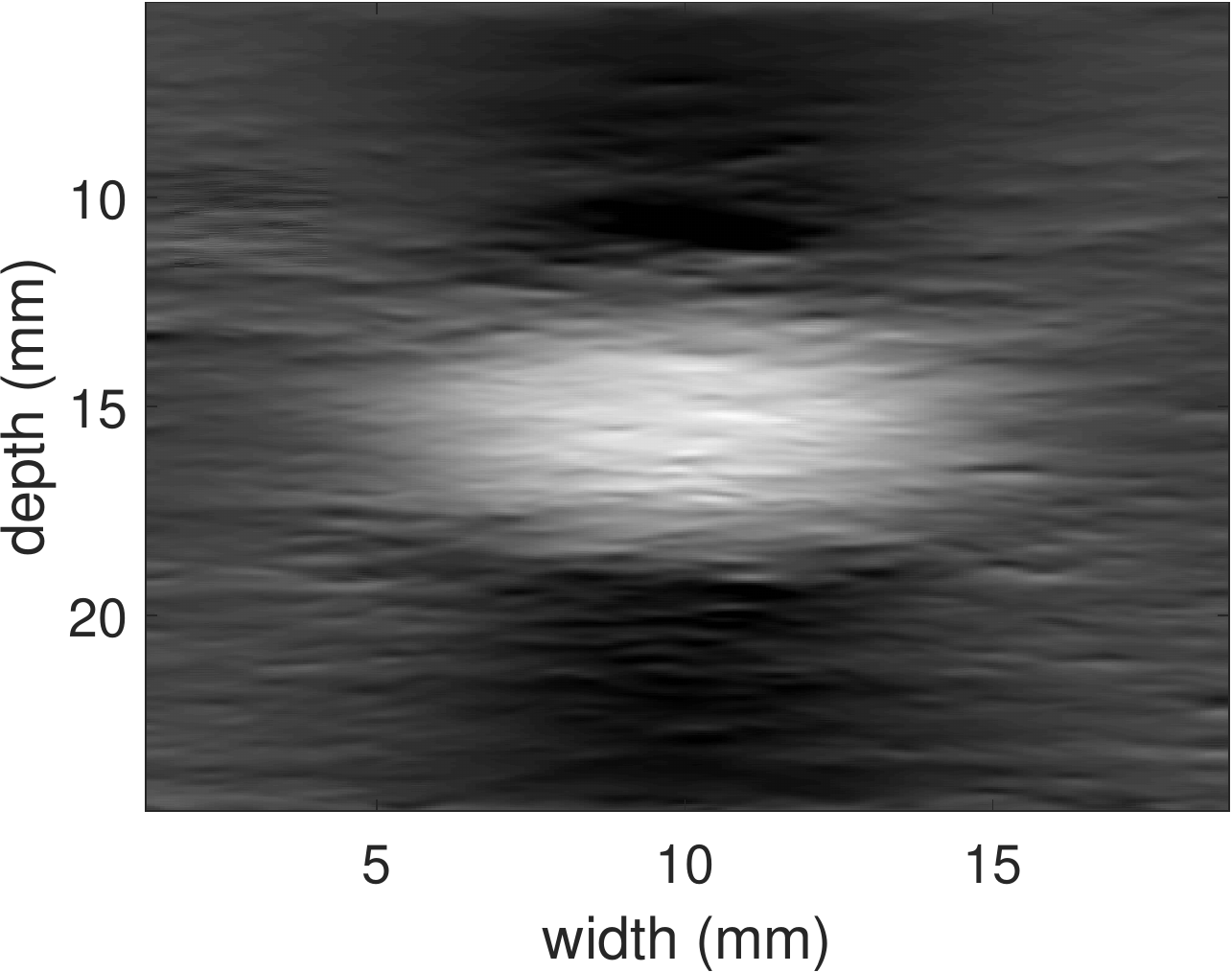} }}
	\subfigure[Strain]{{\includegraphics[width=0.33\textwidth]{results/simulation/color_2} }}%
	\caption{Results for the FEM simulated phantom with an outlier region. Columns 1 and 2 show the pre- and post-compressed B-mode images, whereas columns 3 and 4 show the strain images from GLUE with different levels of regularization. Column 5 represents the strain image obtained by rGLUE. The green arrow marks outline the strain artifacts stemming from the outlier samples.}
	\label{simu_analysis}
\end{figure*}

Fig.~\ref{simu_analysis} depicts the strain images for the soft-inclusion simulated phantom with multiplicative outliers. The outlier samples introduce green arrow marked artifacts in the strain image obtained by GLUE. We obtain the strain images from GLUE by setting \{$\alpha_{1}$, $\alpha_{2}$, $\beta_{1}$, $\beta_{2}$\} to two different combinations: \{4, 0.4, 4, 0.4\} and \{12, 1.2, 12, 1.2\}. It is evident that increasing the spatial regularization weights does not remove the artifacts originating from the outlier samples. On the other hand, the increased values of the continuity parameters oversmoothe the strain image. rGLUE adaptively reduces the weights of the outlier samples to efficiently avoid large strain artifacts while working with optimal parameters values. It is worth mentioning that along with outlier handling, rGLUE maintains a proper contrast between the target and the background.\\

\begin{figure*}
	\centering
	\subfigure[B-mode 1]{{\includegraphics[width=0.2\textwidth,height=0.24\textwidth]{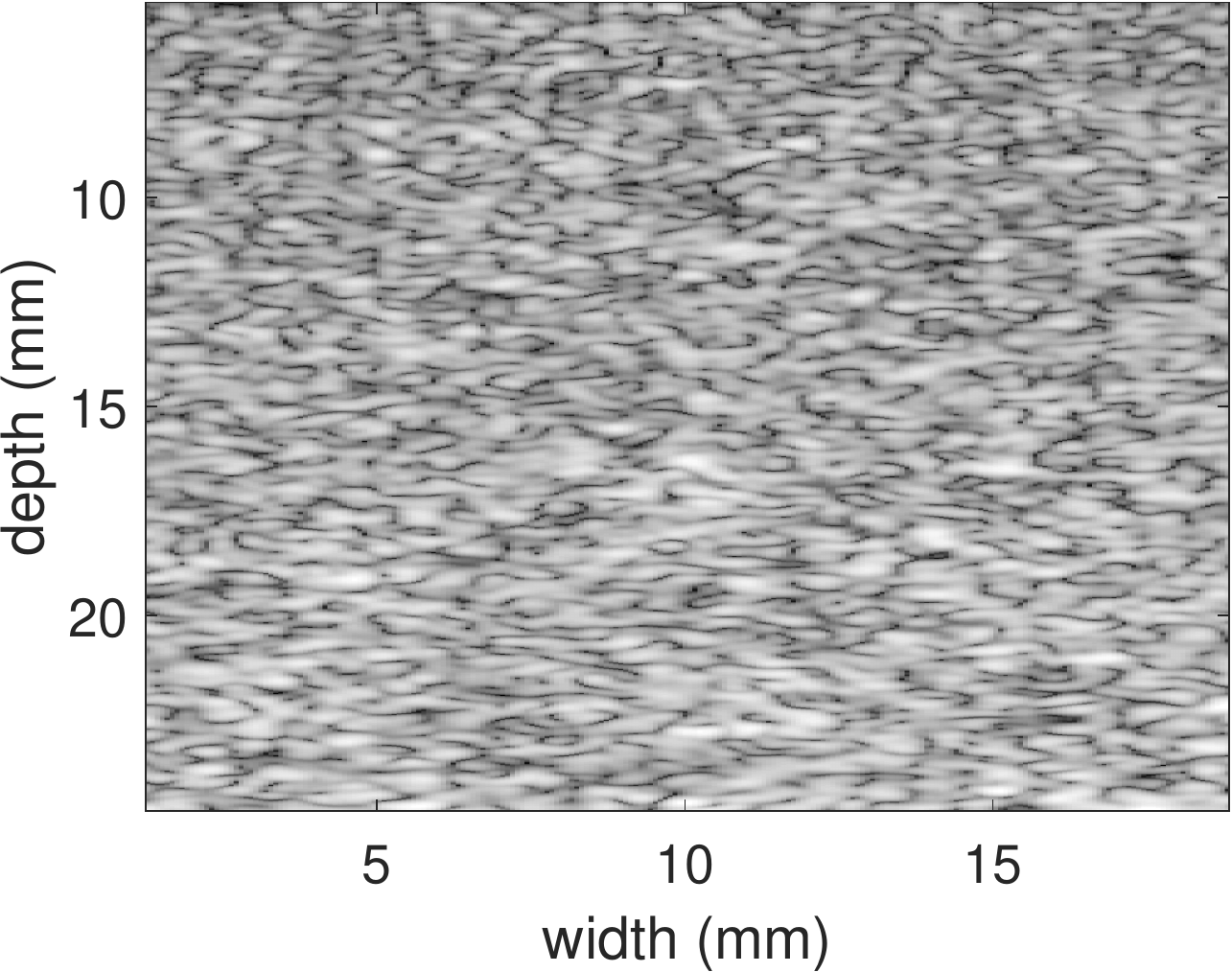} }}%
	\subfigure[B-mode 2]{{\includegraphics[width=0.2\textwidth,height=0.24\textwidth]{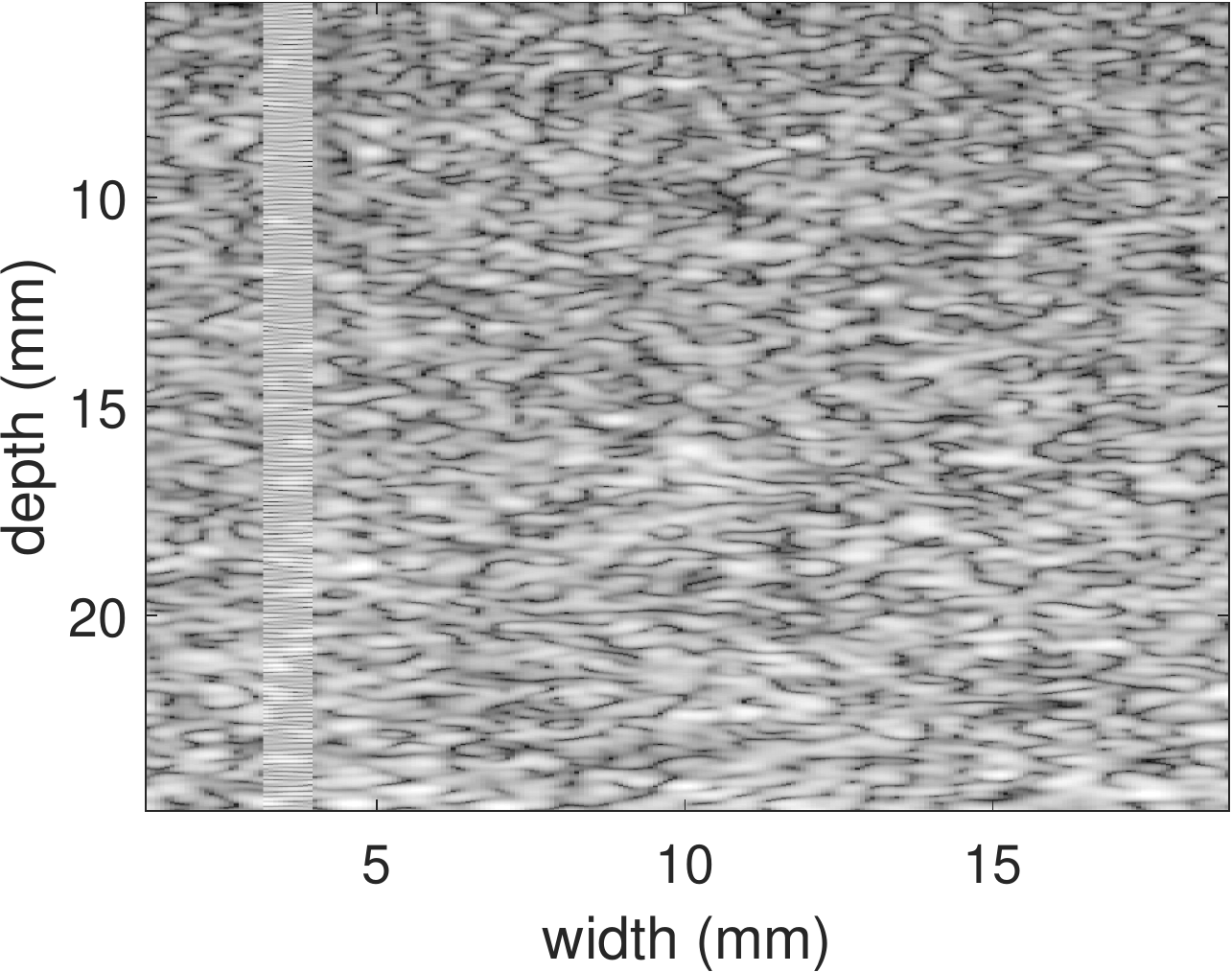} }}%
	\subfigure[Hybrid]{{\includegraphics[width=0.2\textwidth,height=0.24\textwidth]{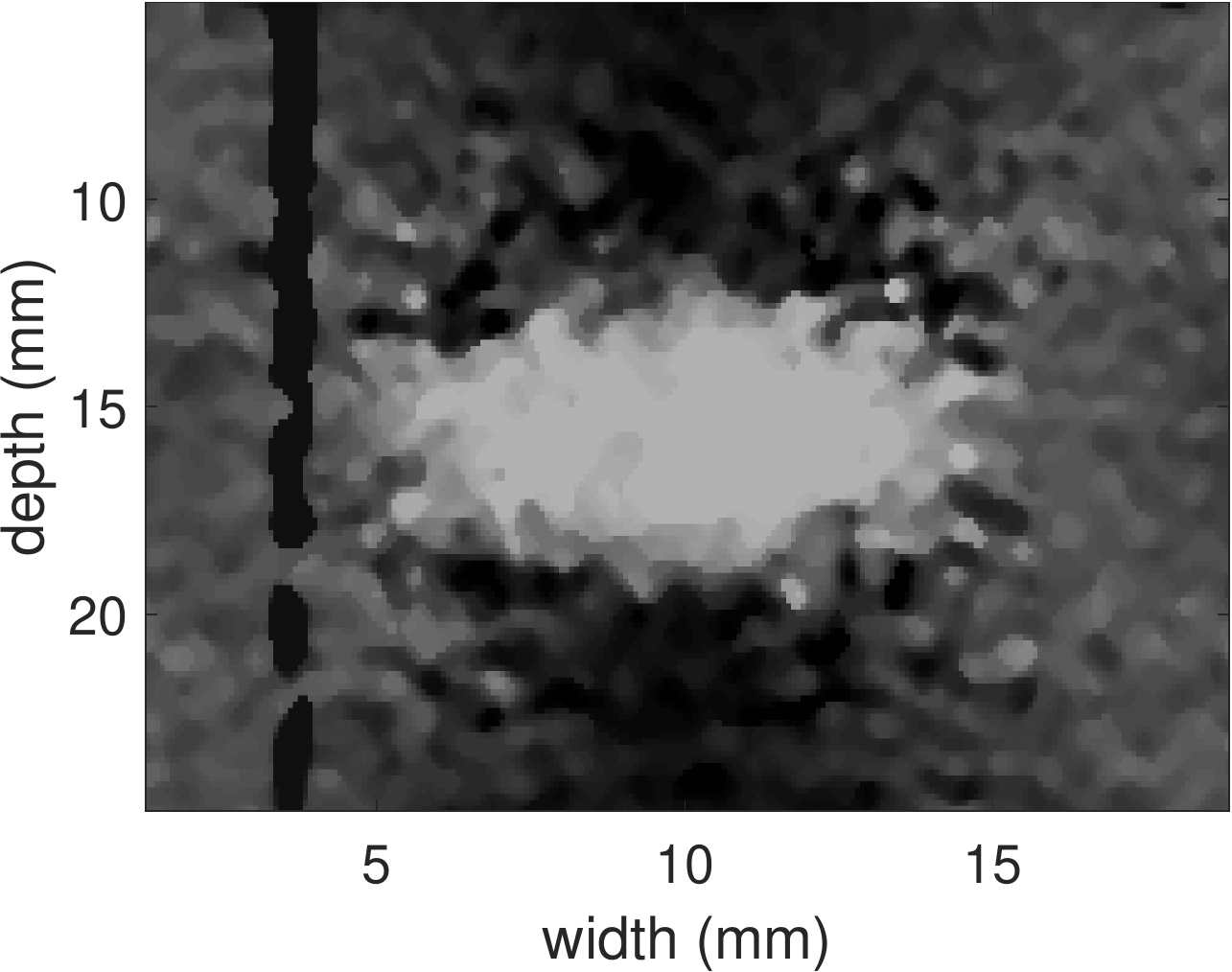} }}%
	\subfigure[GLUE]{{\includegraphics[width=0.2\textwidth,height=0.24\textwidth]{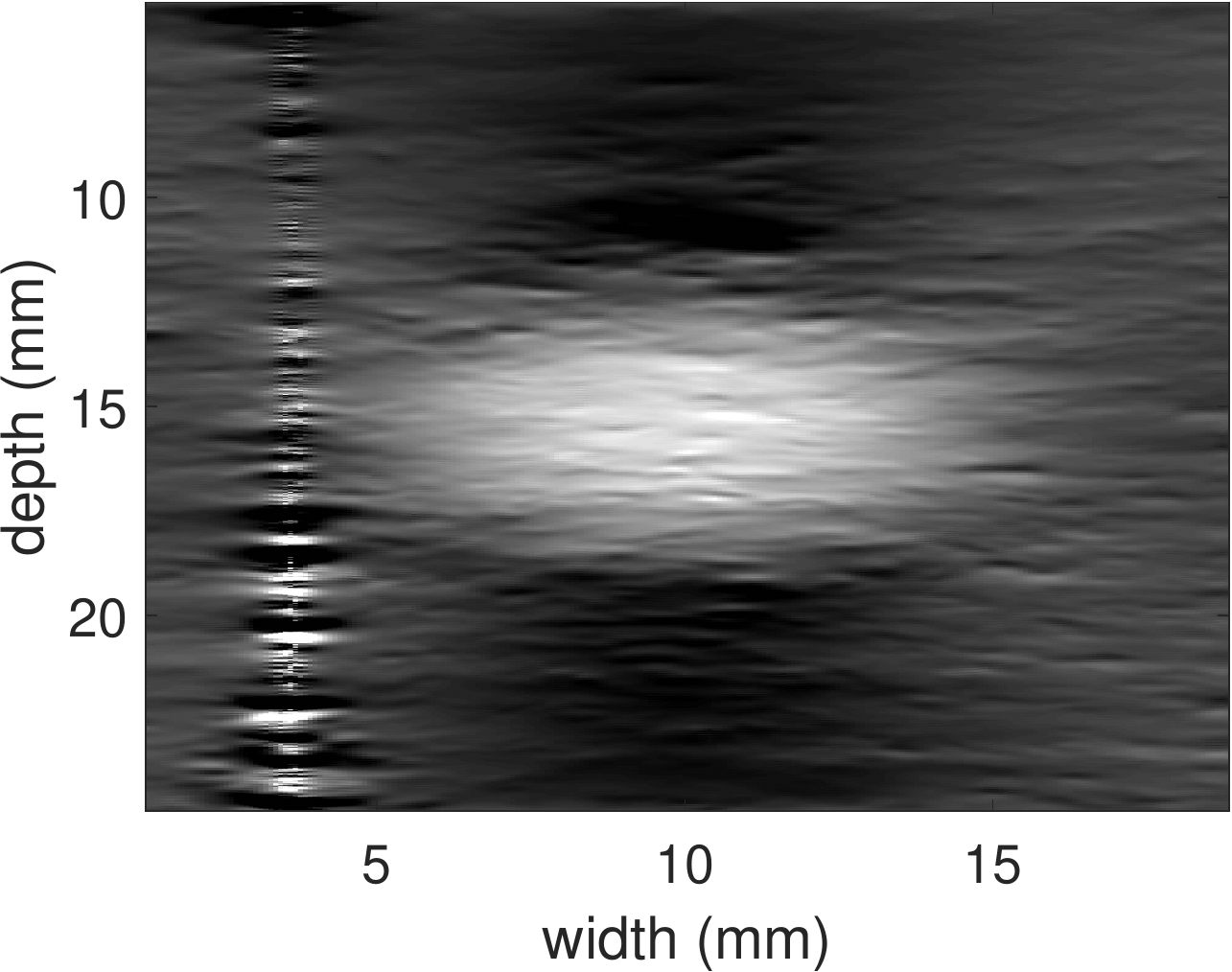} }}%
	\subfigure[rGLUE]{{\includegraphics[width=0.2\textwidth,height=0.24\textwidth]{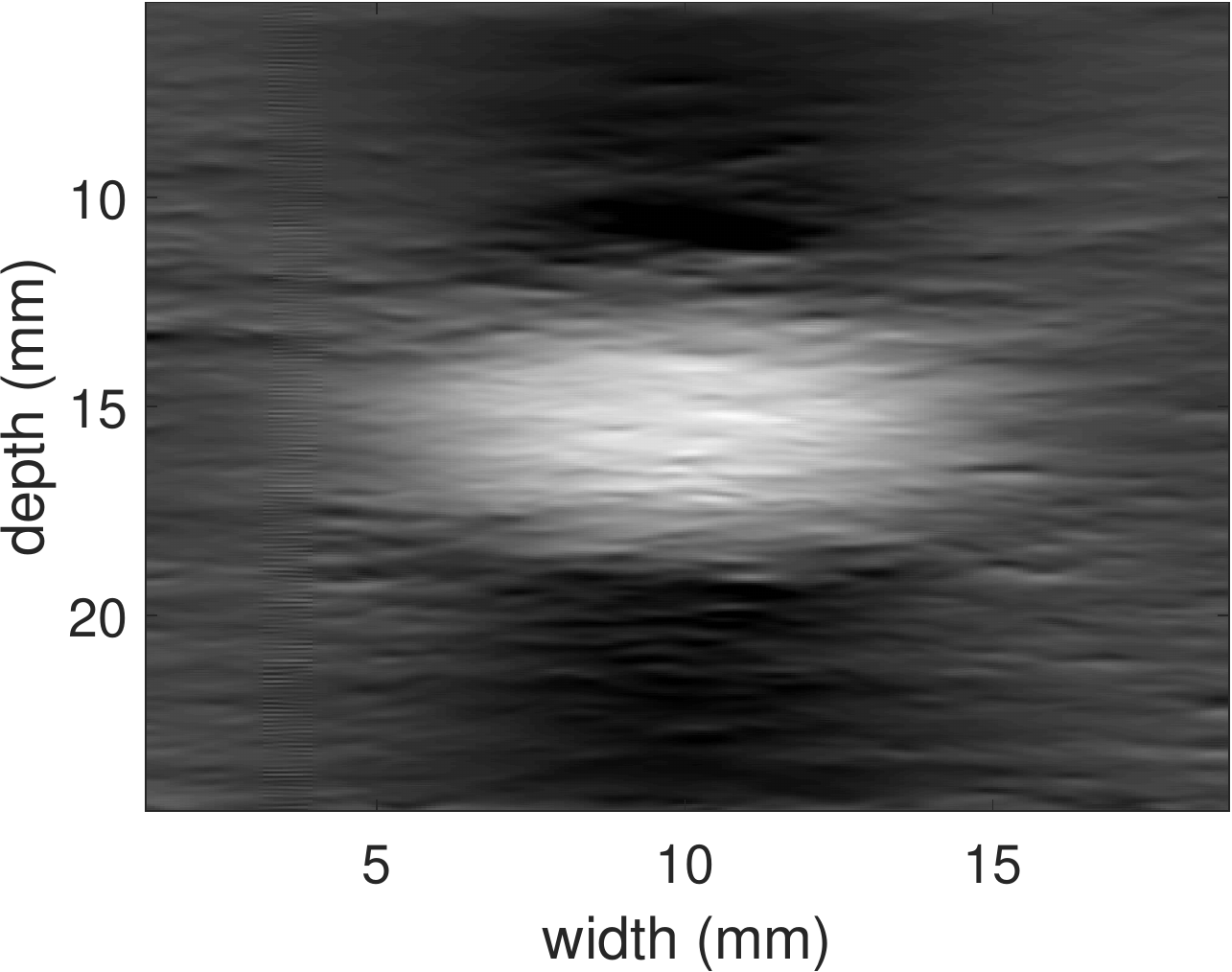} }}
	\subfigure[Strain]{{\includegraphics[width=0.33\textwidth]{results/simulation/color_2} }}
	\caption{Simulation results with additive outliers. Columns 1 and 2 depict the pre- and post-compressed B-mode images. Columns 3-5 show the strain images from Hybrid, GLUE, and rGLUE, respectively.}
	\label{simu_analysis2}
\end{figure*}

Fig.~\ref{simu_analysis2} shows the techniques' performance in the presence of additive outliers. Both Hybrid and GLUE strain images suffer from large artifacts due to the outliers added to few RF lines. rGLUE proves its robustness to outliers by generating a strain image without any apparent artifact.\\



\begin{figure*}
	\centering
	\subfigure[Ground truth strain]{{\includegraphics[width=0.25\textwidth,height=0.39\textwidth]{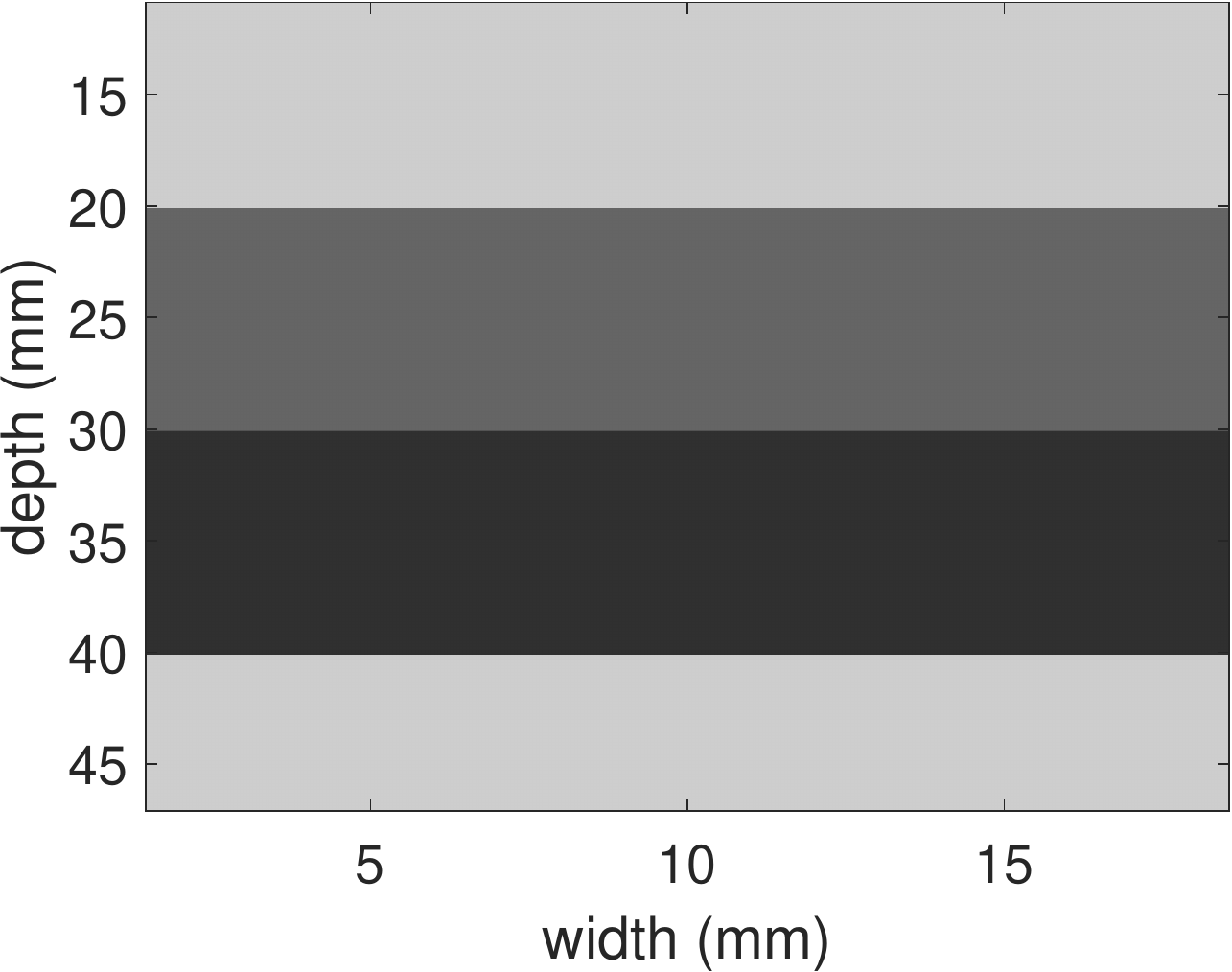} }}%
	\subfigure[Hybrid]{{\includegraphics[width=0.25\textwidth,height=0.39\textwidth]{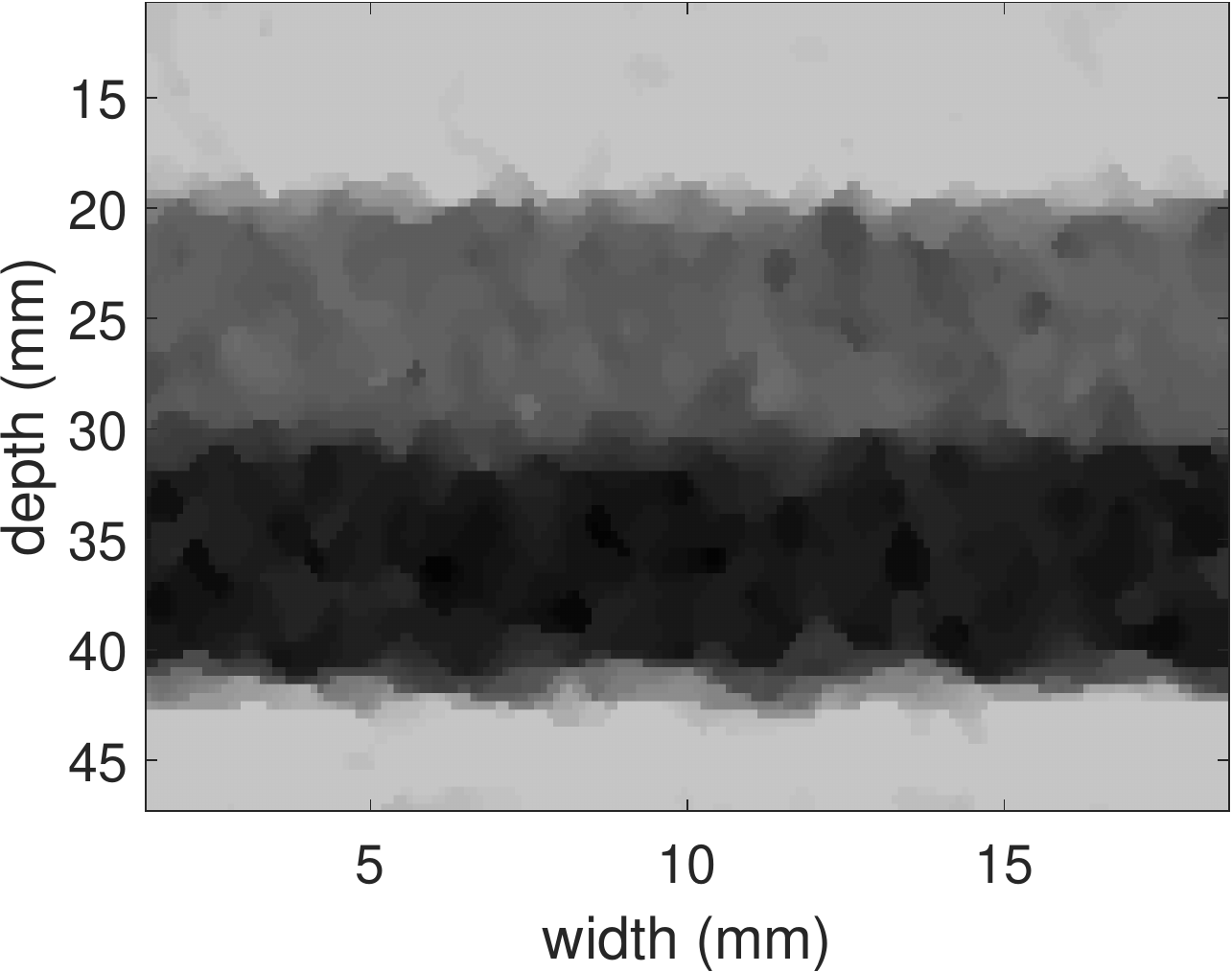} }}%
	\subfigure[GLUE]{{\includegraphics[width=0.25\textwidth,height=0.39\textwidth]{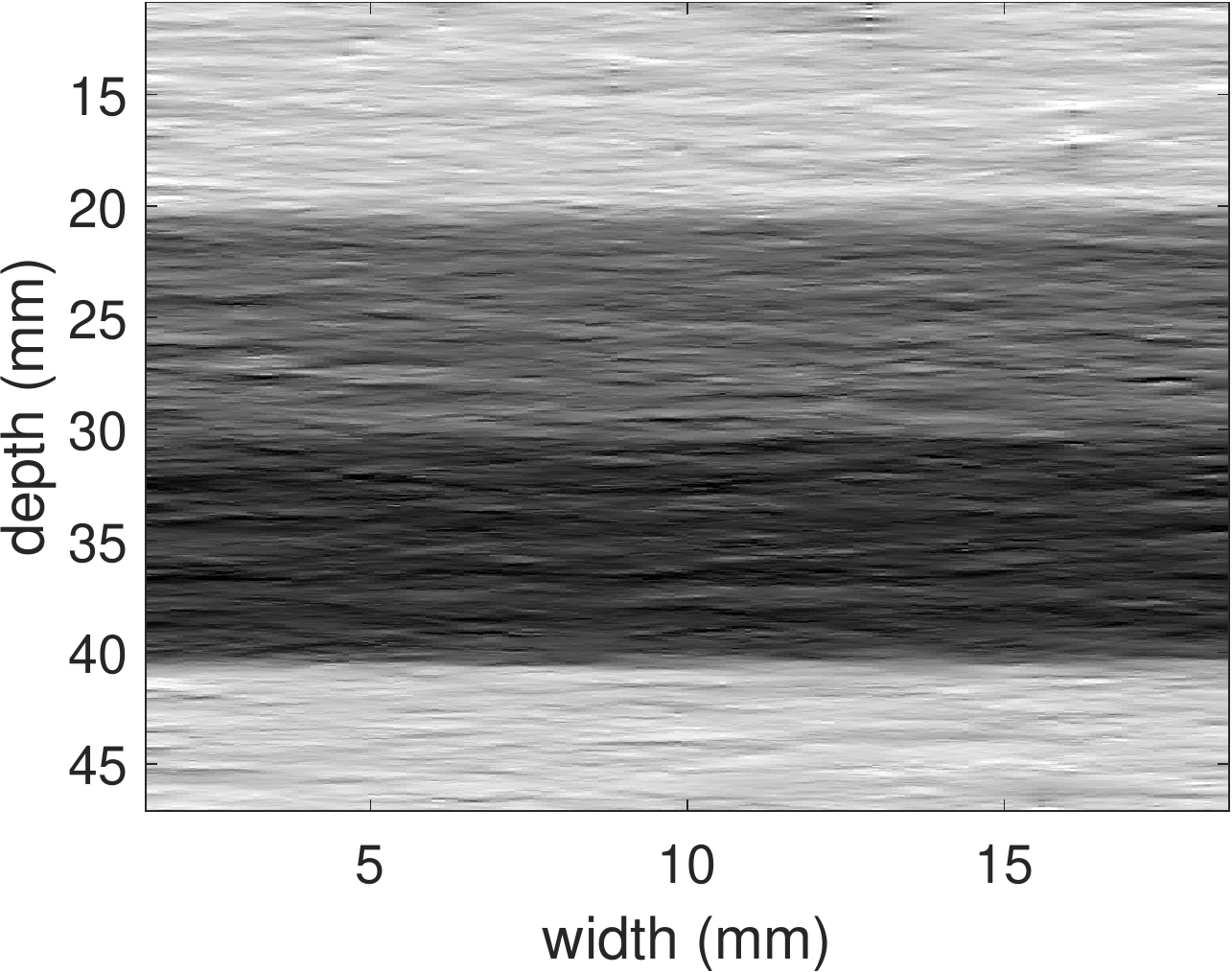} }}%
	\subfigure[rGLUE]{{\includegraphics[width=0.25\textwidth,height=0.39\textwidth]{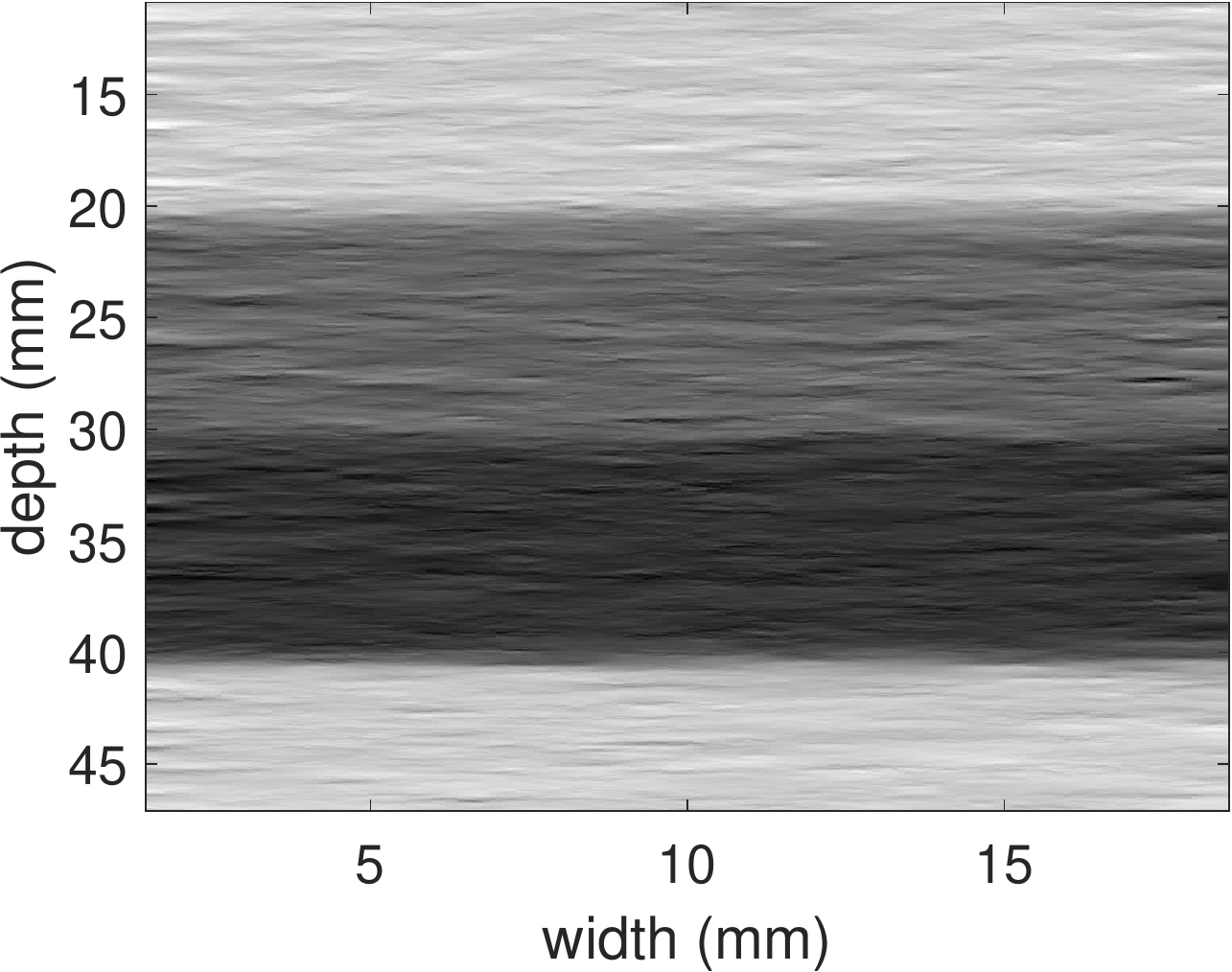} }}
	\subfigure[Strain]{{\includegraphics[width=0.3\textwidth]{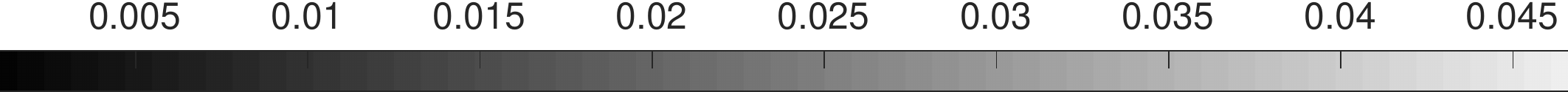} }}
	\caption{Results obtained from the four-layer simulated phantom with added Gaussian noise. Columns 1-4 correspond to ground truth, Hybrid, GLUE, and rGLUE, respectively.}
	\label{simu_distribution}
\end{figure*}

Fig.~\ref{simu_distribution} shows the ground truth and strain estimates for the four-layer phantom. Hybrid exhibits extensive variability in the stiff layer. In addition, the boundaries between different layers are not satisfactorily defined. GLUE yields several artifacts in the background. rGLUE substantially outperforms Hybrid and GLUE by producing an artifact-free strain image. The RMSE values reported in Table~\ref{table_small_low} support our visual inference.\\

\begin{figure*}
	\centering
	\subfigure[B-mode 1]{{\includegraphics[width=0.16\textwidth,height=0.28\textwidth]{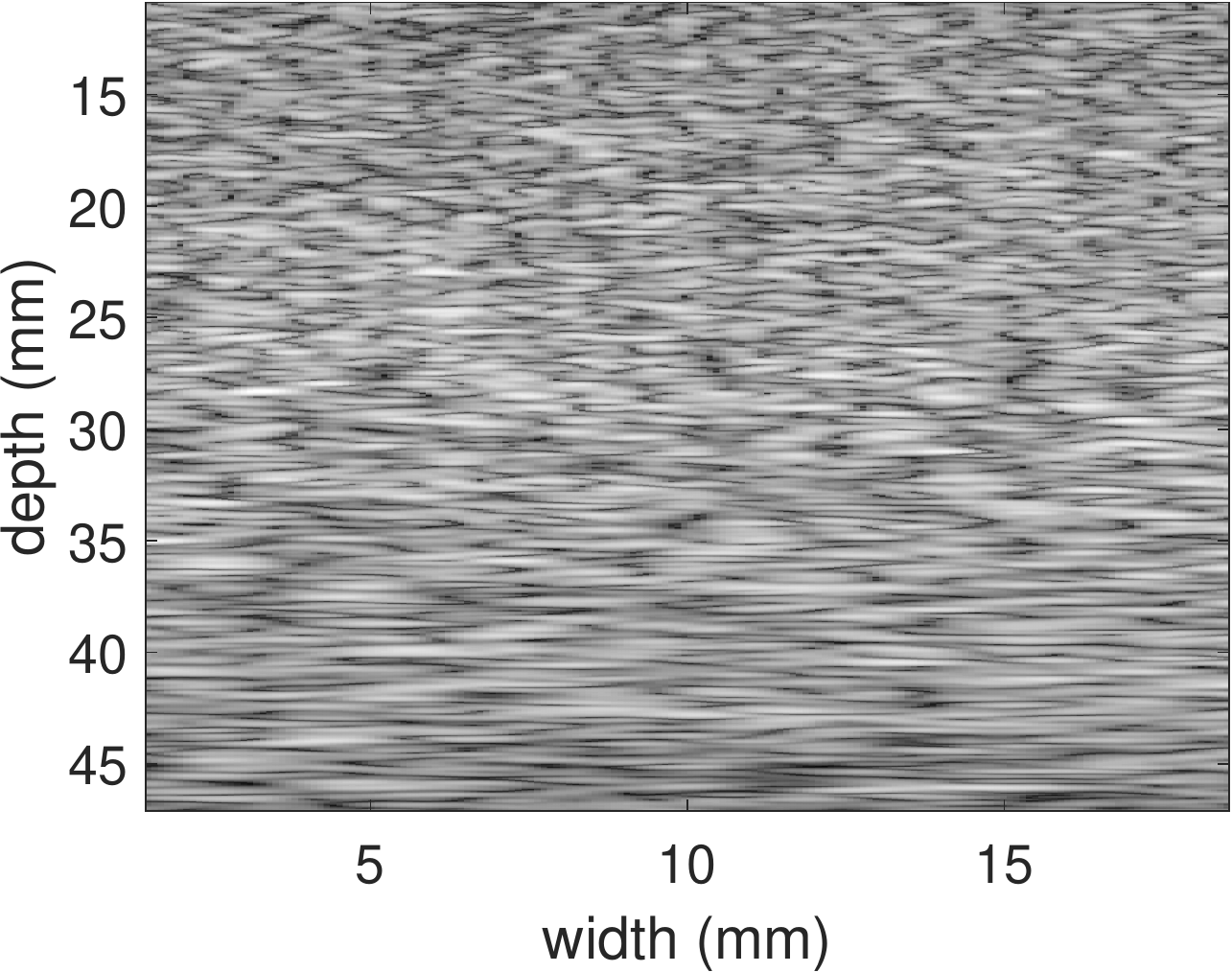} }}%
	\subfigure[B-mode 2]{{\includegraphics[width=0.16\textwidth,height=0.28\textwidth]{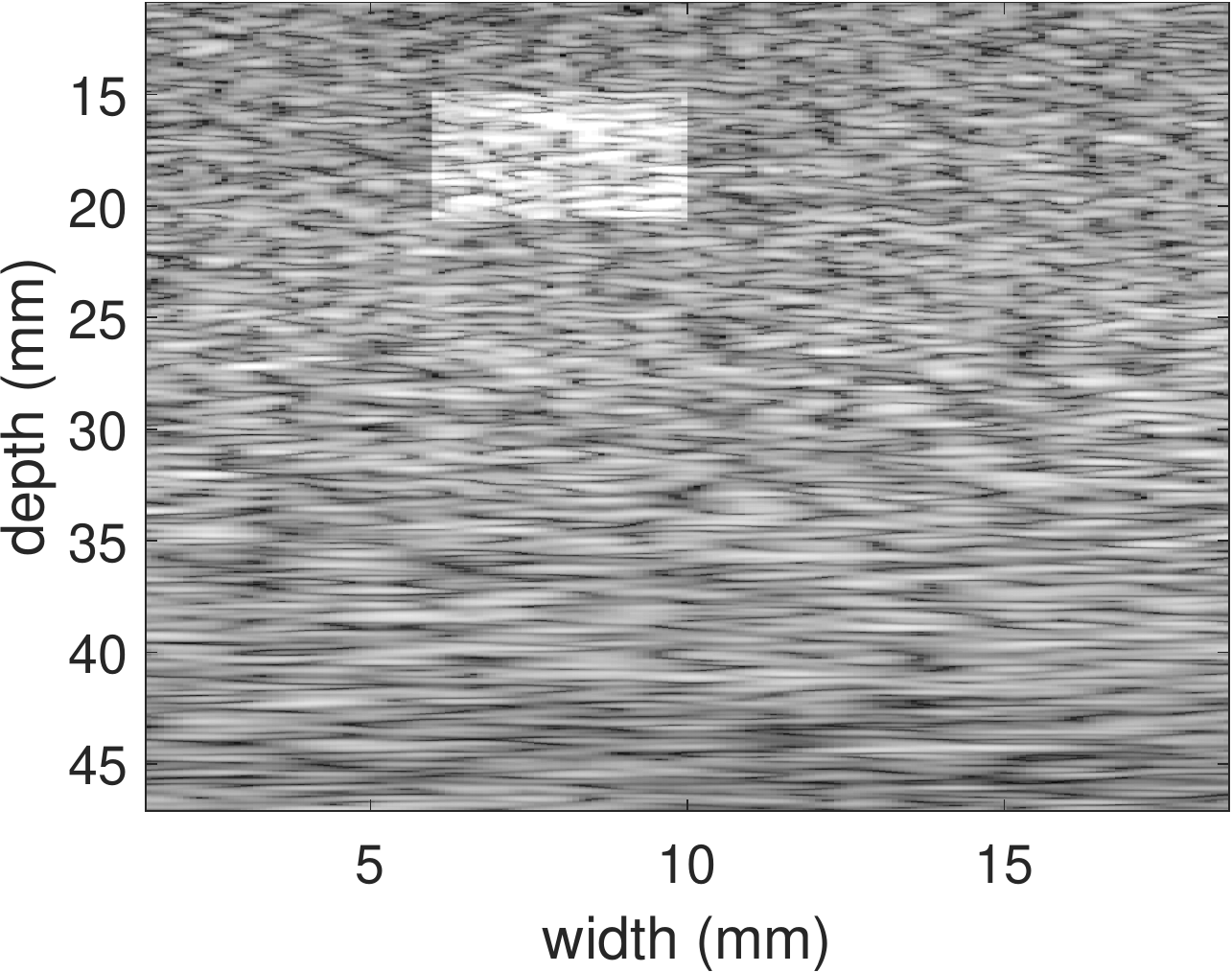} }}%
	\subfigure[Ground truth strain]{{\includegraphics[width=0.16\textwidth,height=0.28\textwidth]{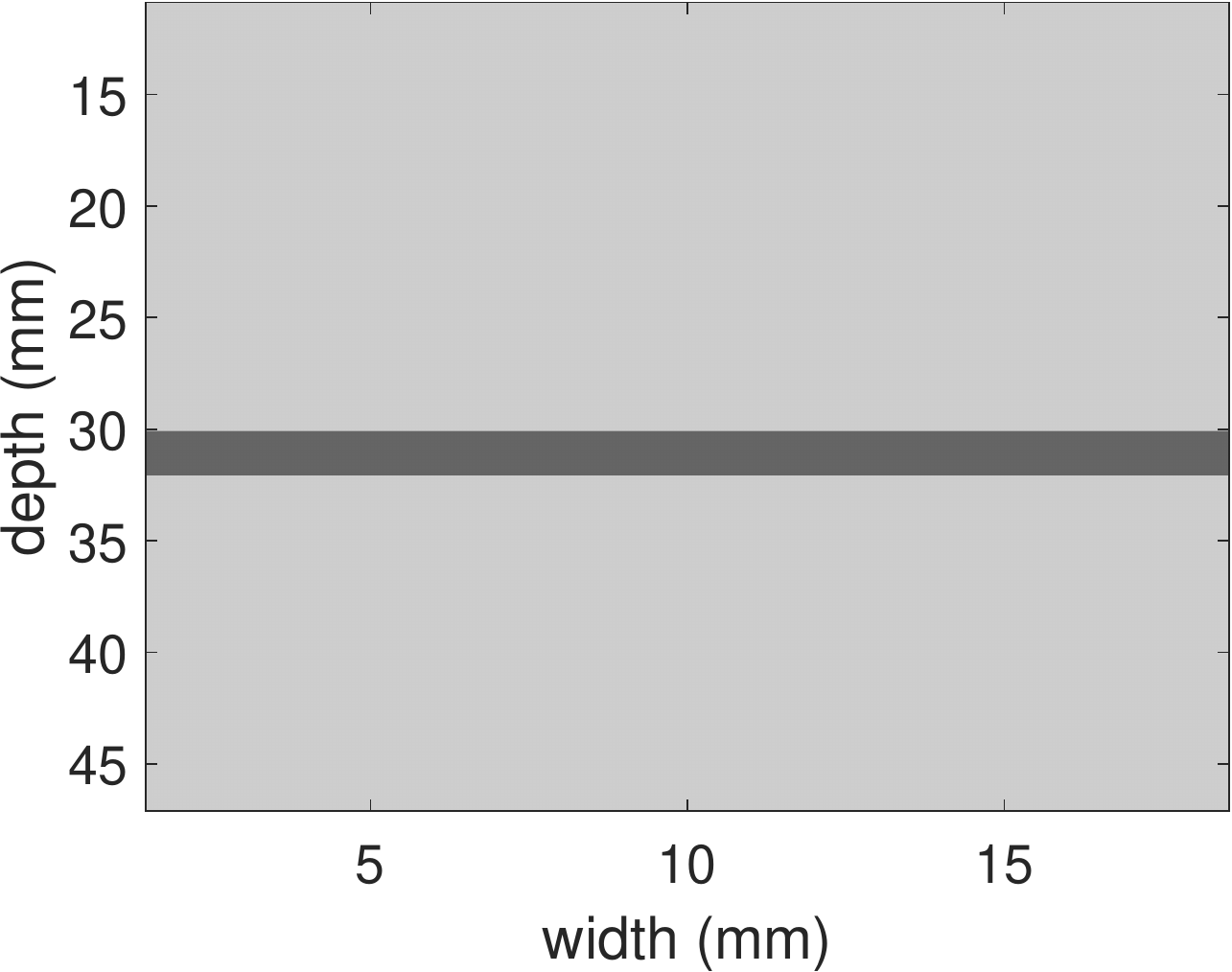} }}%
	\subfigure[Hybrid]{{\includegraphics[width=0.16\textwidth,height=0.28\textwidth]{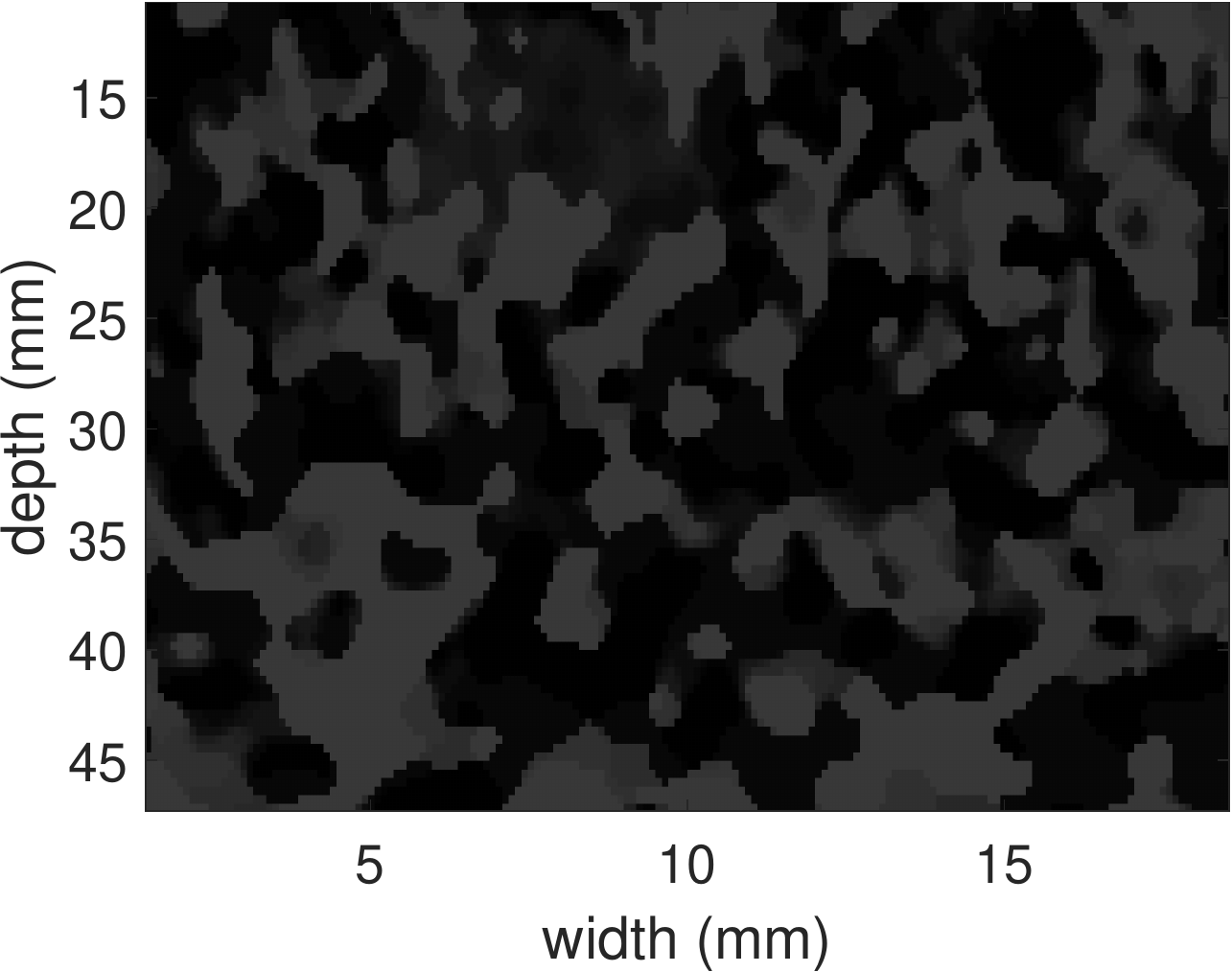} }}%
	\subfigure[GLUE]{{\includegraphics[width=0.16\textwidth,height=0.28\textwidth]{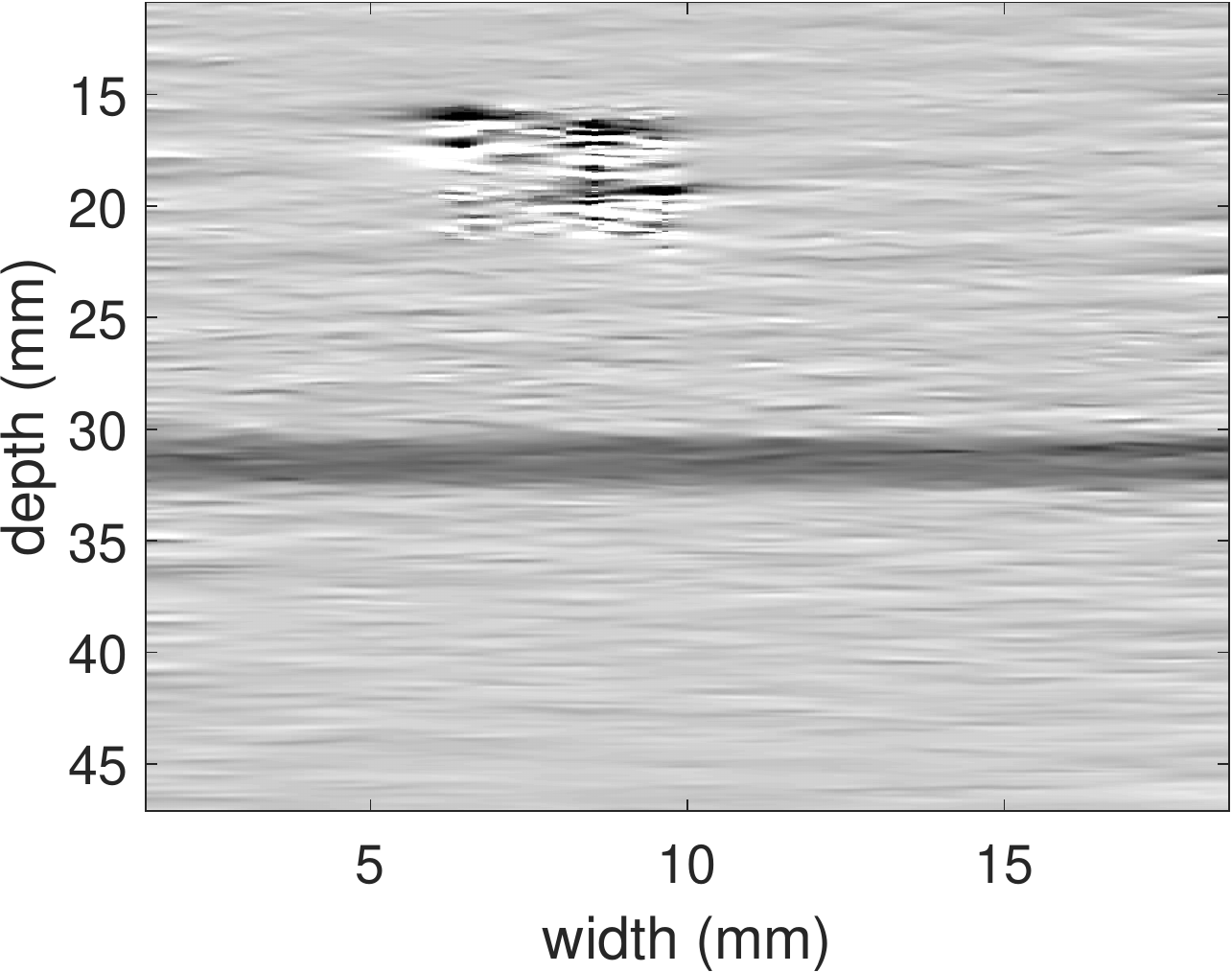} }}%
	\subfigure[rGLUE]{{\includegraphics[width=0.16\textwidth,height=0.28\textwidth]{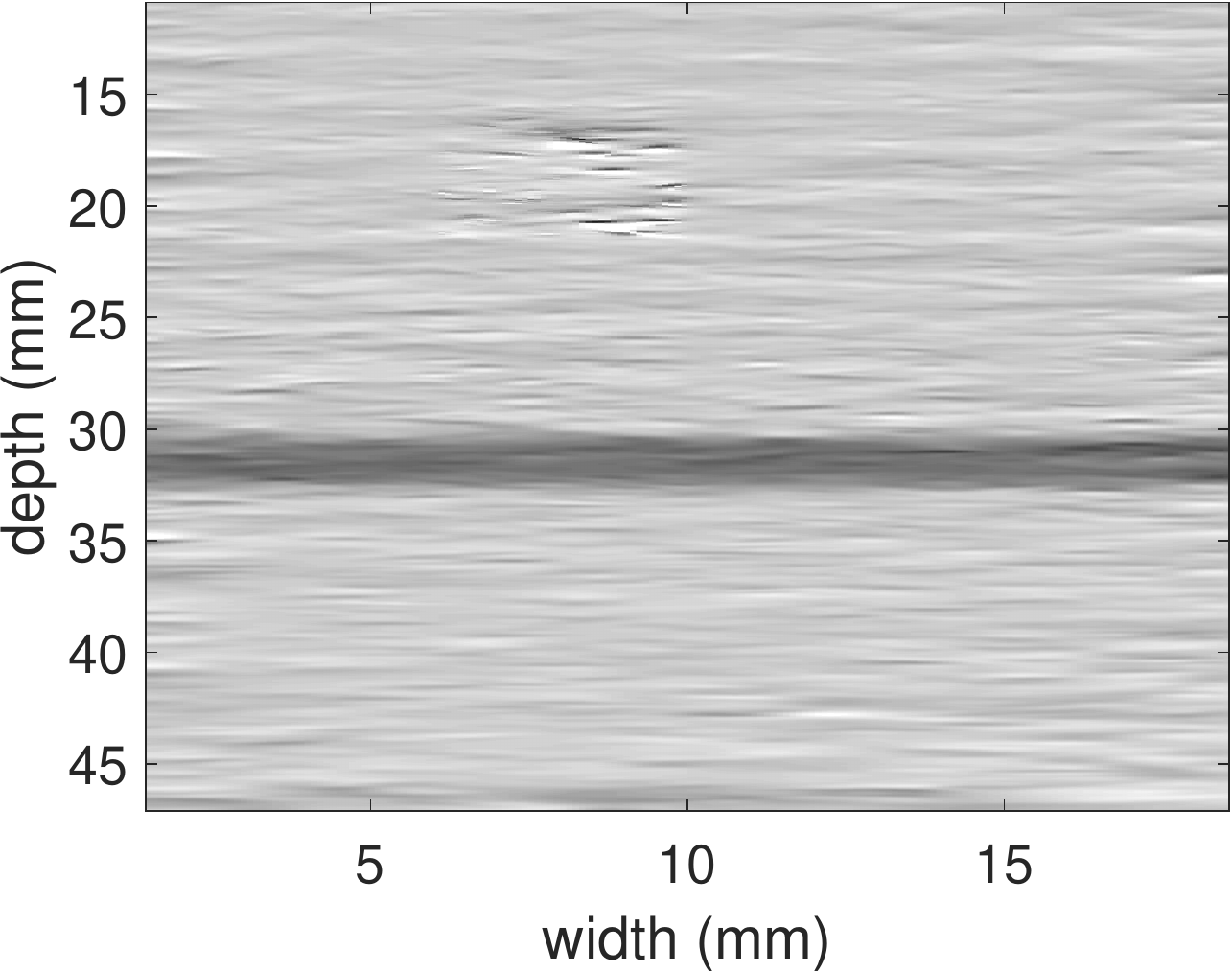} }}
	\subfigure[Strain]{{\includegraphics[width=0.33\textwidth]{results/simu_small/color} }}
	\caption{Simulation results obtained from the thin-layered phantom. Columns 1 and 2 show the pre- and post-compressed B-mode images, respectively whereas column 3 depicts the ground truth strain map. Columns 4-6 show the strain images generated by Hybrid, GLUE, and rGLUE, respectively. Row 2 shows the color bar.}
	\label{simu_small}
\end{figure*}

\begin{figure*}
	\centering
	\subfigure[B-mode 1]{{\includegraphics[width=0.16\textwidth,height=0.28\textwidth]{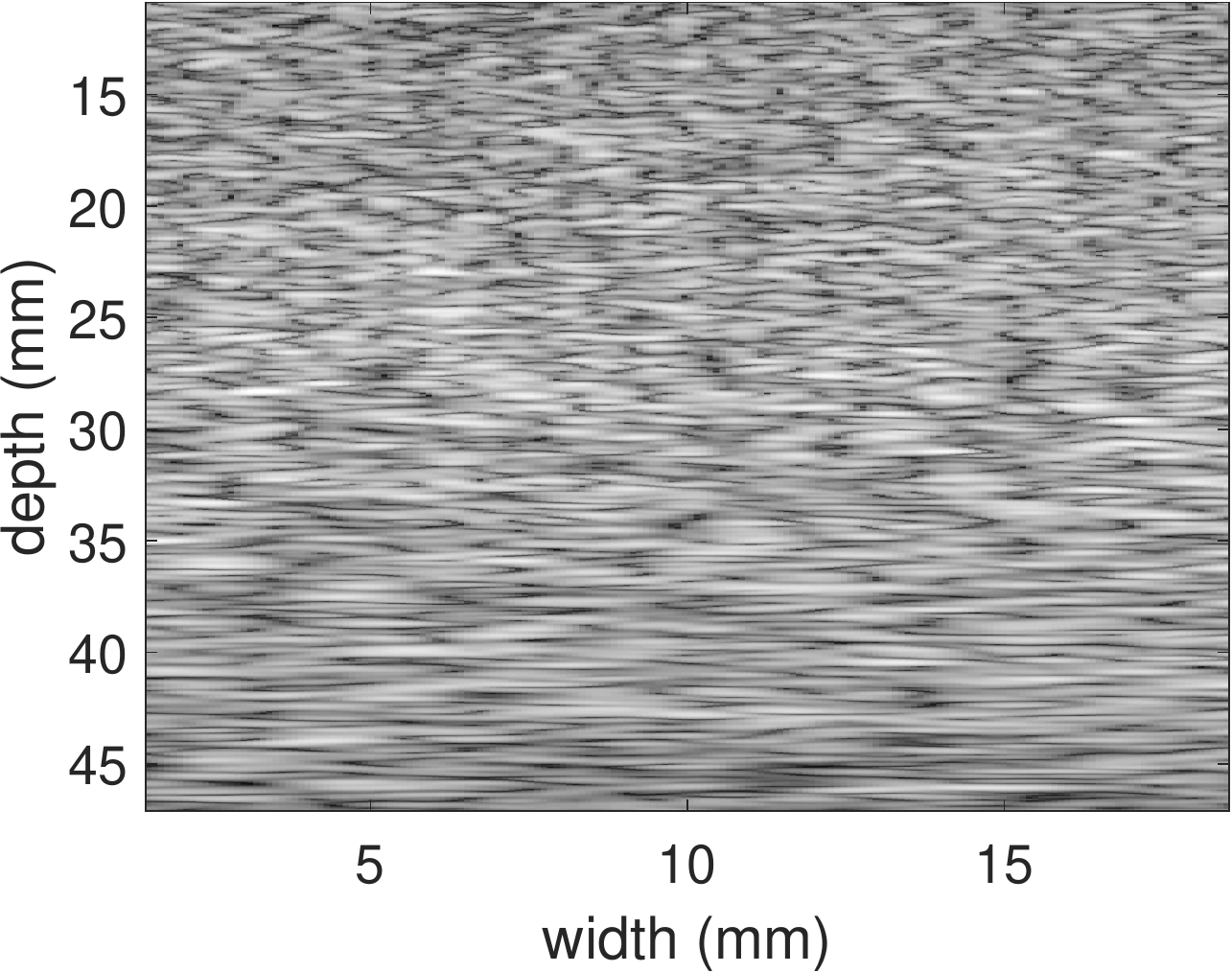} }}%
	\subfigure[B-mode 2]{{\includegraphics[width=0.16\textwidth,height=0.28\textwidth]{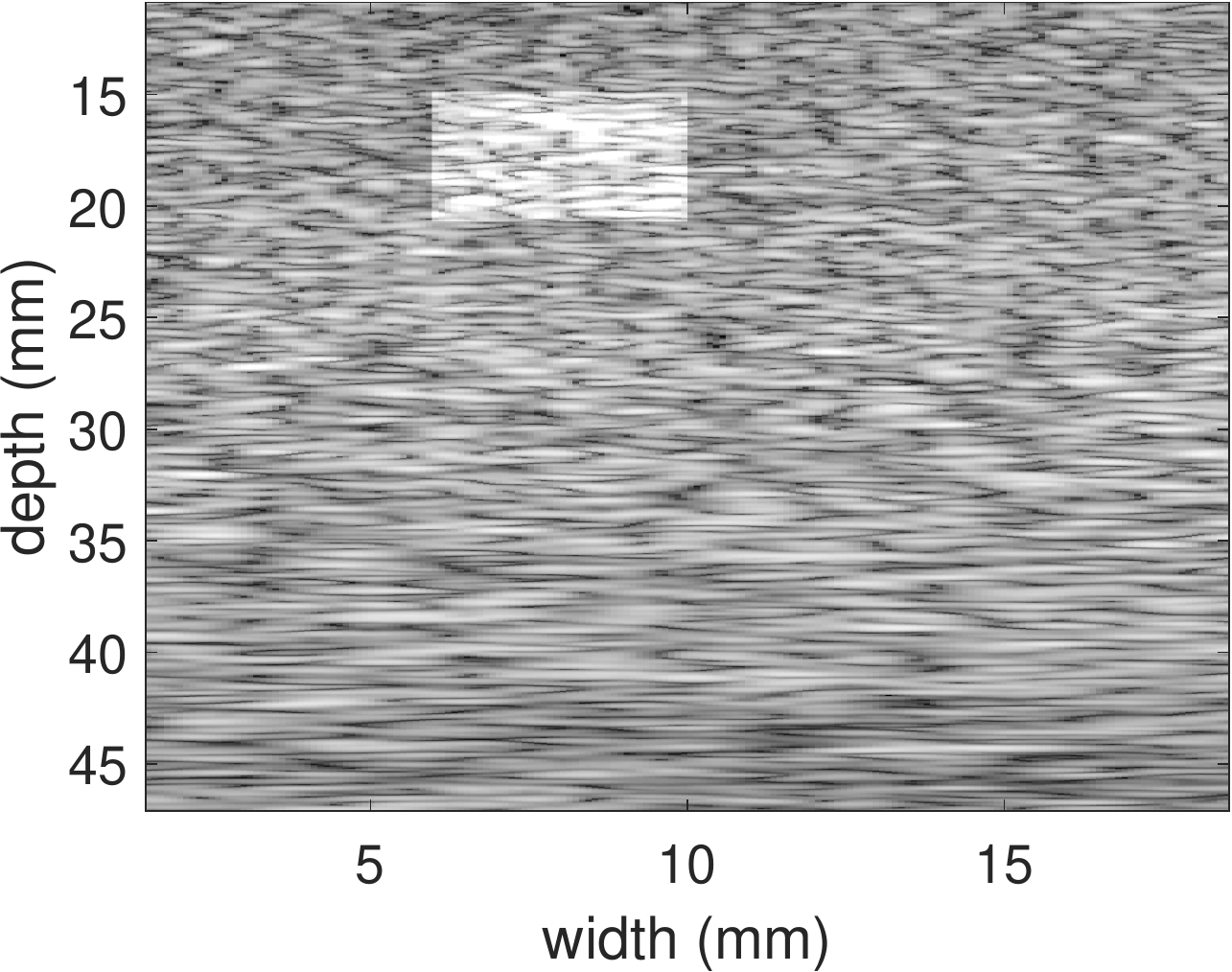} }}%
	\subfigure[Ground truth strain]{{\includegraphics[width=0.16\textwidth,height=0.28\textwidth]{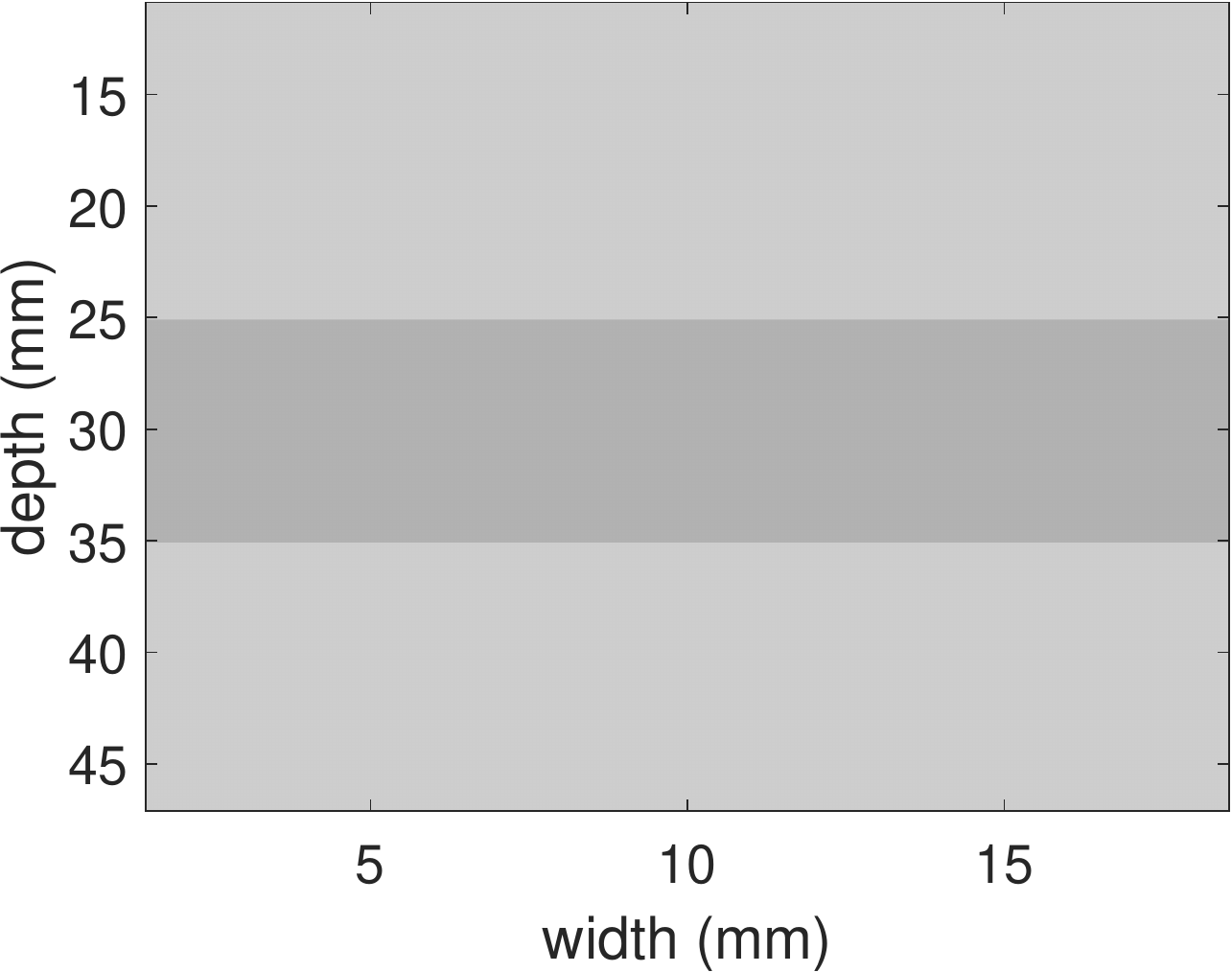} }}%
	\subfigure[Hybrid]{{\includegraphics[width=0.16\textwidth,height=0.28\textwidth]{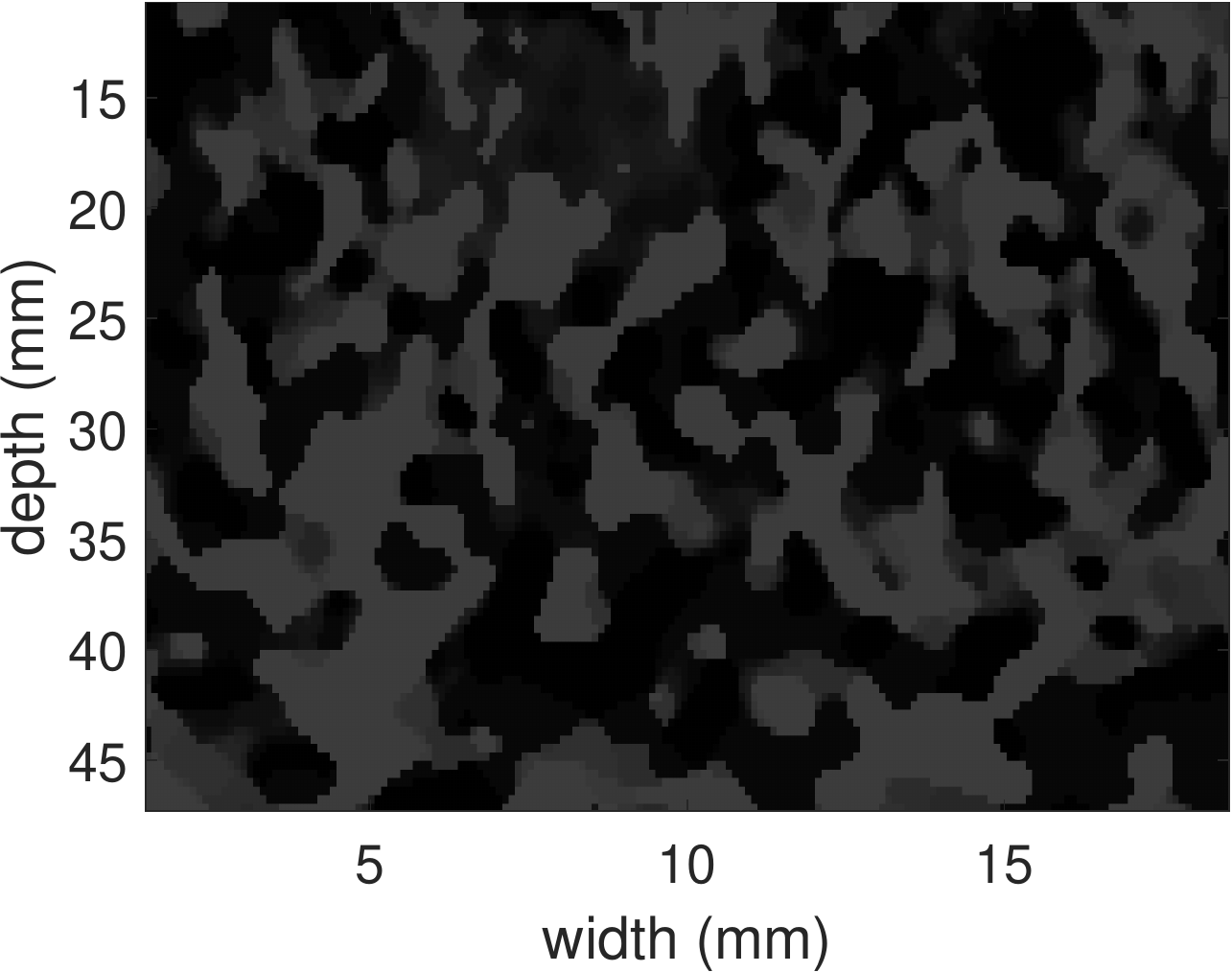} }}%
	\subfigure[GLUE]{{\includegraphics[width=0.16\textwidth,height=0.28\textwidth]{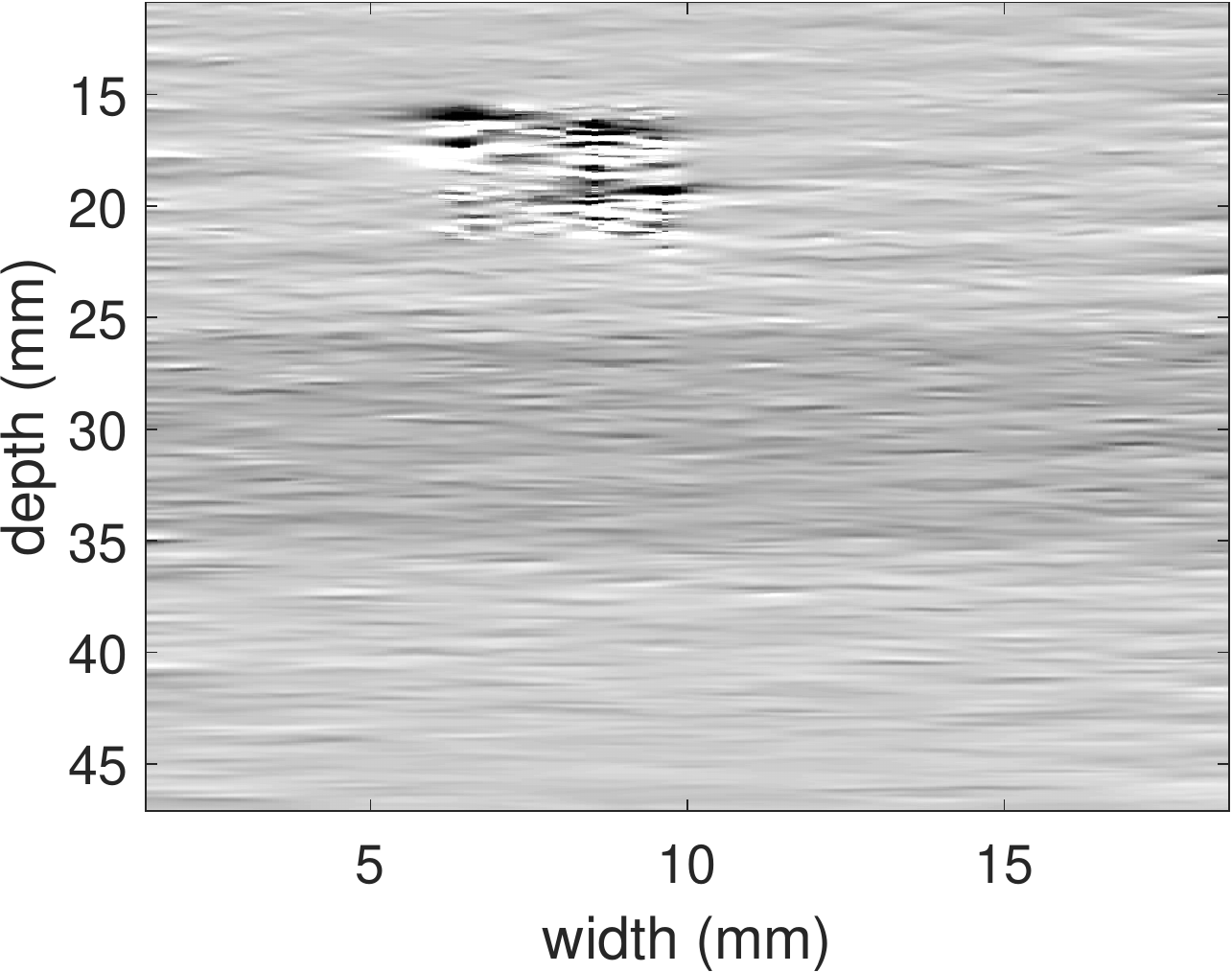} }}%
	\subfigure[rGLUE]{{\includegraphics[width=0.16\textwidth,height=0.28\textwidth]{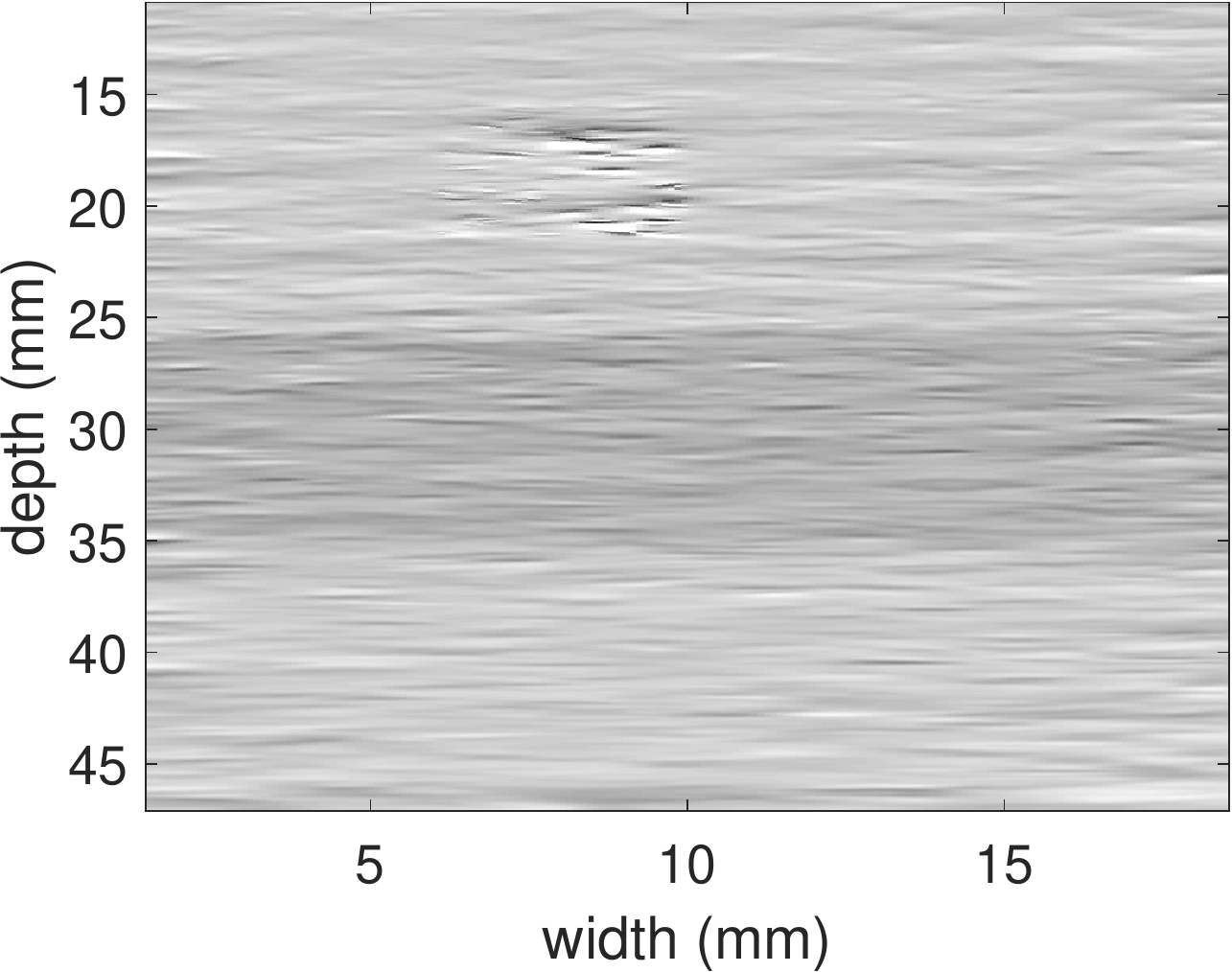} }}
	\subfigure[Strain]{{\includegraphics[width=0.33\textwidth]{results/simu_small/color} }}
	\caption{Results obtained from the low-contrast simulated phantom. Columns 1 and 2 depict the pre- and post-compressed B-mode images, respectively whereas column 3 shows the ground truth strain. Columns 4-6 present the Hybrid, GLUE, and rGLUE strain estimates, respectively. Row 2 shows the color bar.}
	\label{simu_low}
\end{figure*}

The pre- and post-compressed B-mode images along with the strain estimates for the thin-layer and low-contrast phantoms, respectively, have been shown in Figs.~\ref{simu_small} and \ref{simu_low}. Hybrid fails to obtain acceptable results in both cases. The outlier regions are clearly visible in the strain images generated by GLUE. The adaptive weighting technique embedded in rGLUE successfully smoothes out the outlier regions without affecting the target-background contrast. The visual superiority of rGLUE is verified by the RMSE values shown in Table~\ref{table_small_low}. \\

\begin{figure*}
	\centering
	\subfigure[B-mode, pre-deformed]{{\includegraphics[width=0.20\textwidth,height=0.20\textwidth]{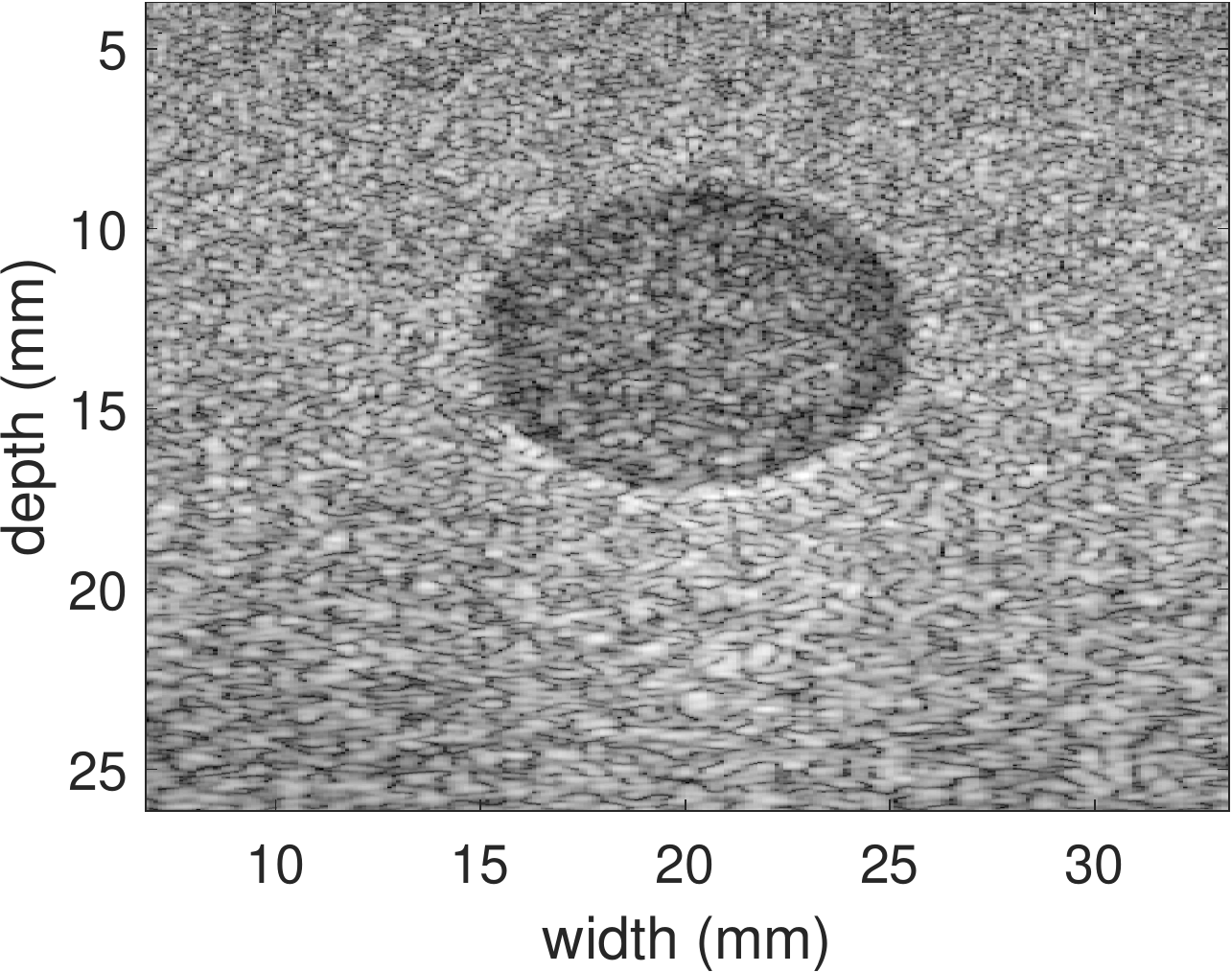} }}%
	\subfigure[B-mode, post-deformed]{{\includegraphics[width=0.20\textwidth,height=0.20\textwidth]{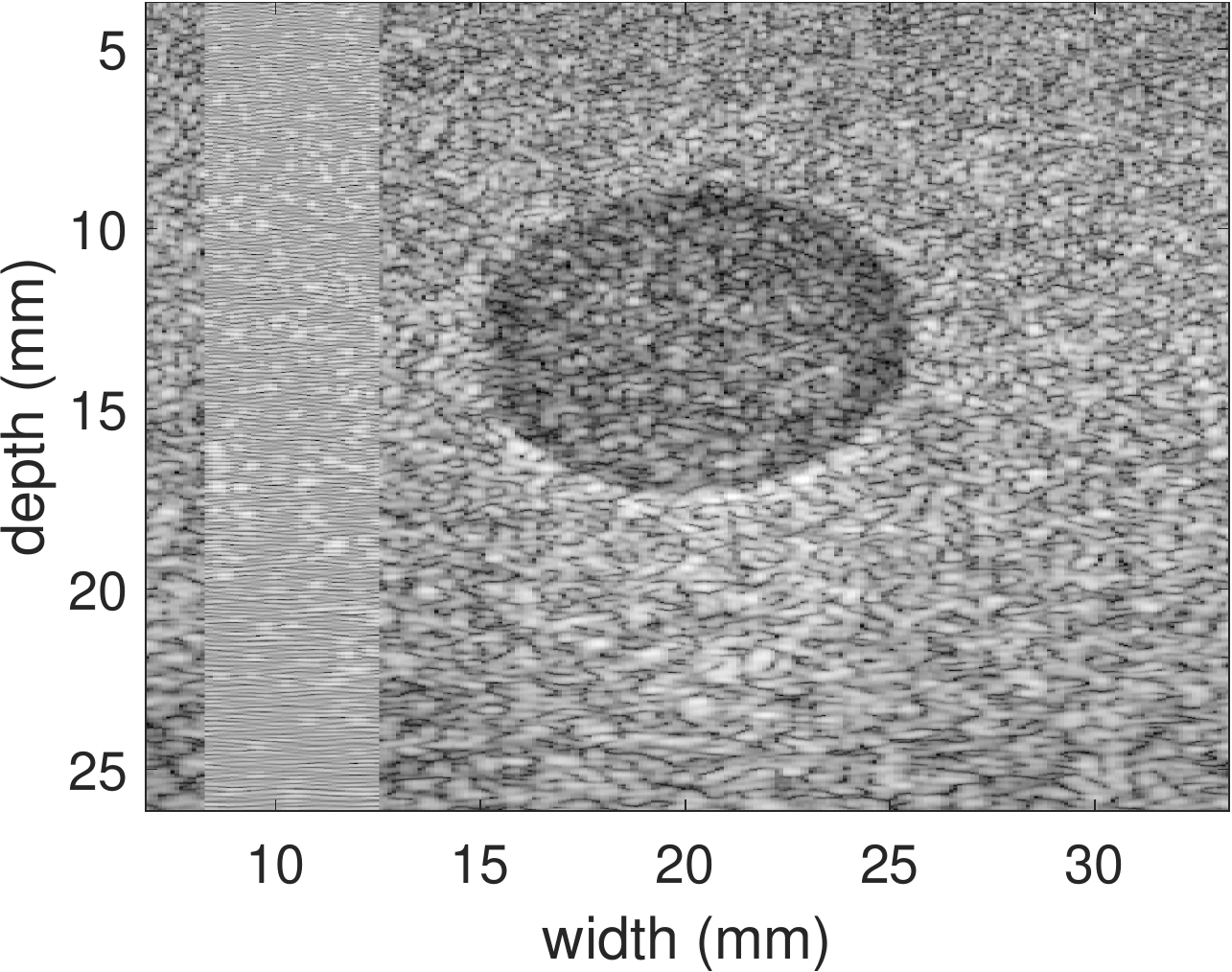} }}%
	\subfigure[Hybrid]{{\includegraphics[width=0.20\textwidth,height=0.20\textwidth]{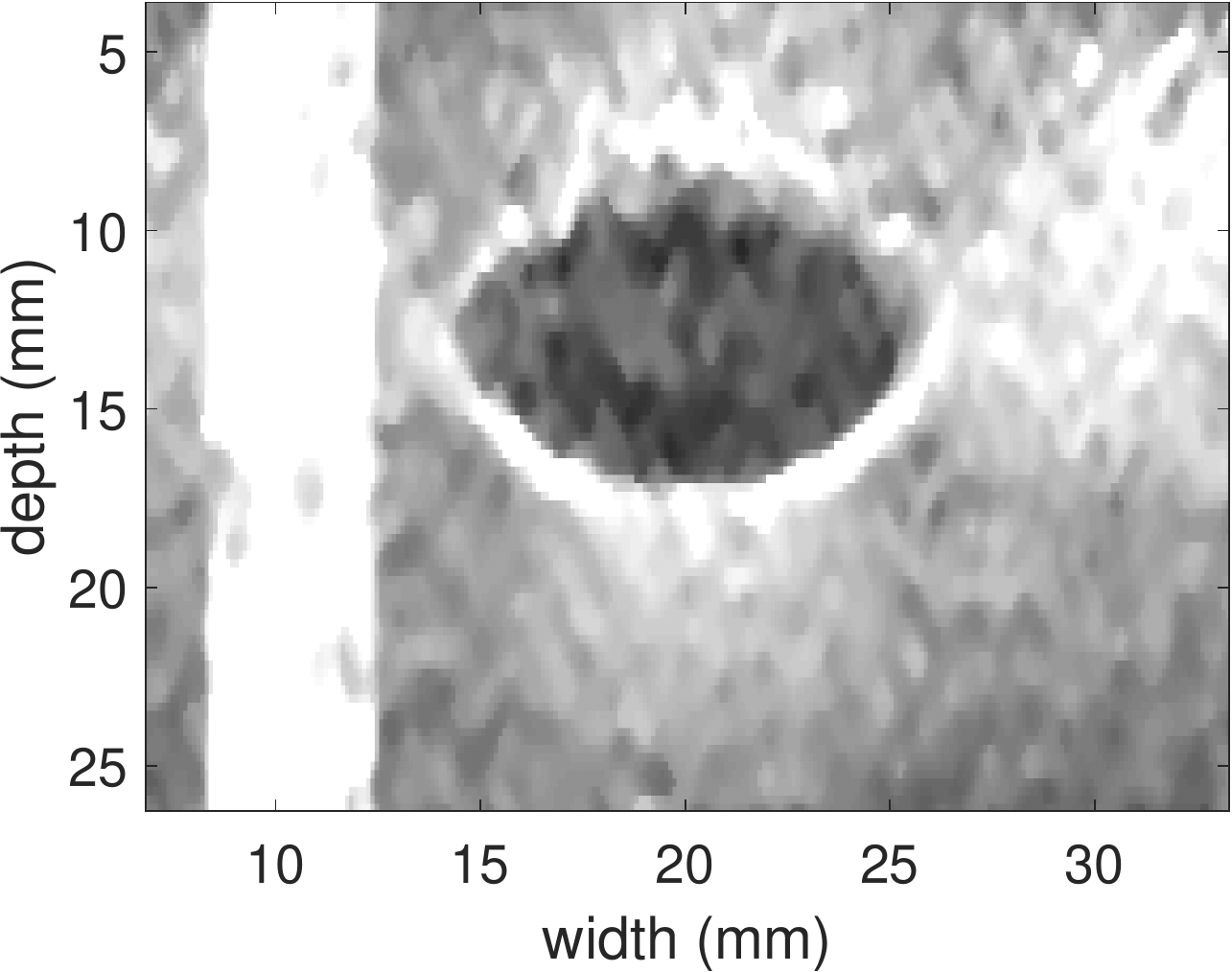} }}%
	\subfigure[GLUE]{{\includegraphics[width=0.20\textwidth,height=0.20\textwidth]{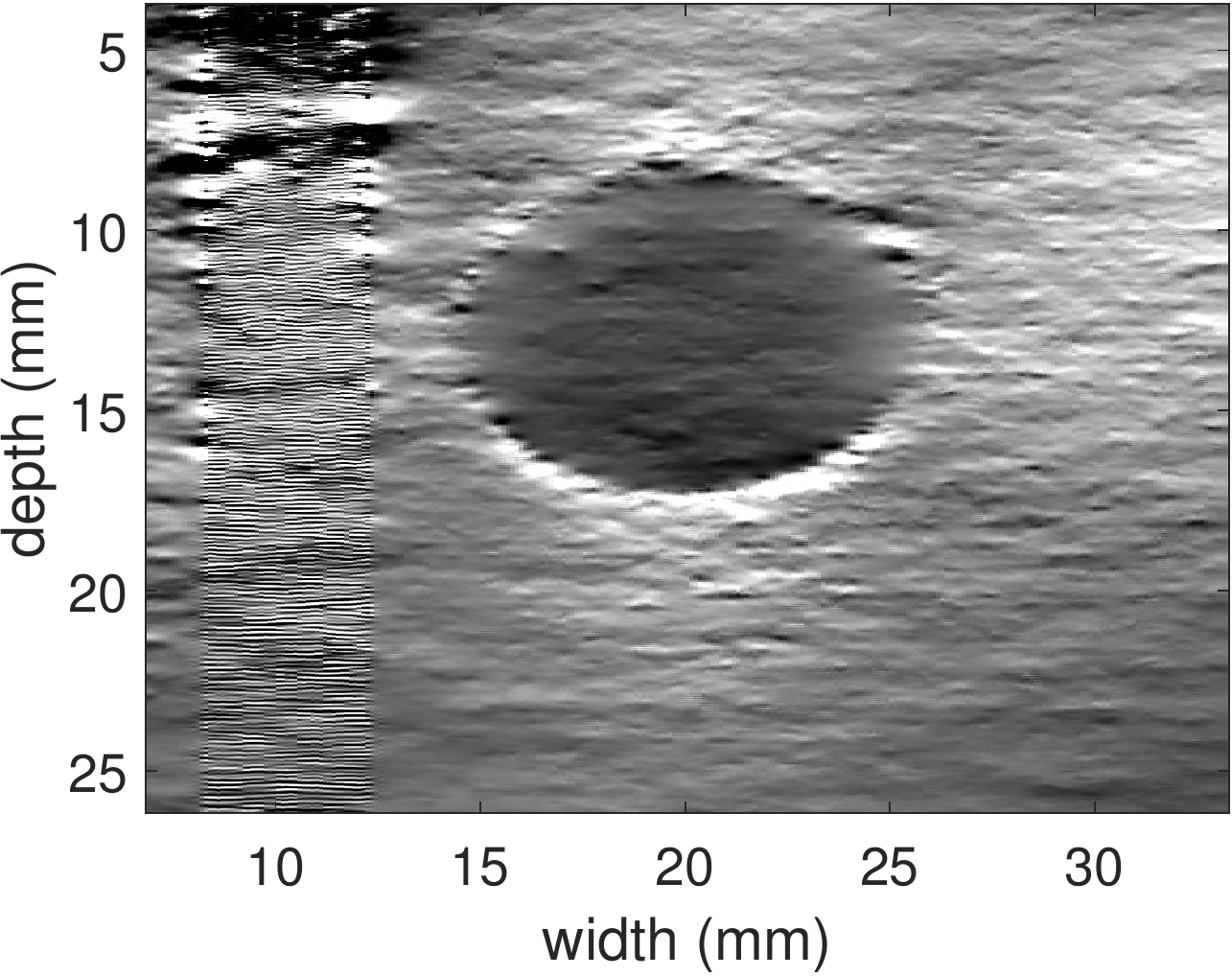} }}%
	\subfigure[rGLUE]{{\includegraphics[width=0.20\textwidth,height=0.20\textwidth]{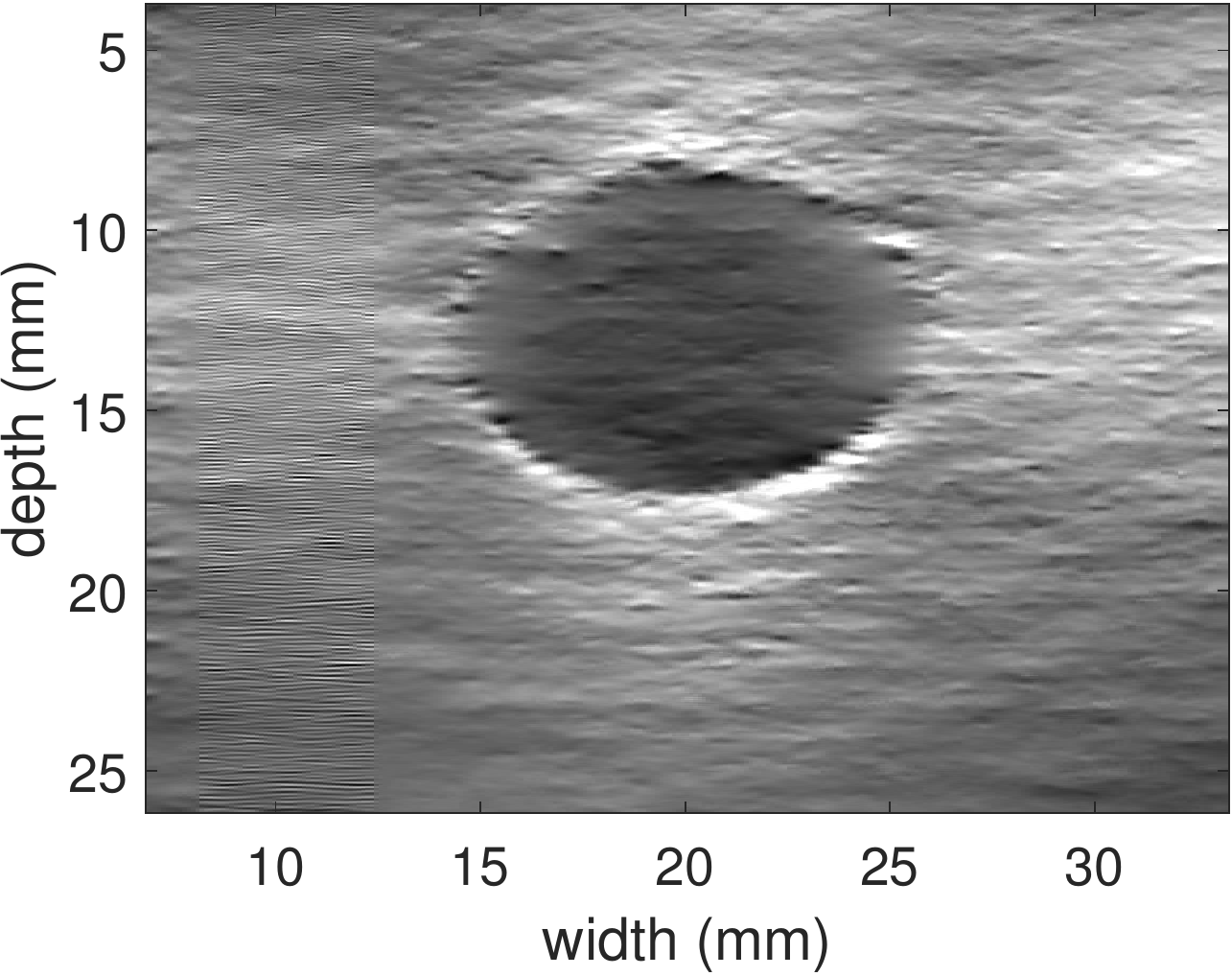} }}
	\subfigure[Strain]{{\includegraphics[width=0.33\textwidth]{results/phantom/color} }}%
	\caption{Phantom results with additive outlier region artificially created by increasing the sample amplitudes of few RF lines. Columns 1 and 2 show the pre- and post-deformed B-mode images, respectively. Columns 3-5 depict the axial strain images obtained from Hybrid, GLUE, and rGLUE, respectively.}
	\label{phan_2}
\end{figure*}

The experimental phantom results in the presence of additive outliers are shown in Fig.~\ref{phan_2} which reveals that both Hybrid and GLUE suffer from noticeable artifacts whereas rGLUE remains almost unaffected by the artificially introduced outliers. A close look at the strain images suggests that Hybrid exhibits strain alteration in different regions of the strain image because of the locally inserted outliers. This might happen due to the consideration of the neighboring samples in displacement estimation. The artifacts in the GLUE strain image appear locally and do not propagate throughout the image. In case of rGLUE, the adaptive weighting of the penalty terms comes into effect and no evident strain artifact is observed.\\

\begin{figure}
	\centering
	\subfigure[GLUE]{{\includegraphics[width=0.24\textwidth,height=0.25\textwidth]{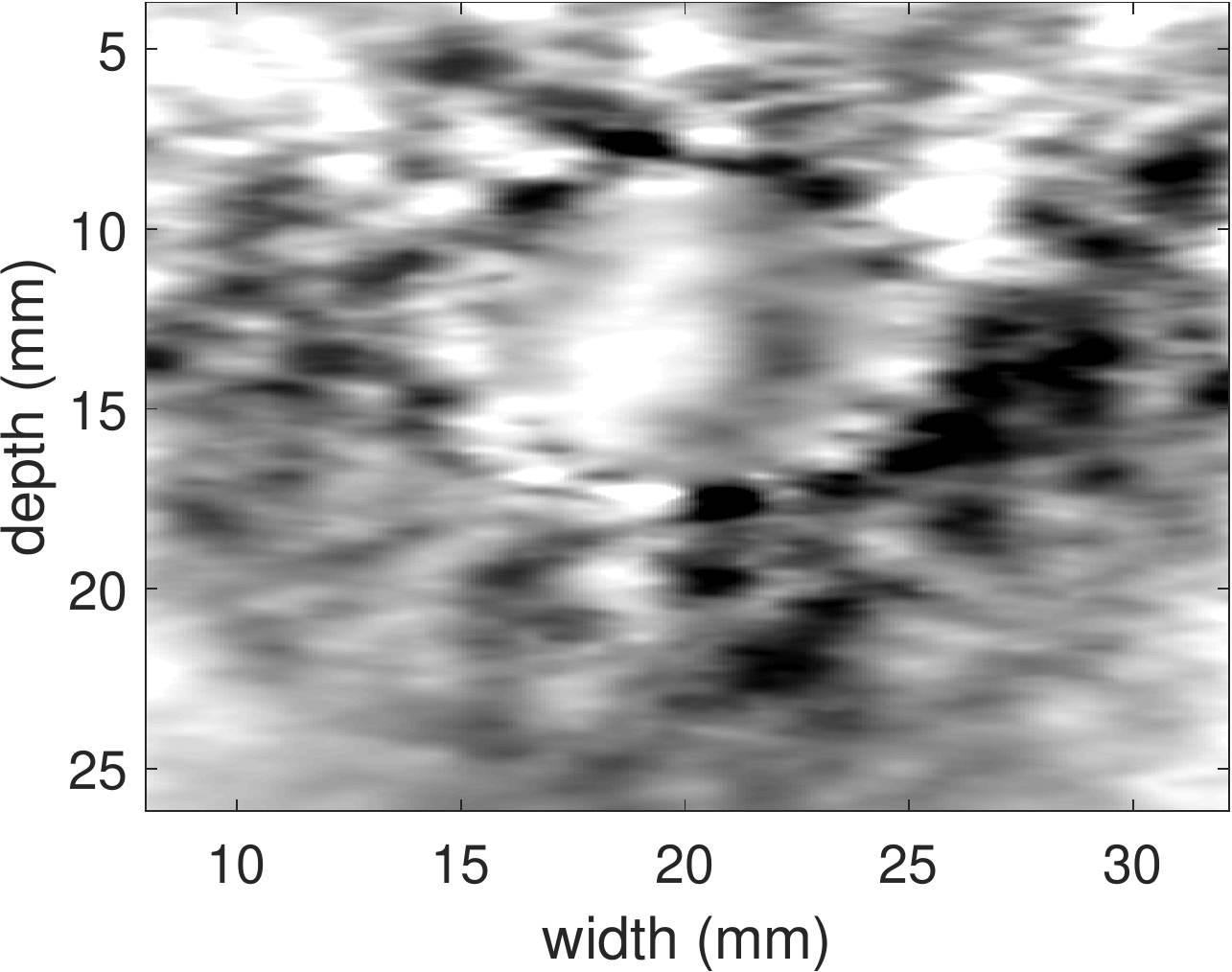}}}%
	\subfigure[rGLUE]{{\includegraphics[width=0.24\textwidth,height=0.25\textwidth]{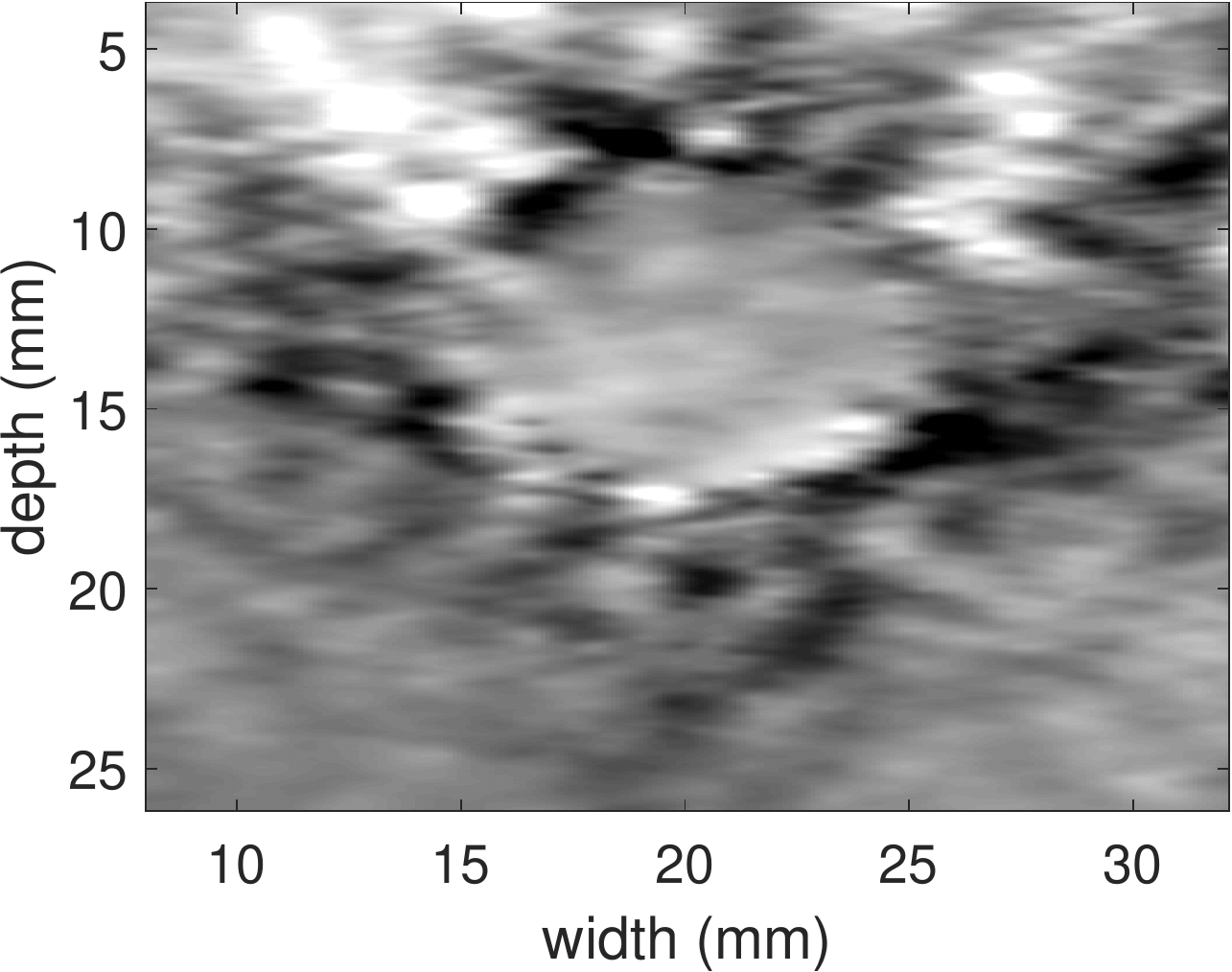}}}
	\subfigure[Strain]{{\includegraphics[width=0.25\textwidth]{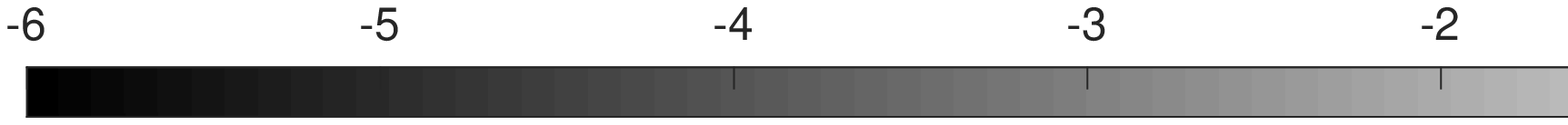}}}
	\caption{Lateral strain images obtained from the breast elastography phantom. Columns 1 and 2 show the strain maps estimated by GLUE and rGLUE, respectively.}
	\label{lateral}
\end{figure}

The lateral strain maps generated from the breast phantom are depicted in Fig.~\ref{lateral}. Both GLUE and rGLUE fail to produce high-quality elastograms due to low resolution in the lateral direction. However, it is evident that rGLUE outperforms GLUE in terms of edge clarity and background smoothness.\\

\begin{figure*}
	\centering
	\subfigure[GLUE, before median filtering]{{\includegraphics[width=0.33\textwidth,height=0.37\textwidth]{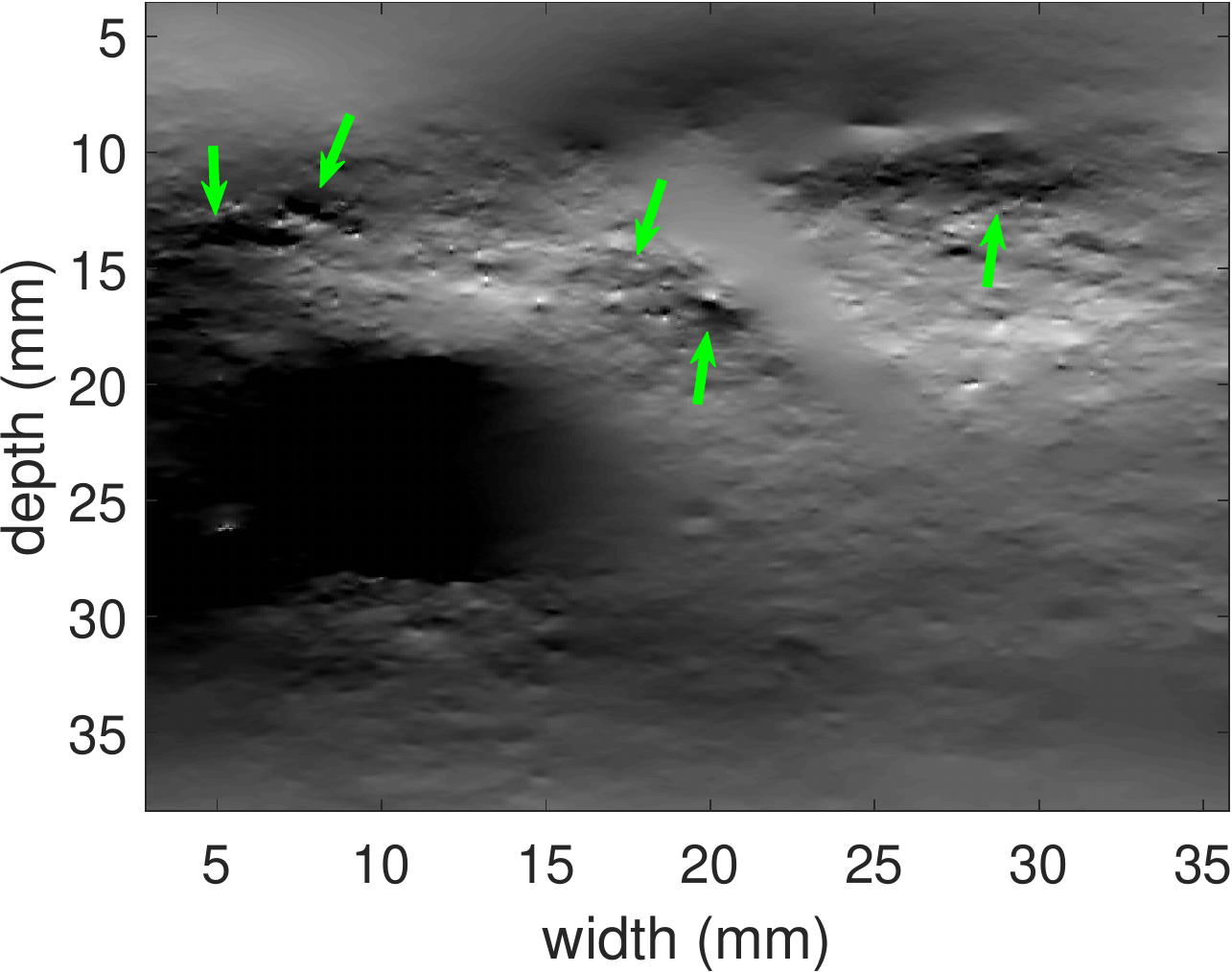} }}%
	\subfigure[GLUE, after median filtering]{{\includegraphics[width=0.33\textwidth,height=0.37\textwidth]{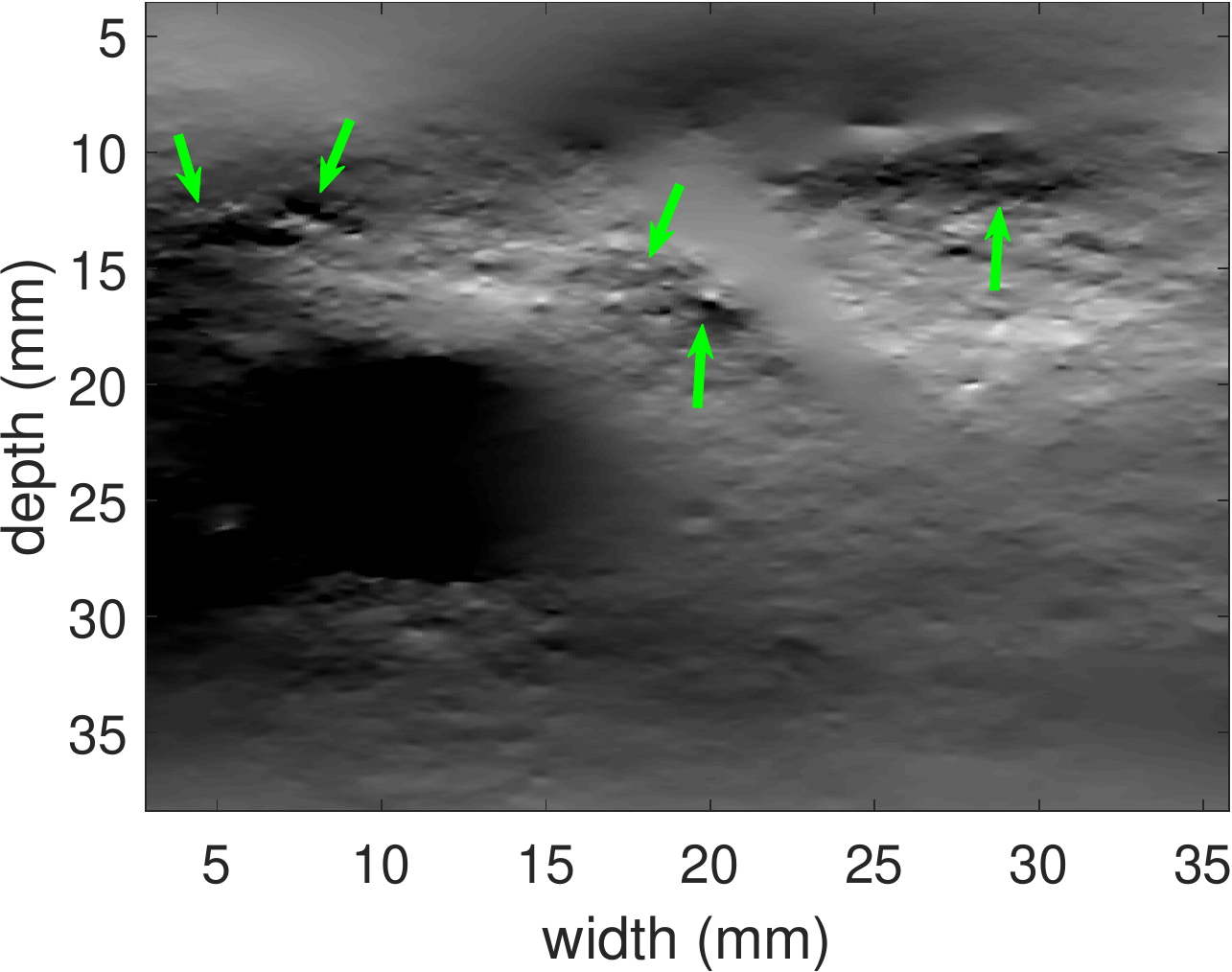} }}%
	\subfigure[rGLUE]{{\includegraphics[width=0.33\textwidth,height=0.37\textwidth]{results/liver_p2/liver_p2_r_new} }}
	\subfigure[Strain]{{\includegraphics[width=0.3\textwidth]{results/liver_p2/color_g}}}
	\caption{Axial strain images for the first liver data. Columns 1-3 correspond to GLUE without and with median filtering, and rGLUE, respectively. The green arrow marks indicate sharp strain fluctuations.}
	\label{median}
\end{figure*}

Fig.~\ref{median} shows the GLUE (before and after median filtering) and rGLUE strain images for the first \textit{in vivo} liver data. We perform a $3 \times 3$ median filtering on the GLUE strain to investigate whether it can suppress the sharp fluctuations and obtain a strain map as good as that of rGLUE. Although the aforementioned median filter smoothens the background, it fails to resolve the issue of sharp strain fluctuation.\\


\begin{table}[tb]  
	\centering
	\caption{RMSE values associated with the strain estimates obtained using the layer phantoms.}
	\label{table_small_low}
	\begin{tabular}{c c c c c c c}
		\hline
		$ $  $ $&  Four-layer & Thin-layer & Low-contrast\\
		\hline
		Hybrid & $6.5 \times 10^{-3}$ & $3.39 \times 10^{-2}$ &  $3.3 \times 10^{-2}$\\
		GLUE & $4.9 \times 10^{-3}$ & $5.6 \times 10^{-3}$ &  $5.8 \times 10^{-3}$\\
		rGLUE & $\mathbf{3.8 \times 10^{-3}}$ & $\mathbf{3.4 \times 10^{-3}}$  & $\mathbf{2.8 \times 10^{-3}}$\\
		\hline
	\end{tabular}
\end{table}

